\documentclass[a4paper, 11pt, reqno]{amsart}
\usepackage{a4wide}
\usepackage[utf8]{inputenc}
\usepackage[english]{babel}
\usepackage{setspace}
\usepackage[euler-digits,euler-hat-accent]{eulervm}

\usepackage{times}
\usepackage{amsmath}
\usepackage{amssymb}
\usepackage{amsthm}
\usepackage{xcolor}
\usepackage{hyperref}
\definecolor{paperlink}{RGB}{0,0,114}
\hypersetup{colorlinks=true,linktoc=all,linkcolor=purple,citecolor=teal,urlcolor=paperlink,pdftitle={Spin--Boson Hamiltonians as Generators of Spin-Dressed Bogoliubov Dynamics},pdfauthor={Nathan Metraud}}
\usepackage{eucal}
\usepackage{esint}

\usepackage{mathtools}
\usepackage{needspace}

\theoremstyle{plain}
\newtheorem{theorem}{Theorem}[section]
\newtheorem{corollary}[theorem]{Corollary}
\newtheorem{proposition}[theorem]{Proposition}
\newtheorem{lemma}[theorem]{Lemma}
\newtheorem{introtheorem}{Theorem}

\theoremstyle{definition}
\newtheorem{definition}[theorem]{Definition}
\newtheorem{introdefinition}{Definition}

\theoremstyle{remark}
\newtheorem{remark}[theorem]{Remark} 
\newtheorem{example}[theorem]{Example} 

\allowdisplaybreaks
\usepackage[a4paper,textwidth=165mm,textheight=230mm,hcentering,headsep=12mm,headheight=14pt]{geometry}

\newcommand{\h}{\mathfrak{h}}
\newcommand{\Hs}{\mathcal{H}}
\newcommand{\Hx}{\widehat{\mathcal{H}}}
\newcommand{\Ar}{\mathfrak{A}}
\newcommand{\Ax}{\widehat{\mathfrak{A}}}
\newcommand{\Sx}{\widehat{S}}
\newcommand{\Fb}{\mathcal{F}_{+}}
\newcommand{\Wa}{\mathcal{W}}
\newcommand{\Sp}{\mathrm{Sp}}
\newcommand{\ASp}{\mathrm{ASp}}
\newcommand{\U}{\mathbf{U}}
\newcommand{\Ui}{\mathbb{U}}
\newcommand{\piF}{\pi_{\mathcal{F}}}
\newcommand{\Bx}{\widehat{B}}
\newcommand{\Kc}{\mathcal{K}}
\newcommand{\Ker}{\operatorname{Ker}}
\newcommand{\piFn}{\pi_{\mathcal{F}}^{(n)}}

\makeatletter
\renewcommand\paragraph{\@startsection{paragraph}{4}{\z@}{.5\linespacing\@plus.7\linespacing}{-.5em}{\normalfont\bfseries}}
\makeatother
\begin{document}
	
\title{Spin--Boson Hamiltonians as Generators of Spin-Dressed Bogoliubov Dynamics}

\author[N. Metraud]{Nathan Metraud}

\address[N. Metraud]{Department of Mathematics, Universidad del Pais Vasco.}
\email{nathan.metraud@ehu.eus}

\begin{abstract}
Following Bach, Lieb and Solovej, quadratic Hamiltonians can be defined as generators of unitary groups implementing one-parameter groups of Bogoliubov transformations. We develop an analogous dynamical definition of generalized spin--boson Hamiltonians. For a single spin level, we review results of Guichardet and Dereziński: the van Hove Hamiltonians, including the renormalized ones, are exactly the generators of affine Bogoliubov dynamics that do not mix creation and annihilation operators. For several spin levels, there is no analogue of this group, and we introduce spin-dressed implementations, that is to say the free flow of the fields is implemented up to bounded operator-valued displacements, which vanish at a rate measured by a so-called regularity scale and have a spin observable, the coupling map, as derivative at time zero. The one-particle Hamiltonian $\Upsilon$ and the coupling map are determined by the Hamiltonian, and the field obeys a quantum Langevin equation. Spin-dressed implementations in Fock space, at the finest scale and for form factors in $D(\Upsilon^{-1/2})$, are exactly the generalized spin--boson Hamiltonians, unique up to the spin Hamiltonian. At the roughest scale, the renormalized spin--boson Hamiltonian of a two-level system with ultraviolet-singular form factor and arbitrary tunnelling is a spin-dressed implementation, and the definition alone yields asymptotic Weyl operators on the test functions along which the coupling map is integrable, a dense set under standard assumptions.

\end{abstract}

\maketitle
\tableofcontents

\section*{Introduction}
\label{sec:intro}

Spin--boson models describe a quantum system with finitely many levels, the ``spin'', linearly coupled to a bosonic field. Up to a rotation of the spin basis \cite{LCD87}, the standard spin--boson Hamiltonian reads
\begin{equation}
	H_{\mathrm{SB}}=\frac{\varepsilon}{2}\,\sigma_z\otimes\mathbf 1+\mathbf 1\otimes\mathrm d\Gamma(\omega)+\sigma_x\otimes\phi(v)\qquad\text{on }\mathbb C^{2}\otimes\Fb .
	\label{eq:intro-SB}
\end{equation}
Here, $\sigma_x,\sigma_y,\sigma_z$ are the Pauli matrices, $\varepsilon\in\mathbb R$, $\Fb$ is the bosonic Fock space over a complex separable Hilbert space $\h$, for instance $L^{2}(\mathbb R^{d})$, $\mathrm d\Gamma(\omega)=\int\omega(k)\,a^{*}(k)a(k)\,\mathrm dk$ is the second quantization of a dispersion relation $\omega\geq0$, and $\phi(v)=\int\big(\overline{v(k)}\,a(k)+v(k)\,a^{*}(k)\big)\,\mathrm dk$ is the field operator depending on a form factor $v$. Note that the free spin Hamiltonian $\frac\varepsilon2\sigma_z$ does not commute with the coupling matrix $\sigma_x$. More generally, we consider
\begin{equation}
	H=E\otimes\mathbf 1+\mathbf 1\otimes\mathrm d\Gamma(\Upsilon)+\sum_{l=1}^{m}\tau_l\otimes\phi(w_l)\qquad\text{on }\mathbb C^{n}\otimes\Fb ,
	\label{eq:intro-gSB}
\end{equation}
where $n,m\in\mathbb N$, the matrices $E,\tau_1,\dots,\tau_m$ belong to the real space $M_n^{\mathrm{sa}}\subset M_n$ of self-adjoint $n\times n$ matrices, $\Upsilon$ is a self-adjoint operator on $\h$, the one-particle Hamiltonian, and $w_1,\dots,w_m\in\h$ are the form factors \cite{AH97,Lon22}. $\Omega$ is the vacuum of $\Fb$, scalar products are antilinear in their first argument, $a(f)$ and $a^{*}(f)$ are the annihilation and creation operators, with $[a(f),a^{*}(g)]=\langle f,g\rangle$, and $\phi(f)$ is the closure of $a(f)+a^{*}(f)$, $f\in\h$, so that $[\phi(f),\phi(g)]=2i\,\mathrm{Im}\langle f,g\rangle$. Here $\mathrm d\Gamma(\Upsilon)$ is the second quantization of $\Upsilon$ and $u_t\doteq e^{it\Upsilon}$, $t\in\mathbb R$, is the free flow generated by $\Upsilon$.

Spin--boson models describe dissipative two-level systems \cite{LCD87} and the coupling between the internal levels of trapped ions and their collective vibrational modes \cite{CZ95,MS99,LBMW03}, it is the basis for the architecture of quantum gates in quantum computing. They have been studied in mathematical physics: equilibrium states and return to equilibrium \cite{FNV88,JP96}, ground states \cite{Spo89,HH11}, scattering theory \cite{DG99,FGS02,DRGK15} and, more recently, ultraviolet-singular form factors and their renormalization \cite{DM20a,Lon22,Lon23,LL25,HLVM25,ALLVM25,FHVM26}. If $\Upsilon\geq0$, $\Ker\Upsilon=\{0\}$ and $w_l\in D(\Upsilon^{-1/2})$ for all $l$, we say that the form factors are \emph{regular}, and \eqref{eq:intro-gSB} is then self-adjoint on $\mathbb C^{n}\otimes D(\mathrm d\Gamma(\Upsilon))$ by the Kato--Rellich theorem. For ultraviolet-singular form factors, $w_l\notin\h$, it is only a formal expression, and a self-adjoint operator has to be constructed by renormalization. For commuting couplings, its dynamics can still be given a meaning (Theorem~\ref{thm:intro-ren} and Proposition~\ref{prop:sd-commuting}).

Quadratic Hamiltonians admit an intrinsic, dynamical definition. Following Bach, Lieb and Solovej \cite{BLS94}, and Bruneau and Dereziński \cite[Definition~3.3]{BD07}, see also \cite{DG13,Der17}, a quadratic Hamiltonian can be defined as the generator of a strongly continuous unitary group implementing a one-parameter group of Bogoliubov transformations. For bosons, this means that $H$ is a quadratic Hamiltonian if it is a self-adjoint operator on $\Fb$ such that
\begin{equation}
	e^{itH}\phi(f)e^{-itH}=\phi(R_tf),\qquad f\in\h,\ t\in\mathbb R,
	\label{eq:intro-BLS}
\end{equation}
for some one-parameter group $(R_t)_{t\in\mathbb R}$ of Bogoliubov transformations, i.e., of bounded invertible real-linear maps of $\h$ preserving the symplectic form $\mathrm{Im}\langle\cdot,\cdot\rangle$\footnote{In this paper, outside this introduction and the main results section, we will adopt the self-dual point of view.}. In \eqref{eq:intro-BLS}, no explicit expression for $H$ is assumed, and $H$ is determined by $(R_t)_{t\in\mathbb R}$ up to an additive constant, by irreducibility of the Fock representation. The same group structure is behind the diagonalization of quadratic Hamiltonians by Bogoliubov transformations \cite{NNS16} and by the Brockett--Wegner flow \cite{BB16}. For spin--boson models this flow does not close, and the available results are perturbative \cite{BMS26}.

The aim of this paper is an analogous definition of the Hamiltonians \eqref{eq:intro-gSB} through their Heisenberg dynamics. For $n\geq2$, the dynamics is not given by a group of transformations of the phase space fixed in advance (see Remark~\ref{rem:sd-cocycle} for the heuristics, and Remark~\ref{rem:sd-nogo} for a precise statement in the case of spin-valued displacements), but the equation of motion keeps its structure: formally, $e^{itH}(\mathbf 1\otimes\phi(f))e^{-itH}$ is the freely evolved field $\mathbf 1\otimes\phi(u_tf)$ plus a bounded displacement $\Delta_t(f)$, built from the spin observables evolved by the interacting dynamics \eqref{eq:intro-langevin}. We take this structure as the definition of a \emph{spin-dressed implementation} of the free flow (Definition~\ref{def:intro-sd}), in which no expression for $H$ is given. A \emph{regularity scale} $L$, a function of $\Upsilon$, measures how fast the displacements vanish as $t\to0$. The scale $L=0$ is the finest, and $L=\Upsilon$ the roughest one considered. We want the definition to
\begin{enumerate}
	\item[(a)] determine the operators \eqref{eq:intro-gSB}, up to a spin Hamiltonian, whenever these are self-adjoint by the Kato--Rellich theorem,
	\item[(b)] remain meaningful for renormalized Hamiltonians.
\end{enumerate}
The spin Hamiltonian in (a) is an explicit ambiguity: it is the self-adjoint part of the commutant of the fields.

\paragraph{Plan.}Section~\ref{sec:main} states the main results, for the Fock representation. Sections~\ref{sec:affine} and~\ref{sec:spin} prove them. Theorem~\ref{thm:intro-vH} is the scalar case $n=1$ and mostly reformulates results of Guichardet and Dereziński (Section~\ref{sec:affine}). Section~\ref{sec:spin} treats $n\geq2$ in the self-dual formalism of Araki \cite{Araki,Quadratic}: Theorems~\ref{thm:intro-structure} and~\ref{thm:E} (structure, quantum Langevin equation and asymptotic Weyl operators) hold in every regular representation of the canonical commutation relations, but are exposed for the Fock representation for clarity. Theorem~\ref{thm:intro-fock} achieves (a), and Theorem~\ref{thm:intro-ren}, extended to all commuting couplings by Proposition~\ref{prop:sd-commuting}, achieves (b) for these models. Appendix~\ref{app:dictionary} recalls the cohomological vocabulary and the notation of \cite{Gui72,Der03}.

\paragraph{Relation to previous work.}
For ultraviolet-singular form factors, the renormalized Hamiltonians with commuting couplings, under an infrared condition, and their cutoff approximation are due to \cite{HLVM25}. At the scale $L=\Upsilon$ of Definition~\ref{def:intro-sd}, the form factors lie in the extrapolation space $\h_{-1}(\Upsilon)$ by Theorem~\ref{thm:intro-structure}(ii), which is exactly the ultraviolet threshold of the self-energy renormalization of spin--boson models with normal couplings \cite[Theorems~2.5 and~2.8]{HLVM25}. Beyond it, for the standard spin--boson model, the self-energy renormalized cutoff Hamiltonians converge, up to unitaries independent of the qubit, to a limit which is independent of the qubit \cite[Corollary~4.4]{DM20a}.

\section{Main Results}
\label{sec:main}

The results are stated in the Fock representation, some of the corresponding results in the technical sections are more general. Namely, Theorems~\ref{thm:intro-structure} and~\ref{thm:E} are proved in Section~\ref{sec:spin} for arbitrary regular representations of the canonical commutation relations, while Theorems~\ref{thm:intro-vH}, \ref{thm:intro-fock} and~\ref{thm:intro-ren} are specific to Fock space.

\subsection{Van Hove Hamiltonians as Affine Bogoliubov Dynamics}
\label{subsec:main-vH}

For $n=1$, \eqref{eq:intro-gSB} is, up to an additive constant, the van Hove Hamiltonian $\mathrm d\Gamma(\Upsilon)+\phi(\theta)$ with $\theta\doteq\sum_l\tau_lw_l$ \cite{VH52,Der03}. The canonical commutation relations give, formally,
\begin{equation}
	e^{itH}\phi(f)e^{-itH}=\phi(u_tf)+2\,\mathrm{Im}\langle f,g_t\rangle\,\mathbf 1,\qquad g_t\doteq\int_0^{t}e^{-ir\Upsilon}\theta\,\mathrm dr,\qquad f\in\h,\qquad t\in\mathbb R^+ .
	\label{eq:intro-vH}
\end{equation}
The dynamics maps fields to fields, up to scalars: it is a group of \emph{affine} Bogoliubov transformations, whose linear part $f\mapsto u_tf$ is complex-linear. We take this as a definition.

\begin{introdefinition}[$N$-diagonal affine quadratic Hamiltonians]
	\label{def:intro-affine}
	A self-adjoint operator $H$ on $\Fb$ is an $N$-diagonal affine quadratic Hamiltonian if, for every $t\in\mathbb R$, there are a unitary operator $R_t$ on $\h$ and a vector $\xi_t\in\h$ such that
	\[
	e^{itH}\phi(f)e^{-itH}=\phi(R_tf)+2\,\mathrm{Im}\langle f,\xi_t\rangle\,\mathbf 1,\qquad f\in\h .
	\]
\end{introdefinition}

The adjective \emph{$N$-diagonal} refers to the fact that the linear part $f\mapsto R_tf$ of the dynamics is implemented by the second quantization of $R_t$, which conserves the number of bosons, whereas the scalar displacements do not. By \eqref{eq:sd-scalar-fields} and \eqref{eq:ASp0}, Definition~\ref{def:intro-affine} is Definition~\ref{def:affine-quad}.

For a self-adjoint operator $A$ on $\h$, let $\h_{-1}(A)$ be the completion of $\h$ for the norm $\|(|A|+\mathbf 1)^{-1}\cdot\|_{\h}$. The scalar product of $\h$ extends by continuity to a sesquilinear pairing between $\h_{-1}(A)$ and $D(A)$, still denoted $\langle \cdot,\cdot\rangle$, with $\langle f,w\rangle \doteq \overline{\langle w,f\rangle}$, and $A$ extends to a bounded map $\h\to\h_{-1}(A)$, whose range we denote by $A\h$ (Subsection~\ref{subsec:affine-ham}). For $\theta\in\h_{-1}(\Upsilon)$, the vector $g_t$ of \eqref{eq:intro-vH} is still a well-defined element of $\h$, see \eqref{eq:ft}.

\begin{introtheorem}[Van Hove Hamiltonians]
	\label{thm:intro-vH}
	\leavevmode
	\begin{enumerate}
		\item[(i)] Let $H$ be an $N$-diagonal affine quadratic Hamiltonian. Then there is a unique pair $(\Upsilon,\theta)$, where $\Upsilon$ is a self-adjoint operator on $\h$ and $\theta\in\h_{-1}(\Upsilon)$, such that $R_t=e^{it\Upsilon}$ and $\xi_t=g_t$, with $g_t$ given by \eqref{eq:intro-vH}, for all $t\in\mathbb R$.
		\item[(ii)] Conversely, for every self-adjoint operator $\Upsilon$ on $\h$ and every $\theta\in\h_{-1}(\Upsilon)$, there is an $N$-diagonal affine quadratic Hamiltonian $H_{\Upsilon,\theta}$ with data $(\Upsilon,\theta)$, unique up to an additive constant. If $\Upsilon\geq0$ and $\Ker\Upsilon=\{0\}$, these are the van Hove Hamiltonians of \cite{Der03}. If, in addition, $\theta\in\h$, then $H_{\Upsilon,\theta}$ can be chosen as the closure of $\mathrm d\Gamma(\Upsilon)+\phi(\theta)$ on $D(\mathrm d\Gamma(\mathbf 1+\Upsilon))$.
		\item[(iii)] Let $\Upsilon\geq0$ with $\Ker\Upsilon=\{0\}$ and $\theta\in\h_{-1}(\Upsilon)$. Then $\Omega\in D(H_{\Upsilon,\theta})$ if and only if $\theta\in\h$, and $H_{\Upsilon,\theta}$ has a ground state if and only if $\theta\in\Upsilon\h$.
	\end{enumerate}
\end{introtheorem}

\begin{proof}
See Theorem~\ref{thm:affine} and Proposition~\ref{prop:hierarchy}(i),(iii)
\end{proof}
Theorem~\ref{thm:intro-vH} mostly reformulates known results: (i) and the existence part of (ii) belong to Guichardet's theory of representations of type~(S) of~$\mathbb R$ \cite{Gui72} (Appendix~\ref{app:dico-Gui}). The identification with the closure of $\mathrm d\Gamma(\Upsilon)+\phi(\theta)$ in (ii) and the ground-state criterion in (iii) are in Dereziński's theory of van Hove Hamiltonians \cite{Der03}, see also \cite{FF24} and Remark~\ref{rem:vanhove}. The criterion for $\Omega\in D(H_{\Upsilon,\theta})$ in (iii) is proved here (Proposition~\ref{prop:hierarchy}(i) and Remark~\ref{rem:hierarchy}).

Two features of Theorem~\ref{thm:intro-vH} contribute to the philosophy of the case $n\geq2$. First, $H_{\Upsilon,\theta}$ is defined for every $\theta\in\h_{-1}(\Upsilon)$, without cutoff or counterterm, although for $\theta\notin\h+\Upsilon^{1/2}\h$ the formal expression $\mathrm d\Gamma(\Upsilon)+\phi(\theta)$ requires an infinite renormalization \cite[Sects.~1.1 and~3.4]{Der03} (Remarks~\ref{rem:vanhove} and~\ref{rem:hierarchy}). Second, ultraviolet singularities are removed by a Weyl dressing, in the spirit of Gross and Nelson \cite{Gro62,Nel64}: if $\Upsilon\geq0$ and $\Ker\Upsilon=\{0\}$, then, for every $\kappa>0$, $\theta=\mathbf 1_{[0,\kappa]}(\Upsilon)\theta+\Upsilon q$ with $q\doteq\Upsilon^{-1}\mathbf 1_{(\kappa,\infty)}(\Upsilon)\theta\in\h$ (Step~1 of the proof of Proposition~\ref{prop:sd-commuting}(ii)), and the part $\Upsilon q$ is a coboundary in the sense of Lemma~\ref{lem:cocycle}(iii), so that the cohomology class of $\theta$ in $\h_{-1}(\Upsilon)/\Upsilon\h$, which decides the existence of a ground state by Proposition \ref{prop:hierarchy} (iii), only depends on its infrared part (Remark~\ref{rem:vanhove}). This is the mechanism we use for renormalized spin--boson Hamiltonians (Proposition~\ref{prop:sd-commuting}(ii) and Remark~\ref{rem:sd-dressing}).

\subsection{Spin-Dressed Implementations and Spin--Boson Hamiltonians}
\label{subsec:main-sd}

For $n\geq2$, the same formal computation gives, with
\[
\Theta(f)\doteq\sum_{l=1}^{m}2\,\mathrm{Im}\langle f,w_l\rangle\,\tau_l\in M_n^{\mathrm{sa}},\qquad f\in\h,
\]
the identity $i[H,\mathbf 1\otimes\phi(f)]=\mathbf 1\otimes\phi(i\Upsilon f)+\Theta(f)\otimes\mathbf 1$, hence, after integration,
\begin{equation}
	e^{itH}\big(\mathbf 1\otimes\phi(f)\big)e^{-itH}=\mathbf 1\otimes\phi(u_tf)+\int_0^{t}e^{isH}\big(\Theta(u_{t-s}f)\otimes\mathbf 1\big)e^{-isH}\,\mathrm ds .
	\label{eq:intro-langevin}
\end{equation}
The field evolves freely, up to a displacement generated by the spin observables along the trajectory: \eqref{eq:intro-langevin} is a quantum Langevin equation for the field, see \cite{FKM65,GC85}. The displacement is a scalar for $n=1$, as in \eqref{eq:intro-vH}, and a bounded operator which, in general, does not commute with the fields for $n\geq2$. The dynamics is then no longer a group of transformations in the phase space. Instead, it incorporates a displacement that satisfies a cocycle identity twisted by the interacting dynamics itself (Theorem~\ref{thm:intro-structure}(i) and Remark~\ref{rem:sd-cocycle}). We take this structure as the definition.

\begin{introdefinition}[Spin-dressed implementations]
	\label{def:intro-sd}
	Let $\Upsilon$ be a self-adjoint operator on $\h$, and let $L\doteq F(\Upsilon)$ for a real Borel function $F$ such that $|F(x)|\leq c(1+|x|)$, $c<\infty$. Let $\Theta:D(L)\to M_n^{\mathrm{sa}}$ be a map. A self-adjoint operator $H$ on $\mathbb C^{n}\otimes\Fb$ \emph{$\Theta$-implements the free flow at scale $L$} if, for every $t\in\mathbb R$, there is a bounded real-linear map $\Delta_t:\h\to\mathcal B(\mathbb C^{n}\otimes\Fb)$ with self-adjoint values such that:
	\begin{enumerate}
		\item[(D1)] $e^{itH}(\mathbf 1\otimes\phi(f))e^{-itH}=\mathbf 1\otimes\phi(u_tf)+\Delta_t(f)$ for all $f\in\h$,
		\item[(D2$_L$)] there is $C_\Delta<\infty$ such that $\|\Delta_t(f)\|\leq C_\Delta|t|\,\|(|L|+\mathbf 1)f\|$ for all $f\in D(L)$ and $|t|\leq1$, and $t^{-1}\Delta_t(f)\to\Theta(f)\otimes\mathbf 1$ strongly as $t\to0$, for every $f\in D(L)$.
	\end{enumerate}
	We then call $H$ a \emph{spin-dressed implementation} of the free flow at scale $L$, with \emph{coupling map} $\Theta$ and \emph{displacements} $\Delta_t$. At the scale $L=0$, we simply say that $H$ \emph{$\Theta$-implements the free flow}.
\end{introdefinition}

The displacements may not commute with the fields nor take values in $M_n\otimes\mathbf 1$. The growth condition on $F$ ensures $D(\Upsilon)\subseteq D(L)$, hence $\h_{-1}(L)\subseteq\h_{-1}(\Upsilon)$ (Corollary~\ref{cor:sd-scalar-scale}). For $n=1$, $\Upsilon\geq0$ and $\Ker\Upsilon=\{0\}$, the scales $L=0$, $L=\Upsilon^{1/2}$ and $L=\Upsilon$ encode the van Hove Hamiltonians $H_{\Upsilon,\theta}$ with $\theta$ in $\h$, $\h+\Upsilon^{1/2}\h$ and $\h_{-1}(\Upsilon)$, respectively (Remark~\ref{rem:sd-scales}).

In Section~\ref{sec:spin} we use the self-dual point of view: identifying $f\in\h$ with $(f,Cf)$, where $C$ is a fixed antiunitary involution of $\h$, in the real phase space $\Hs_{\Ar}$ of Subsection~\ref{subsec:affine-setting}, Definition~\ref{def:intro-sd} is Definition~\ref{def:sd-impl} for the Fock representation and the free flow $e^{it\Upsilon}\oplus e^{-it\Upsilon^{\top}}$ (with $\Upsilon^{\top}\doteq C\Upsilon C$), with the same coupling map and displacements, see \eqref{eq:sd-D1} and \eqref{eq:sd-sigma-ext}. This flow is strongly continuous and, since $L$ is a function of $\Upsilon$, it commutes with the operator $\Lambda_L$ of Subsection~\ref{subsec:sd-def}, as required in Theorem~\ref{thm:sd-structure}(iv) and Proposition~\ref{prop:sd-asymptotic}. Definition~\ref{def:sd-impl} allows arbitrary regular representations of the canonical commutation relations, such as the Araki--Woods representations \cite{AW63}, and arbitrary families of Bogoliubov transformations. Let $\alpha_t\doteq\mathrm{Ad}\,e^{itH}$, $t\in\mathbb R$, denote the Heisenberg dynamics.

\begin{introtheorem}[Structure and quantum Langevin equation]
	\label{thm:intro-structure}
	Let $H$ be a spin-dressed implementation of the free flow at scale $L$, with coupling map $\Theta$ and displacements $(\Delta_t)_{t\in\mathbb R}$.
	\begin{enumerate}
		\item[(i)] The data are determined by $H$: if $H$ also $\Theta'$-implements the free flow of $\Upsilon'$ at a scale $L'=F'(\Upsilon')$, then $\Upsilon'=\Upsilon$ and $\Theta'=\Theta$ on $D(L)\cap D(L')$. Moreover, $\Delta_{t+s}(f)=\Delta_t(u_sf)+\alpha_t(\Delta_s(f))$ for all $s,t\in\mathbb R$ and $f\in\h$.
		\item[(ii)] $\Theta$ is real-linear, and, for every basis $(\tau_l)_{l=1}^{n^{2}}$ of $M_n^{\mathrm{sa}}$, there are unique $w_1,\dots,w_{n^{2}}\in\h_{-1}(L)$ such that
		\[
		\Theta(f)=\sum_{l=1}^{n^{2}}2\,\mathrm{Im}\langle f,w_l\rangle\,\tau_l,\qquad f\in D(L).
		\]
		\item[(iii)] For all $t\in\mathbb R$ and $f\in D(L)$, the quantum Langevin equation
		\[
		e^{itH}\big(\mathbf 1\otimes\phi(f)\big)e^{-itH}=\mathbf 1\otimes\phi(u_tf)+\int_0^{t}\alpha_s\big(\Theta(u_{t-s}f)\otimes\mathbf 1\big)\,\mathrm ds
		\]
		holds, the integral converging in the strong operator topology. In particular, for every $f\in\h$, $\Delta_t(f)$ belongs to the von Neumann algebra generated by the evolved spin observables $\alpha_s(M_n\otimes\mathbf 1)$, with $s\in[0,t]$ (or $s\in[t,0]$).
	\end{enumerate}
\end{introtheorem}

\begin{proof} 
	The first assertion of (i) is deduced from Theorem~\ref{thm:sd-structure}(i) and the paragraph following that theorem. The cocycle identity, (ii) and (iii) are Theorem~\ref{thm:sd-structure}(ii), (iii) and (iv). Part~(iii) is the rigorous form of \eqref{eq:intro-langevin}.
\end{proof}

Theorem~\ref{thm:intro-structure} describes every spin-dressed implementation while Theorem~\ref{thm:intro-fock} tells which operators are such at the scale $0$, for regular form factors. Let $\Upsilon\geq0$ with $\Ker\Upsilon=\{0\}$, let $\tau_1,\dots,\tau_m\in M_n^{\mathrm{sa}}$, let $w_1,\dots,w_m\in D(\Upsilon^{-1/2})$ be regular form factors, let $\Theta$ be defined from these data as above, and let
\[
H_{\Upsilon,\Theta}\doteq\mathbf 1\otimes\mathrm d\Gamma(\Upsilon)+\sum_{l=1}^{m}\tau_l\otimes\phi(w_l)\qquad\text{on }\mathbb C^{n}\otimes D(\mathrm d\Gamma(\Upsilon)),
\]
which only depends on $\Theta$ (see after \eqref{eq:sd-Hhat}). For $n=1$, it is the van Hove Hamiltonian $H_{\Upsilon,\theta}$ of Theorem~\ref{thm:intro-vH} with $\theta=\sum_l\tau_lw_l$ and $\langle\Omega,H_{\Upsilon,\theta}\Omega\rangle=0$: the subscript $\Theta$ is a coupling map, the subscript $\theta$ a vector.

\begin{introtheorem}[Spin--boson Hamiltonians in Fock space]
	\label{thm:intro-fock}
	\leavevmode
	\begin{enumerate}
		\item[(i)] For every $E\in M_n^{\mathrm{sa}}$, the operator $H_{\Upsilon,\Theta}+E\otimes\mathbf 1$ is self-adjoint on $\mathbb C^{n}\otimes D(\mathrm d\Gamma(\Upsilon))$, bounded from below, and it $\Theta$-implements the free flow.
		\item[(ii)] Conversely, every self-adjoint operator $H$ which $\Theta$-implements the free flow is of the form $H=H_{\Upsilon,\Theta}+E\otimes\mathbf 1$ for a unique $E\in M_n^{\mathrm{sa}}$.
	\end{enumerate}
\end{introtheorem}

\begin{proof}
	 See Theorem~\ref{thm:sd-fock}.
\end{proof}

Thus, generalized spin--boson Hamiltonians with regular form factors are characterized by their Heisenberg dynamics, through (D1)--(D2$_0$), without any domain assumption. Indeed, (D1)--(D2$_0$) force the vacuum and the coherent vectors with arguments in $D(\Upsilon)$ into $D(H)$ (Proposition~\ref{prop:sd-vacuum}), where $H$ acts by the formal expression \eqref{eq:intro-gSB} (Proposition~\ref{prop:sd-universal}). This is the analogue of the equivalence between Berezin's and Bach--Lieb--Solovej's definitions of quadratic Hamiltonians \cite{Ber66,BLS94,Quadratic}, the Lipschitz condition (D2$_0$) playing the role of the Shale--Stinespring-like condition (Subsection~\ref{subsec:sd-fock}). Theorem~\ref{thm:intro-fock} covers the standard and the rotating-wave spin-boson models (Example~\ref{ex:sd-SB}).

The ambiguity in Theorem~\ref{thm:intro-fock}(ii) is the self-adjoint part of the commutant $M_n\otimes\mathbf 1$ of the Weyl operators $\mathbf 1\otimes e^{i\phi(f)}$, $f\in\h$. Unlike the additive constant of the case $n=1$, the spin Hamiltonian $E$ modifies the dynamics of the fields as soon as it does not commute with the coupling (discussion after Theorem~\ref{thm:sd-fock}). Conversely, every self-adjoint element of the commutant can be added to a spin-dressed implementation with locally bounded displacements, i.e., $\sup_{|t|\leq T}\|\Delta_t\|<\infty$ for all $T>0$, without changing its data, in any regular representation and at any scale (Lemma~\ref{lem:sd-commutant}).

For ultraviolet-singular form factors, the vacuum is no longer in the domain of the Hamiltonian, and the proof of Theorem~\ref{thm:intro-fock}(ii) breaks down. The definition still applies if the regularity scale is raised to $L=\Upsilon$. For $\varepsilon=0$, the formal Hamiltonian \eqref{eq:intro-SB} is the direct sum of the van Hove Hamiltonians $H_{\omega,\pm v}$ over the eigenspaces of $\sigma_x$, and the tunnelling term $\frac\varepsilon2\sigma_z\otimes\mathbf 1$ is bounded and commutes with the fields $\mathbf 1\otimes\phi(f)$, so that Lemma~\ref{lem:sd-commutant} applies. Let $P\doteq(-1)^{\mathrm d\Gamma(\mathbf 1)}$ be the parity operator, so that $P\phi(f)P=-\phi(f)$ for $f\in\h$.

\begin{introtheorem}[Renormalized spin--boson model]
	\label{thm:intro-ren}
	Let $n=2$, $\Upsilon\geq0$ with $\Ker\Upsilon=\{0\}$, $\varepsilon\in\mathbb R$, $v\in\h_{-1}(\Upsilon)$, and $\Theta(f)\doteq2\,\mathrm{Im}\langle f,v\rangle\,\sigma_x$ for $f\in D(\Upsilon)$. Let $H_{\Upsilon,v}$ be a van Hove Hamiltonian as in Theorem~\ref{thm:intro-vH}(ii), let $H_{\Upsilon,-v}\doteq PH_{\Upsilon,v}P$, and let $\Pi_\pm\doteq\frac12(\mathbf 1\pm\sigma_x)$.
	\begin{enumerate}
		\item[(i)] The self-adjoint operator
		\[
		H^{\mathrm{ren}}_{\mathrm{SB}}\doteq\Pi_+\otimes H_{\Upsilon,v}+\Pi_-\otimes H_{\Upsilon,-v}+\frac{\varepsilon}{2}\,\sigma_z\otimes\mathbf 1
		\]
		$\Theta$-implements the free flow at the scale $L=\Upsilon$. In particular, it satisfies the quantum Langevin equation of Theorem~\ref{thm:intro-structure}(iii).
		\item[(ii)] If $\mathbf 1_{[0,\kappa]}(\Upsilon)v\in D(\Upsilon^{-1/2})$ for some $\kappa>0$, then, for a suitable choice of the additive constant of $H_{\Upsilon,v}$, the operator $H^{\mathrm{ren}}_{\mathrm{SB}}$ is the limit in the norm resolvent sense, as $\Lambda\to\infty$, of the cutoff Hamiltonians
		\[
		\frac{\varepsilon}{2}\,\sigma_z\otimes\mathbf 1+\mathbf 1\otimes\mathrm d\Gamma(\Upsilon)+\sigma_x\otimes\phi\big(\mathbf 1_{[0,\Lambda]}(\Upsilon)v\big)+\big\|\Upsilon^{-1/2}\mathbf 1_{(\kappa,\Lambda]}(\Upsilon)v\big\|^{2}\,\mathbf 1 .
		\]
		Up to the spectral representation of $\Upsilon$, it is thus the renormalized spin--boson Hamiltonian of \cite[Theorem~2.5]{HLVM25}.
	\end{enumerate}
\end{introtheorem}

\begin{proof}
	See Proposition~\ref{prop:sd-commuting} and Example~\ref{ex:sd-SB-ren}. The compatibility of the normalization $H_{\Upsilon,-v}=PH_{\Upsilon,v}P$ with the additive constants of Proposition~\ref{prop:sd-commuting}(ii) is checked in Example~\ref{ex:sd-SB-ren}.
\end{proof}

 What Theorem~\ref{thm:intro-ren} adds to the construction of \cite{HLVM25} is that $H^{\mathrm{ren}}_{\mathrm{SB}}$ has the same dynamical structure as the regular models, with the bare coupling map $\Theta$: the counterterm, a scalar here, commutes with the fields and does not show up in (D1)--(D2$_\Upsilon$).

Finally, our self-dynamical definition yields asymptotic Weyl operators, in any regular representation and without any expression for $H$, on the test functions along which the coupling map is integrable. This applies in particular to the renormalized spin--boson model of Theorem~\ref{thm:intro-ren}, with its singular coupling map and an arbitrary tunnelling term (Example~\ref{ex:sd-SB-ren}). Let $\mathbb R_+\doteq[0,\infty)$ and $\mathbb R_-\doteq(-\infty,0]$.

\begin{introtheorem}[Asymptotic fields]\label{thm:E}
Let $H$ be a spin-dressed implementation of the free flow at scale $L$, with coupling map $\Theta$, and let $f\in D(L)$ satisfy $\int_{\mathbb R_\pm}\|\Theta(u_{-s}f)\|\,\mathrm ds<\infty$. Then the limit
\[
W_\pm(f)\doteq\lim_{t\to\pm\infty}\alpha_t\big(\mathbf 1\otimes e^{i\phi(u_{-t}f)}\big)
=\exp\left(i\left(\mathbf 1\otimes\phi(f)+\int_0^{\pm\infty}\alpha_s\big(\Theta(u_{-s}f)\otimes\mathbf 1\big)\,\mathrm ds\right)\right)
\]
exists in norm. The $W_\pm(f)$ satisfy the Weyl relations, and $\alpha_r(W_\pm(f))=W_\pm(u_rf)$ for $r\in\mathbb R$. If $\h=L^2(\mathbb R^d)$, $\Upsilon$ is the multiplication by a real function $\omega\in C^\infty(\mathbb R^d\setminus\{0\})$ with $\nabla\omega\neq0$ on $\mathbb R^d\setminus\{0\}$, and the form factors $w_l$ of Theorem~\ref{thm:intro-structure}(ii) are smooth on $\mathbb R^d\setminus\{0\}$, then the set of such $f$, for both signs, is dense in $\h$.
\end{introtheorem}

\begin{proof}
See Proposition~\ref{prop:sd-asymptotic} and Lemma~\ref{lem:sd-density}.
\end{proof}

\section{\texorpdfstring{$N$}{N}-Diagonal Affine Bogoliubov Transformations}
\label{sec:affine}

This section treats the scalar case $n=1$, the template for Section~\ref{sec:spin}. Definition~\ref{def:affine-quad}, Theorem~\ref{thm:affine} and Proposition~\ref{prop:hierarchy} are the technical-section versions of Definition~\ref{def:intro-affine} and Theorem~\ref{thm:intro-vH}. 

\subsection{Self-Dual Weyl Algebra and Affine Bogoliubov Transformations}
\label{subsec:affine-setting}

Let $\h$ be a complex separable Hilbert space endowed with an antiunitary involution $C$. Throughout, scalar products are antilinear in the first argument. We define the self-dual one-particle space as $\Hs\doteq\h\oplus\h$, which is endowed with the antiunitary involution

\[ \Ar =
\begin{pmatrix}
	0 & C \\
	C & 0
\end{pmatrix}.\]
 Set $S\doteq\mathbf 1\oplus(-\mathbf 1)$, so that $\Ar S=-S\Ar$. Then

\[
\Hs_{\Ar}\doteq\{\psi\in\Hs:\Ar\psi=\psi\}=\{(f,Cf):f\in\h\}
\]
is a real Hilbert space for the restriction of $\langle\cdot,\cdot\rangle_{\Hs}$, which is real-valued on it, and $\sigma(\psi,\varphi)\doteq-i\langle\psi,S\varphi\rangle_{\Hs}$ is a non-degenerate symplectic form on it: $\sigma((f,Cf),(g,Cg))=2\,\mathrm{Im}\langle f,g\rangle_{\h}$ for any $\psi=(f,Cf)$ and $\varphi = (g,Cg)$ in $\Hs_\Ar$.

\begin{definition}[Bogoliubov transformations]
	\label{def:bogo}
	A (bosonic) Bogoliubov transformation is an operator $U\in\mathcal B(\Hs)$ such that
	\begin{equation}
		U\Ar=\Ar U\qquad\text{and}\qquad U^{*}SU=S=USU^{*}.
		\label{eq:bogo}
	\end{equation}
	They form a group, denoted $\Sp(\Hs)$.
\end{definition}
Note that $U^{-1}=SU^{*}S$. By Shale's theorem \cite[Theorem~24]{DerezinskiCCR}, $U\in\Sp(\Hs)$ is implementable in the Fock representation below if and only if its off-diagonal blocks are Hilbert--Schmidt.

Let $\Wa$ be the Weyl $C^{*}$-algebra of $(\Hs_{\Ar},\sigma)$. It is the universal $C^*$-algebra \cite[Proposition~3.4]{Slawny} generated by unitaries $W(\psi)$ indexed by $\psi\in\Hs_{\Ar}$ as well as $\mathbf 1$, subject to the relations $W(\psi)^{*}=W(-\psi)$ and
\begin{equation}
	W(\psi)W(\varphi)=e^{-\frac i2\sigma(\psi,\varphi)}W(\psi+\varphi),\qquad \psi,\varphi\in\Hs_\Ar.
	\label{eq:weyl}
\end{equation}
Since $\sigma$ is non-degenerate, $\Wa$ is simple, and any $C^*$-algebra generated by unitaries satisfying these relations is isomorphic to $\Wa$ \cite[Theorem~3.7]{Slawny}.

\paragraph{Self-dual fields.}
A representation $\pi$ of $\Wa$ on a Hilbert space $\Kc_\pi$ is regular if $t\mapsto\pi(W(t\psi))$ is strongly continuous for every $\psi\in\Hs_{\Ar}$. By \eqref{eq:weyl}, $(\pi(W(t\psi)))_{t\in\mathbb R}$ is then a strongly continuous unitary group. Stone's theorem yields a unique self-adjoint operator $\Phi_\pi(\psi)$, the Segal field, such that
\begin{equation}
	\pi\big(W(t\psi)\big)=e^{it\Phi_\pi(\psi)},\qquad t\in\mathbb R,\ \psi\in\Hs_{\Ar}.
	\label{eq:segal}
\end{equation}
Since $W(\psi)W(t\varphi)W(\psi)^{*}=e^{-it\sigma(\psi,\varphi)}W(t\varphi)$ by \eqref{eq:weyl}, one has
\begin{equation}
	\pi\big(W(\psi)\big)\,\Phi_\pi(\varphi)\,\pi\big(W(\psi)\big)^{*}=\Phi_\pi(\varphi)-\sigma(\psi,\varphi)\mathbf 1,\qquad \psi,\varphi\in\Hs_{\Ar}.
	\label{eq:segal-weyl}
\end{equation}
Every $\psi\in\Hs$ decomposes uniquely as $\psi=\psi_R+i\psi_I$ with $\psi_R,\psi_I\in\Hs_{\Ar}$, namely
\[
\psi_R\doteq\tfrac12(\psi+\Ar\psi),\qquad \psi_I\doteq\tfrac1{2i}(\psi-\Ar\psi)\,.
\]
The fields are extended antilinearly to $\Hs$ by
\begin{equation}
	B_\pi(\psi)\doteq\Phi_\pi(\psi_R)-i\,\Phi_\pi(\psi_I),\qquad D\big(B_\pi(\psi)\big)\doteq D\big(\Phi_\pi(\psi_R)\big)\cap D\big(\Phi_\pi(\psi_I)\big).
	\label{eq:B-def}
\end{equation}
Since $(i\psi)_R=-\psi_I$ and $(i\psi)_I=\psi_R$, this gives $B_\pi(i\psi)=-iB_\pi(\psi)$. Moreover, $B_\pi(\psi)+B_\pi(\varphi)\subseteq B_\pi(\psi+\varphi)$ by \cite[Theorem~5]{DerezinskiCCR}, so that $B_\pi$ is antilinear in the sense of unbounded operators. By \eqref{eq:B-def}, $B_\pi(\psi)=\Phi_\pi(\psi)$ for $\psi\in\Hs_{\Ar}$. Note that $\langle\psi,S\varphi\rangle_{\Hs}=i\sigma(\psi,\varphi)$ for $\psi,\varphi\in\Hs_{\Ar}$.

\begin{remark}[Self-dual CCR]
	\label{rem:selfdual-ccr}
	For every regular representation $\pi$ and every $\psi\in\Hs$, the operator $B_\pi(\psi)$ is closed and densely defined, $B_\pi(\psi)^{*}=B_\pi(\Ar\psi)$, and
	\[
	[B_\pi(\psi),B_\pi(\varphi)^{*}]=\langle\psi,S\varphi\rangle_{\Hs}\mathbf 1,\qquad
	[B_\pi(\psi),B_\pi(\varphi)]=\langle\psi,S\Ar\varphi\rangle_{\Hs}\mathbf 1,
	\]
	in the sense of forms on common cores. These are the standard properties of fields and of creation and annihilation operators in regular CCR representations, here written in the self-dual form. Closedness is \cite[Theorem~5(3)]{DerezinskiCCR}, and the commutation relations follow from \cite[Theorem~5(2)]{DerezinskiCCR}; density of the domain and the adjoint relation follow from the Stone--von Neumann theorem \cite[Theorem~4]{DerezinskiCCR}, applied to the restriction of $\pi$ to a finite-dimensional symplectic subspace of $\Hs_\Ar$ containing $\psi_R$ and $\psi_I$.
\end{remark}

Automorphisms of $\Wa$ of the form $W(\psi)\mapsto e^{i\sigma(\psi,\nu)}W(U\psi)$ change the fields by a Bogoliubov transformation of the argument and a scalar displacement. The next lemma records this: it gives \eqref{eq:chi-fields} below and is used in Subsection~\ref{subsec:affine-ham} and throughout Section~\ref{sec:spin}.

\begin{lemma}[Covariance of fields]
	\label{lem:field-cov}
	Let $\pi$ be a regular representation of $\Wa$, let $U\in\Sp(\Hs)$ and let $\nu\in\Hs_{\Ar}$. Let $\chi$ be the $*$-automorphism of $\Wa$ such that $\chi(W(\psi))=e^{i\sigma(\psi,\nu)}W(U\psi)$ (it exists and is unique by Steps~1 and~2 of the proof of Proposition~\ref{prop:affine-aut}, which do not use the present lemma). Then $\pi\circ\chi$ is regular, and for all $\psi\in\Hs$,
	\begin{equation}
		B_{\pi\circ\chi}(\psi)=B_\pi(U\psi)-i\langle\psi,S\nu\rangle_{\Hs}\mathbf 1 ,\qquad\text{with }D\big(B_{\pi\circ\chi}(\psi)\big)=D\big(B_\pi(U\psi)\big).
		\label{eq:B-cov}
	\end{equation}
	In particular, taking $U=\mathbf 1$, so that $\chi=\mathrm{Ad}\,W(\nu)$, gives
	\[
	\pi\big(W(\nu)\big)\,B_\pi(\psi)\,\pi\big(W(\nu)\big)^{*}=B_\pi(\psi)-i\langle\psi,S\nu\rangle_{\Hs}\mathbf 1 .
	\]
\end{lemma}

\begin{proof}
	For $\psi\in\Hs_{\Ar}$,
	\[
	\pi\circ\chi\big(W(t\psi)\big)=e^{it\sigma(\psi,\nu)}e^{it\Phi_\pi(U\psi)} .
	\]
	By uniqueness in Stone's theorem, and because $\pi\circ \chi$ is a regular representation, $\Phi_{\pi\circ\chi}(\psi)=\Phi_\pi(U\psi)+\sigma(\psi,\nu)\mathbf 1$. Now insert this into \eqref{eq:B-def}.
	Since $U$ commutes with $\Ar$, $(U\psi)_{R,I}=U\psi_{R,I}$. In particular, both sides of \eqref{eq:B-cov} have the domain $D(\Phi_\pi(U\psi_R))\cap D(\Phi_\pi(U\psi_I))$. Moreover, $\sigma(\psi_R,\nu)-i\sigma(\psi_I,\nu)=-i\langle\psi,S\nu\rangle_{\Hs}$.
\end{proof}

\paragraph{Fock representation.}
Let $\Fb$ be the bosonic Fock space over $\h$, with vacuum $\Omega$, annihilation and creation operators $a(f)$, $a^{*}(f)$, and finite-particle (dense) subspace $\mathcal D_{0}$. The Fock representation $\piF$ of $\Wa$ is the irreducible regular representation \cite[Theorem~21]{DerezinskiCCR} defined by
\[
\piF\big(W(\psi)\big)\doteq\exp\!\Big(i\,\overline{a(f)+a^{*}(f)}\Big),\qquad \psi=(f,Cf)\in\Hs_{\Ar}.
\]
In this representation, the self-dual fields are the usual Fock fields: for $f,g\in\h$, $\Phi_{\piF}((f,Cf))=\overline{a(f)+a^{*}(f)}$ (the field $\phi(f)$ of the Introduction) and $B_{\piF}((f,g))=a(f)+a^{*}(Cg)$ on $\mathcal D_0$. In particular, $B_{\piF}((f,0))=a(f)$ and $B_{\piF}((0,Cf))=a^{*}(f)$.

\paragraph{Second quantization.}
The linear part $u\oplus\bar u$ of the $N$-diagonal transformations below (see~\eqref{eq:ASp0}) acts on $\Fb$ through the second quantization of a unitary $u$ acting on $\h$, which is the unitary
\[
\Gamma(u)\doteq\bigoplus_{p\in\mathbb N_0}u^{\otimes p}\big|_{\vee^{p}\h}\quad\text{on }\Fb,\qquad u^{\otimes0}\doteq1 .
\]
where $\vee$ is the symmetrized tensor product. Then for any unitary operators $u_1,u_2$ acting on $\h$, and with $\#$ denoting either $*$ or the empty symbol,
\[\Gamma(u_1u_2)=\Gamma(u_1)\Gamma(u_2),\qquad \Gamma(u)\Omega=\Omega\quad \mbox{and}\quad \Gamma(u)a^{\#}(f)\Gamma(u)^{*}=a^{\#}(uf).\] Since $(u\oplus\bar u)(f,Cf)=(uf,Cuf)$, where $\bar u\doteq CuC$, the last identity, on $\mathcal D_0$, gives
\begin{equation}\label{eq:bogotransfo}
	\Gamma(u)\piF(W(\psi))\Gamma(u)^{*}=\piF(W((u\oplus\bar u)\psi)),\qquad \psi\in\Hs_\Ar .
\end{equation} 
For a self-adjoint operator $\Upsilon$ on $\h$, $(\Gamma(e^{it\Upsilon}))_{t\in\mathbb R}$ is a strongly continuous unitary group, and $\mathrm d\Gamma(\Upsilon)$ denotes its self-adjoint generator: $\Gamma(e^{it\Upsilon})=e^{it\,\mathrm d\Gamma(\Upsilon)}$.
On the algebraic symmetric tensor power $\vee^{p}D(\Upsilon)$, $\mathrm d\Gamma(\Upsilon)$ acts as 
\[\sum_{k=1}^{p}\mathbf 1^{\otimes(k-1)}\otimes\Upsilon\otimes\mathbf 1^{\otimes(p-k)},\] and the finite sums of such vectors form a core for $\mathrm d\Gamma(\Upsilon)$. See, e.g., \cite[Sect.~5.2.1]{BR97}. The operator $\mathrm d\Gamma(\Upsilon)$ is the free part of the Hamiltonians of Subsections~\ref{subsec:affine-ham} and~\ref{subsec:sd-fock}.

\paragraph{Exponential vectors.}
For $x\in\h$, the \emph{exponential vector} is
\begin{equation}\label{eq:defexpvector}
	\epsilon(x)\doteq\bigoplus_{k\in\mathbb N_0}(k!)^{-1/2}\,x^{\otimes k}\in\Fb .
\end{equation}
The exponential vectors are total in $\Fb$, the map $x\mapsto\epsilon(x)$ is continuous \cite[Props.~2.1 and~2.2]{Gui72}, and the Weyl operators act on them by
\begin{equation}\label{eq:Weyl-exp}
	\piF\big(W((g,Cg))\big)\,\epsilon(x)=e^{-\|g\|^{2}/2-\langle ig,x\rangle}\,\epsilon(x+ig),\qquad g,x\in\h .
\end{equation}
In particular, the normalized vectors $e^{-\|x\|^{2}/2}\epsilon(x)=\piF(W((-ix,C(-ix))))\Omega$, $x\in\h$, are the \emph{coherent vectors}. The definition of $\Gamma(u)$ gives, for every unitary $u$ on $\h$,
\begin{equation}\label{eq:exp-facts}
	\Gamma(u)\,\epsilon(x)=\epsilon(ux),\qquad x\in\h ,
\end{equation}
and the case $x=0$ of \eqref{eq:Weyl-exp}, with $\epsilon(0)=\Omega$, gives the vacuum expectation of Weyl operators,
\begin{equation}\label{eq:vac-weyl}
	\big\langle\Omega,\piF\big(W((g,Cg))\big)\Omega\big\rangle=e^{-\|g\|^{2}/2},\qquad g\in\h .
\end{equation}
The map $\xi\mapsto\piF(W(\xi))$ is strongly continuous on $\Hs_{\Ar}$. Indeed, on exponential vectors this follows from \eqref{eq:Weyl-exp} and the continuity of $x\mapsto\epsilon(x)$ \cite[Prop.~2.1]{Gui72}. Since the exponential vectors are total \cite[Prop.~2.2]{Gui72} and the operators $\piF(W(\xi))$ are unitary, hence uniformly bounded, strong continuity on exponential vectors extends to all of $\Fb$. These facts are used in the proofs of Theorem~\ref{thm:affine}(iii), Proposition~\ref{prop:hierarchy}(i) and Theorem~\ref{thm:sd-structure}(i), and in Subsection~\ref{subsec:sd-fock}.

\paragraph{Homogeneous coordinates.}
A displacement, that is, a translation by $\nu$ in phase space, does not move the argument $\psi$ of a field $B_\pi(\psi)$, but rather adds to the field a form that is antilinear in $\psi$, like $B_\pi$ itself (Lemma~\ref{lem:field-cov}), so the transformation of fields is affine and not linear. To make it linear, we adjoin a coordinate $z$, which enters the field as $\bar z\mathbf 1$. We mimic homogeneous coordinates for affine maps: we introduce the space $\Hx\doteq\Hs\oplus\mathbb C$, endowed with the antiunitary involution $\Ax(\psi,z)\doteq(\Ar\psi,\bar z)$, with the (degenerate) self-adjoint operator $\Sx\doteq S\oplus0$, and with the vector $e\doteq(0,1)$. The extended fields are
\begin{equation}\label{eq:defextendfields}
	\Bx_\pi(\psi,z)\doteq B_\pi(\psi)+\bar z\mathbf 1,\qquad \Psi\doteq(\psi,z)\in\Hx
\end{equation}
for every regular representation $\pi$, so that $\Bx_\pi(\Ax\Psi)=\Bx_\pi(\Psi)^{*}$ and $\Bx_\pi(e)=\mathbf 1$. They transform linearly under the maps $\U$ below (Proposition~\ref{prop:affine-aut}). This self-dual formalism follows Araki \cite{Araki} in the bosonic case and \cite{Quadratic} in the fermionic case. Unlike in \cite{Araki}, our fields $B_\pi$ are antilinear.

\begin{definition}[Affine Bogoliubov transformations]
	\label{def:affine-bogo}
	An affine Bogoliubov transformation is an invertible operator $\U\in\mathcal B(\Hx)$ such that
	\[
	\U\Ax=\Ax\U,\qquad \U^{*}\Sx\U=\Sx,\qquad \U e=e .
	\]
	They form a group, denoted $\ASp(\Hs)$.
\end{definition}

For $\varphi\in\Hs$, write $\varphi^{*}\doteq\langle\varphi,\cdot\,\rangle_{\Hs}$. A direct verification using the relations above and the Riesz theorem yields the following lemma.

\begin{lemma}[Block structure]
	\label{lem:block}
	The elements of $\ASp(\Hs)$ are exactly the operators
	\begin{equation}
		\U_{(U,\nu)}=\begin{pmatrix}U&0\\ i(S\nu)^{*}&1\end{pmatrix},
		\label{eq:U-block}
	\end{equation}
	where $\nu\in\Hs_{\Ar}$ and $U$ is a bosonic Bogoliubov transformation, i.e., $U\Ar=\Ar U$ and $U^{*}SU=S=USU^{*}$. Moreover, using $U^{-1}=SU^*S$,
	\[
	\U_{(U_1,\nu_1)}\U_{(U_2,\nu_2)}=\U_{(U_1U_2,\,U_2^{-1}\nu_1+\nu_2)} .
	\]
\end{lemma}

In particular, $(U,\nu)\mapsto(U\nu,U)$ identifies $\ASp(\Hs)$ with the affine symplectic group $\Hs_{\Ar}\rtimes\Sp(\Hs)$ where the composition law is $(a,U)(a',U')=(a+Ua',UU')$ for all $(a,U),(a',U')\in \Hs_{\Ar}\rtimes\Sp(\Hs)$.

In this section, we only need the \emph{$N$-diagonal} subgroup
\begin{equation}
	\ASp_0(\Hs)\doteq\big\{\U_{(U,\nu)}\in\ASp(\Hs):\ US=SU\big\}
	=\big\{\U_{(u\oplus\bar u,\,\nu)}:\ u\in\mathcal U(\h),\ \nu\in\Hs_{\Ar}\big\}.
	\label{eq:ASp0}
\end{equation}
Here $\mathcal U(\h)$ is the unitary group of $\h$ and $\bar u\doteq CuC$: indeed, $US=SU$ means $U=u_1\oplus u_2$, then $U\Ar=\Ar U$ means $u_2=Cu_1C$, and $U^{*}SU=S=USU^{*}$ means that $u_1$ is unitary. These are the affine Bogoliubov transformations that do not mix the two sectors $\h\oplus0$ and $0\oplus\h$ of $\Hs$. Their linear part is implemented by $\Gamma(u)$. Unlike general elements of $\ASp(\Hs)$, they are all implementable in the Fock representation (Corollary~\ref{cor:implementation}).
Elements of this subgroup are represented by the following operator block matrices on $\h\oplus\h\oplus\mathbb C$, where $\nu=(g,Cg)$:
\[
\U_{(u\oplus\bar u,\nu)}=
\begin{pmatrix}
	u&0&0\\
	0&\bar u&0\\
	i\langle g,\cdot\,\rangle&-i\langle Cg,\cdot\,\rangle&1
\end{pmatrix}.
\]

\begin{proposition}[Affine Bogoliubov automorphisms]
	\label{prop:affine-aut}
	For every $\U=\U_{(U,\nu)}\in\ASp(\Hs)$, there is a unique $*$-automorphism $\chi_{\U}$ of $\Wa$ such that
	\begin{equation}
		\chi_{\U}\big(W(\psi)\big)=e^{i\sigma(\psi,\nu)}\,W(U\psi),\qquad \psi\in\Hs_{\Ar} .
		\label{eq:chi}
	\end{equation}
	The map $\U\mapsto\chi_{\U}$ is an injective group homomorphism from $\ASp(\Hs)$ into $\mathrm{Aut}(\Wa)$. Moreover, for every regular representation $\pi$ of $\Wa$, the representation $\pi\circ\chi_{\U}$ is regular and
	\begin{equation}
		\Bx_{\pi\circ\chi_{\U}}(\Psi)=\Bx_\pi(\U\Psi),\qquad \Psi\in\Hx .
		\label{eq:chi-fields}
	\end{equation}
	Thus the extended fields transform linearly under $\U$, compare with \cite[Eq.~(96)]{Quadratic}.
\end{proposition}

\begin{proof}
	\emph{Step 1: existence and uniqueness of $\chi_{\U}$.} The family $\psi\mapsto e^{i\sigma(\psi,\nu)}W(U\psi)$ satisfies the Weyl relations \eqref{eq:weyl}, because $U$ preserves $\Hs_{\Ar}$ and $\sigma$ (Definition~\ref{def:bogo}) and $\sigma(\cdot,\nu)$ is additive. Hence, by the universal property of $\Wa$ \cite[Proposition~3.4]{Slawny}, there is a unique $*$-homomorphism $\chi_{\U}:\Wa\to\Wa$ satisfying \eqref{eq:chi}, and it is injective since $\Wa$ is simple \cite[Theorem~3.7]{Slawny}.

	\emph{Step 2: homomorphism and surjectivity.} The homomorphism property follows from
	\[
	\chi_{\U_1}\circ\chi_{\U_2}\big(W(\psi)\big)=e^{i\sigma(\psi,\,U_2^{-1}\nu_1+\nu_2)}\,W(U_1U_2\psi)
	\]
	and Lemma~\ref{lem:block}. Hence $\chi_{\U^{-1}}$ is the inverse of $\chi_{\U}$, which is therefore surjective.

	\emph{Step 3: injectivity of $\U\mapsto\chi_{\U}$.} If $\chi_{\U}=\mathrm{id}$, then $W(U\psi)\in\mathbb C\,W(\psi)$, i.e., $W(U\psi-\psi)\in\mathbb C\mathbf 1$, for all $\psi\in\Hs_{\Ar}$. If $W(\xi)=\lambda\mathbf 1$ for some $\xi\in\Hs_{\Ar}$ and $\lambda\in\mathbb C$, then, by \eqref{eq:weyl}, \[\lambda\mathbf 1 = W(s\varphi)W(\xi)W(s\varphi)^{*}=e^{-is\sigma(\varphi,\xi)}W(\xi),\qquad s\in\mathbb R,\quad \varphi\in\Hs_{\Ar},\] that is, $e^{-is\sigma(\varphi,\xi)}=1,$ for all $s\in\mathbb R$ and $\varphi\in\Hs_{\Ar}$, hence $\xi=0$ by non-degeneracy. Therefore $U=\mathbf 1$ on $\Hs_{\Ar}$, hence on $\Hs$. Then, $e^{i\sigma(\psi,\nu)}=1$ for all $\psi\in\Hs_{\Ar}$ gives $\nu=0$ by non-degeneracy. This proves the injectivity of $\U\mapsto\chi_{\U}$.

	\emph{Step 4: extended fields.} Finally, \eqref{eq:chi-fields} follows from Lemma \ref{lem:field-cov}, since $\Bx_\pi(\U(\psi,z))=B_\pi(U\psi)+\overline{(i\langle S\nu,\psi\rangle_{\Hs}+z)}\,\mathbf 1=B_\pi(U\psi)+(-i\langle\psi,S\nu\rangle_{\Hs}+\bar z)\mathbf 1$.
\end{proof}

\begin{corollary}[Implementation of $N$-diagonal affine transformations]
	\label{cor:implementation}
	Let $\U=\U_{(u\oplus\bar u,\,\nu)}\in\ASp_0(\Hs)$. Then
	\[
	\Ui_{\U}\doteq\Gamma(u)\,\piF\big(W(\nu)\big)
	\]
	is a unitary on $\Fb$ such that $\Ui_{\U}\,\piF(A)\,\Ui_{\U}^{*}=\piF\circ\chi_{\U}(A)$ for all $A\in\Wa$. Every other unitary with this property is of the form $c\,\Ui_{\U}$ with $|c|=1$.
\end{corollary}

\begin{proof}
	On the one hand, \eqref{eq:bogotransfo} holds. On the other hand, \[\piF(W(\nu))\piF(W(\psi))\piF(W(\nu))^{*}=e^{i\sigma(\psi,\nu)}\piF(W(\psi)),\qquad\psi\in\Hs_\Ar\] by \eqref{eq:weyl}. Combining these and comparing with \eqref{eq:chi} gives the implementation on the generators $W(\psi)$, hence on $\Wa$ by linearity and continuity. Uniqueness up to a phase follows from the irreducibility of $\piF$.
\end{proof}

\subsection{Affine Quadratic Hamiltonians}
\label{subsec:affine-ham}

This subsection classifies the self-adjoint operators $H$ on $\Fb$ whose Heisenberg dynamics maps each field to a field plus a scalar, without mixing creation and annihilation (Definition~\ref{def:affine-quad}). The model is $\mathrm d\Gamma(\Upsilon)+a(\theta)+a^{*}(\theta)$, with $\Upsilon$ self-adjoint on $\h$ and $\theta\in\h$. It displaces fields by $g_t=\int_0^{t}e^{-ir\Upsilon}\theta\,\mathrm dr$, see \eqref{eq:intro-vH}.

The difficulty is that, for an arbitrary affine quadratic Hamiltonian, the form factor need not be a vector of $\h$: the vectors $g_t\in\h$ are determined by the dynamics, but $\theta=\lim_{t\to0}t^{-1}g_t$ lies, in general, only in the extrapolation space $\h_{-1}(\Upsilon)$, and $\mathrm d\Gamma(\Upsilon)+\phi(\theta)$ may then require renormalization. itis more convenient to look at the dynamic, i.e., $g$ instead, where composition law survives. Upon $u_t\doteq e^{it\Upsilon}$, the vectors $b_t\doteq iu_tg_t$ satisfy $b_{t+s}=b_t+u_tb_s$: they form a $1$-cocycle of $\mathbb R$. The phases $c_t$ of $e^{itH}$ solve $\delta c=\mathsf m$, where the multiplier $\mathsf m$ determined by $b$ is the obstruction to $t\mapsto c_t\,\Gamma(u_t)\piF(W(\nu_t))$ being a group (Appendix~\ref{app:cohomology}).

These are the data of Guichardet's representations of type~(S) \cite{Gui72}, which give Theorem~\ref{thm:affine}(i) and the existence part of~(ii) through Lemma~\ref{lem:type-S}. Dereziński's theory of van Hove Hamiltonians \cite{Der03} supplies the identification of $H$ with $\mathrm d\Gamma(\Upsilon)+a(\theta)+a^{*}(\theta)$ and the ground-state criterion (Theorem~\ref{thm:affine}(iii), Proposition~\ref{prop:hierarchy}(ii)--(iii)); the criterion for $\Omega\in D(H)$, Proposition~\ref{prop:hierarchy}(i), is proved here (see also Remark~\ref{rem:hierarchy}).

\begin{definition}[$N$-diagonal affine quadratic Hamiltonians]
	\label{def:affine-quad}
	A self-adjoint operator $H$ on $\Fb$ is an ($N$-diagonal) affine quadratic Hamiltonian if, for every $t\in\mathbb R$, there is $\U_t\in\ASp_0(\Hs)$ such that
	\[
	e^{itH}\piF(A)e^{-itH}=\piF\circ\chi_{\U_t}(A),\qquad A\in\Wa .
	\]
\end{definition}

Let $\Phi\doteq\Phi_{\piF}$ be the Segal field \eqref{eq:segal} of the Fock representation. By Definition~\ref{def:affine-quad} and Lemma~\ref{lem:field-cov}, a self-adjoint operator $H$ on $\Fb$ is an $N$-diagonal affine quadratic Hamiltonian if and only if, for every $t\in\mathbb R$, there are $U_t\in\Sp(\Hs)$ commuting with $S$ and $\nu_t\in\Hs_{\Ar}$ such that
\begin{equation}
	e^{itH}\,\Phi(\psi)\,e^{-itH}=\Phi(U_t\psi)+\sigma(\psi,\nu_t)\mathbf 1,\qquad\psi\in\Hs_{\Ar}.
	\label{eq:sd-scalar-fields}
\end{equation}
Definition~\ref{def:affine-quad} is the analogue of \cite[Definition 3.4]{Quadratic}, where the implementing group convention is $e^{itH/2}$ instead of $e^{itH}$. Neither the continuity nor the group property of $t\mapsto\U_t$ is assumed, they both follow from Proposition~\ref{prop:affine-aut} and Theorem~\ref{thm:affine}(i)

\paragraph{Cocycles and coboundaries.}
Let $(u_t)_{t\in\mathbb R}$ be a strongly continuous unitary group on $\h$. A (continuous) \emph{$1$-cocycle} of $\mathbb R$ with values in $u$ is a continuous map $b:\mathbb R\to\h$ such that
\begin{equation}
	b_{t+s}=b_t+u_tb_s,\qquad s,t\in\mathbb R .
	\label{eq:cocycle-std}
\end{equation}
Equivalently, the maps $x\mapsto u_tx+b_t$, $t\in\mathbb R$, define an action of $\mathbb R$ on $\h$ by affine isometries. A $1$-cocycle $b$ is a \emph{$1$-coboundary} if there is $w\in\h$ such that $b_t=u_tw-w$ for all $t\in\mathbb R$, that is, if this affine action is conjugate to the linear action $x\mapsto u_tx$ by the time-independent translation $x\mapsto x+w$. The $1$-cocycles form a vector space $Z^{1}(\mathbb R,u)$, the $1$-coboundaries a subspace $B^{1}(\mathbb R,u)$, and
\[
H^{1}(\mathbb R,u)\doteq Z^{1}(\mathbb R,u)/B^{1}(\mathbb R,u)
\]
is the first cohomology group of $\mathbb R$ with values in $u$. For $u_t=e^{it\Upsilon}$, a cocycle is, formally, the solution of the inhomogeneous Cauchy problem $\partial_tb_t=i\Upsilon b_t+i\theta$, $b_0=0$, namely $b_t=(e^{it\Upsilon}-1)\Upsilon^{-1}\theta$, where the constant forcing $\theta$ lies, in general, outside $\h$ (Lemma~\ref{lem:cocycle} below). The cocycle is a coboundary if and only if the forcing can be removed by a time-independent change of origin $x\mapsto x+w$ with $w\in\h$, that is, if and only if $\theta=\Upsilon w$ (Lemma~\ref{lem:cocycle}(iii)). In the same way, the phases of the unitary implementers are governed by $\mathbb T$-valued $2$-cocycles, the \emph{multipliers}, see Appendix~\ref{app:cohomology}.

\paragraph{Representations of type (S)} For $A\in\mathcal U(\h)$, $b\in\h$ and $c\in\mathbb T$, set
\[
U_{A,b,c}\doteq c\,\mathrm D_b\,\Gamma(A),\qquad \mathrm D_b\doteq e^{\overline{a^{*}(b)-a(b)}}=\piF\big(W((-ib,C(-ib)))\big),
\]
where $\mathrm D_b$ is the displacement operator (the second expression follows from \eqref{eq:Weyl-exp} with $g=-ib$, see Appendix~\ref{app:dico-Gui}). $U_{A,b,c}$ is Guichardet's operator \cite[Sect.~2.2]{Gui72}. A \emph{representation of type~(S)} of $\mathbb R$ \cite[Def.~3.3]{Gui72} is a strongly continuous unitary group $(V_t)_{t\in\mathbb R}$ on $\Fb$ such that, for every $t\in\mathbb R$, $V_t=U_{A_t,b_t,c_t}$ for some $A_t\in\mathcal U(\h)$, $b_t\in\h$, $c_t\in\mathbb T$.

\paragraph{Extrapolation space.}
Let $\Upsilon$ be a self-adjoint operator on $\h$. We denote by $\h_{-1}(\Upsilon)$ the completion of $\h$ with respect to the norm $\|\cdot\|_{-1}\doteq\|(|\Upsilon|+1)^{-1}\cdot\|_{\h}$. The operator $(|\Upsilon|+1)^{-1}$ extends to a unitary map $J:\h_{-1}(\Upsilon)\to\h$, and the inner product of $\h$ extends by continuity to the non-degenerate sesquilinear pairing
\[
\langle\theta,x\rangle\doteq\langle J\theta,(|\Upsilon|+1)x\rangle_{\h},\qquad \theta\in\h_{-1}(\Upsilon),\ x\in D(\Upsilon).
\]
It is antilinear in $\theta$, linear in $x$, and coincides with $\langle\cdot,\cdot\rangle_{\h}$ on $\h\times D(\Upsilon)$.

The space $\h$ is dense in $\h_{-1}(\Upsilon)$, and since $\|\Upsilon x\|_{-1}\leq\|x\|_{\h}$, $\Upsilon$ extends to a bounded operator $\h\to\h_{-1}(\Upsilon)$. We denote its range by $\Upsilon\h$. Note that $\Upsilon w$ determines $w\in\h$ only modulo $\Ker\Upsilon$. Indeed, if $\Ker\Upsilon=\{0\}$, we write $\Upsilon^{-1}\theta\doteq w$ for $\theta=\Upsilon w\in\Upsilon\h$, so that $\Upsilon^{-1}\theta\in\h$ means $\theta\in\Upsilon\h$. For $\Upsilon\geq0$, every $\theta\in\h_{-1}(\Upsilon)$ reads $\theta=\xi+\Upsilon\xi$ with $\xi\doteq J\theta\in\h$, so that $\h_{-1}(\Upsilon)$ is the space $\h+\Upsilon\h=D(\Upsilon)^{*}$ of \cite[Sect.~2.2]{Der03}.

If $\beta$ is a Borel function such that $x\mapsto \beta(x)(|x|+1)$ is bounded, then $\beta(\Upsilon)$ extends to a bounded operator $\h_{-1}(\Upsilon)\to\h$, namely $\beta(\Upsilon)\theta\doteq[\beta(\Upsilon)(|\Upsilon|+1)]J\theta$. We apply this to
\begin{equation}
	f_t(x)\doteq\int_0^{t}e^{-irx}\,\mathrm dr=\frac{1-e^{-itx}}{ix},\qquad t\in\mathbb R,
	\label{eq:ft}
\end{equation}
for which $|f_t(x)|(|x|+1)\leq|t|+2$. Formally, $f_t(\Upsilon)\theta=\int_0^{t}e^{-ir\Upsilon}\theta\,\mathrm dr$.

Lemma~\ref{lem:cocycle} is stated for $g_t=-iu_t^{*}b_t$, the vector in $\piF(W((g_t,Cg_t)))$. $g$ satisfies \eqref{eq:cocycle} if and only if $b$ satisfies \eqref{eq:cocycle-std}.

\begin{lemma}[Cocycles of $\mathbb R$]
	\label{lem:cocycle}
	Let $\Upsilon$ be self-adjoint on the separable Hilbert space $\h$, and $u_t\doteq e^{it\Upsilon}$.
	\begin{enumerate}
		\item[(i)] For every $\theta\in\h_{-1}(\Upsilon)$, the map $t\mapsto g_t\doteq f_t(\Upsilon)\theta\in\h$ is norm-continuous, $t^{-1}g_t\to\theta$ in $\h_{-1}(\Upsilon)$ as $t\to0$, and
		\begin{equation}
			g_{t+s}=u_s^{*}g_t+g_s,\qquad s,t\in\mathbb R .
			\label{eq:cocycle}
		\end{equation}
		\item[(ii)] Conversely, every continuous $g:\mathbb R\to\h$ satisfying \eqref{eq:cocycle} is of the form $g_t=f_t(\Upsilon)\theta$ for a unique $\theta\in\h_{-1}(\Upsilon)$.
		\item[(iii)] $g$ is of the form $g_t=-i(1-u_t^{*})w$ with $w\in\h$ iff $\theta\in\Upsilon\h\doteq\{\Upsilon w:w\in\h\}\subseteq\h_{-1}(\Upsilon)$, with $\theta=\Upsilon w$.
	\end{enumerate}
\end{lemma}

\begin{proof}
	Let $t,s\in\mathbb R$. Part (i) follows by dominated convergence in the spectral representation of $\Upsilon$, using $|f_t(x)|(|x|+1)\leq|t|+2$ and $|t^{-1}f_t(x)|\leq1$, $t^{-1}f_t(x)\to1$. \eqref{eq:cocycle} is the identity $f_{t+s}(x)=e^{-isx}f_t(x)+f_s(x)$.

	\emph{Translation.} The cocycles of \cite{Gui72} are normalized differently. For a map $g:\mathbb R\to\h$, set $b_t\doteq iu_tg_t$. Then $g$ satisfies \eqref{eq:cocycle} if and only if $b_{t+s}=b_t+u_tb_s$ for all $s,t\in\mathbb R$, i.e., $b\in Z^{1}(\mathbb R,u)$ (see \eqref{eq:cocycle-std}). Moreover:
\begin{itemize}
	\item For $\theta\in\h_{-1}(\Upsilon)$, one has $g_t=f_t(\Upsilon)\theta$ if and only if $b_t=F_t(\Upsilon)\theta$, where $F_t(x)\doteq ie^{itx}f_t(x)=(e^{itx}-1)/x$, extended by $F_t(0)=it$.
	\item For $w\in\h$, $g_t=-i(\mathbf 1-u_t^{*})w$ for all $t\in\mathbb R$ is equivalent to $b_t=u_tw-w$ for all $t$, i.e., $b\in B^{1}(\mathbb R,u)$ ($b=\delta_0w$ in the notation of Appendix~\ref{app:cohomology}).
\end{itemize}

	\emph{Part (ii).} This is the description of $Z^{1}(\mathbb R,u)$ in \cite[App.~B.2]{Gui72}, combined with \cite[Sect.~3.3, Cases~2 and~3]{Gui72} for the invariant vectors of $u$, i.e., $\Ker\Upsilon$. In a spectral representation of $\Upsilon|_{(\Ker\Upsilon)^{\perp}}$ as a multiplication operator, with spectral measure $\mu$, cocycles $b_t$ become $\lambda\mapsto(e^{it\lambda}-1)w_\lambda$ with $\int\frac{\lambda^{2}}{1+\lambda^{2}}\|w_\lambda\|^{2}\,\mathrm d\mu(\lambda)<\infty$. With $w_\lambda=\lambda^{-1}\theta_\lambda$, this is exactly the condition $\theta\in\h_{-1}(\Upsilon)$. On $\Ker\Upsilon$, cocycles are continuous additive maps $\mathbb R\to\Ker\Upsilon$, i.e., $b_t=t\,b_1$, so that $\theta=-ib_1\in\Ker\Upsilon$. Uniqueness of $\theta$ follows from (i), since $\theta=\lim_{t\to0}t^{-1}g_t$.

	\emph{Part (iii).} This is the corresponding description of $B^{1}$: for $w\in\h$, $f_t(\Upsilon)\Upsilon w=-i(1-u_t^{*})w$, since $xf_t(x)=-i(1-e^{-itx})$. Conversely, if $g_t=-i(1-u_t^{*})w$, then $\theta=\Upsilon w$ by the uniqueness in (ii).
\end{proof}

\begin{lemma}[Affine quadratic Hamiltonians and representations of type (S)]
	\label{lem:type-S}
	A self-adjoint operator $H$ on $\Fb$ is an $N$-diagonal affine quadratic Hamiltonian if and only if $(e^{itH})_{t\in\mathbb R}$ is a representation of type~(S) of $\mathbb R$ (see Appendix~\ref{app:dico-Gui}). In this case, writing $\U_t=\U_{(u_t\oplus\bar u_t,\,\nu_t)}$ as in \eqref{eq:ASp0}, with $u_t\in\mathcal U(\h)$ and $\nu_t\doteq(g_t,Cg_t)\in\Hs_{\Ar}$,
	\begin{equation}
		e^{itH}=c_t\,\Gamma(u_t)\,\piF\big(W((g_t,Cg_t))\big)=U_{u_t,b_t,c_t},\qquad b_t\doteq iu_tg_t,\quad t\in\mathbb R,
		\label{eq:type-S}
	\end{equation}
	for some $c_t\in\mathbb T$, where $U_{A,b,c}$ is the notation of \cite{Gui72}, see \eqref{eq:dictionary}.
\end{lemma}

\begin{proof}
	If $H$ is an affine quadratic Hamiltonian, then $e^{itH}$ and $\Gamma(u_t)\piF(W(\nu_t))$ implement the same automorphism $\chi_{\U_t}$  (Corollary~\ref{cor:implementation}). Hence, $\piF$ being irreducible, they differ by a phase $c_t$, and \eqref{eq:dictionary} gives \eqref{eq:type-S}. The continuity required in \cite[Def.~3.3]{Gui72} is the one of $t\mapsto e^{itH}$.

	Conversely, every operator $U_{A,b,c}$ (where $A\in\mathcal U(\h)$) is of the form \eqref{eq:dictionary} up to a phase, with $u=A$ and $g=-iA^{*}b$. By Corollary~\ref{cor:implementation}, it therefore implements $\chi_{\U}$ with $\U=\U_{(A\oplus\bar A,\,(g,Cg))}\in\ASp_0(\Hs)$.
\end{proof}

By \cite[Def.~2.3 and Lemma~2.1]{Gui72}, the operators $U_{A,b,c}$ are exactly the unitaries mapping the set $\{\lambda\,\epsilon(x):\lambda\in\mathbb C\setminus\{0\},\ x\in\h\}$ of non-zero multiples of exponential vectors \eqref{eq:defexpvector} onto itself.
Hence, by Lemma~\ref{lem:type-S}, affine quadratic Hamiltonians are also the self-adjoint operators whose unitary groups preserve the set of (multiples of) coherent vectors.

\Needspace{10\baselineskip}
\begin{theorem}[Affine quadratic Hamiltonians]
	\label{thm:affine}
	\leavevmode
	\begin{enumerate}
		\item[(i)] Let $H$ be an $N$-diagonal affine quadratic Hamiltonian, with $\U_t\in\ASp_0(\Hs)$ as in Definition~\ref{def:affine-quad}. Then there is a unique pair $(\Upsilon,\theta)$, with $\Upsilon$ self-adjoint on $\h$ and $\theta\in\h_{-1}(\Upsilon)$, such that $\U_t=\U_{(U_t,\nu_t)}$ for all $t\in\mathbb R$, where
		\begin{equation}
			U_t=e^{it\Upsilon}\oplus e^{-it\Upsilon^{\top}},\qquad
			\nu_t=(g_t,Cg_t),\qquad g_t\doteq f_t(\Upsilon)\theta .
			\label{eq:affine-data}
		\end{equation}
		Here $\Upsilon^{\top}\doteq C\Upsilon C$, so that $Ce^{it\Upsilon}C=e^{-it\Upsilon^{\top}}$. Moreover, $(\U_t)_{t\in\mathbb R}$ is a strongly continuous group on $\Hx$ whose generator is $i\mathbf K$, where
		\begin{equation}
			\mathbf K=
			\begin{pmatrix}
				\Upsilon&0&0\\
				0&-\Upsilon^{\top}&0\\
				\langle\theta,\cdot\,\rangle&-\langle C\theta,\cdot\,\rangle&0
			\end{pmatrix}
			\quad\text{on }\ D(\Upsilon)\oplus D(\Upsilon^{\top})\oplus\mathbb C ,
			\label{eq:K-block}
		\end{equation}
		and $\langle C\theta,y\rangle\doteq\overline{\langle\theta,Cy\rangle}$ for $y\in D(\Upsilon^{\top})=C\,D(\Upsilon)$ (this defines $C\theta\in\h_{-1}(\Upsilon^{\top})$).

		\item[(ii)] Conversely, for every self-adjoint $\Upsilon$ and every $\theta\in\h_{-1}(\Upsilon)$, there is an $N$-diagonal affine quadratic Hamiltonian $H_{\Upsilon,\theta}$ with data $(\Upsilon,\theta)$, unique up to an additive constant. It satisfies
		\begin{equation}
			e^{itH_{\Upsilon,\theta}}=e^{i\gamma_t}\,\Gamma(e^{it\Upsilon})\,\piF\big(W(\nu_t)\big),\qquad t\in\mathbb R,
			\label{eq:implementation}
		\end{equation}
		for a continuous $\gamma:\mathbb R\to\mathbb R$ with $\gamma_0=0$, which is unique up to $t\mapsto\gamma_t+Et$, $E\in\mathbb R$. For $\Upsilon\geq0$ with $\mathrm{Ker}\Upsilon=\{0\}$, the operators $H_{\Upsilon,\theta}$ are the van Hove Hamiltonians of \cite{Der03}, with $h=\Upsilon$ and $z=\theta$ (see Appendix~\ref{app:dico-Der}).
		\item[(iii)] Let $\Upsilon\geq0$ with $\mathrm{Ker}\Upsilon=\{0\}$, and $\theta\in\h$. Then one can choose $\gamma_t=-\int_0^{t}\operatorname{Im}\langle g_r,\theta\rangle_{\h}\,\mathrm dr$. The corresponding $H_{\Upsilon,\theta}$ satisfies $\Omega\in D(H_{\Upsilon,\theta})$, $\langle\Omega,H_{\Upsilon,\theta}\Omega\rangle=0$, and it is essentially self-adjoint on $D(\mathrm d\Gamma(\mathbf 1+\Upsilon))$, where it equals $\mathrm d\Gamma(\Upsilon)+a(\theta)+a^{*}(\theta)$.
	\end{enumerate}
\end{theorem}

\begin{proof}
	Part (i) applies Guichardet's structure theorem and computes the generator on a core. Part (ii) shows that the multiplier is a coboundary. Part (iii) checks the multiplier identity by hand and identifies $H$ with the closure of the formal sum.

	(i) \emph{Step 1: the data $(\Upsilon,\theta)$.} By Lemma~\ref{lem:type-S}, $(e^{itH})$ is a representation of type~(S), and $e^{itH}=U_{A_t,b_t,c_t}$. By \cite[Thm.~2.1]{Gui72}, $(A,b,c)\mapsto U_{A,b,c}$ is a bicontinuous group isomorphism onto its range, for the group law \cite[(2.8)]{Gui72} and the strong topologies. Hence $t\mapsto(A_t,b_t,c_t)$ is a continuous homomorphism, i.e., \cite[(3.1)--(3.3)]{Gui72} hold: $(A_t)$ is a strongly continuous unitary group, $b$ is a continuous $1$-cocycle, and $c$ solves $\delta c=\mathsf m$ for the multiplier (see Appendix~\ref{app:cohomology}) $\mathsf m(s,t)=e^{i\operatorname{Im}(b_s|A_sb_t)}$, where $(\cdot|\cdot)$ is the scalar product of \cite{Gui72}, linear in the first variable (Appendix~\ref{app:dico-Gui}). Stone's theorem gives $\Upsilon$, and Lemma~\ref{lem:cocycle} applied to $g_t=-iA_t^{*}b_t$ gives $\theta$. Uniqueness follows from the faithfulness of $\piF$ (since $\Wa$ is simple), the injectivity in Proposition~\ref{prop:affine-aut} and the uniqueness parts of Stone's theorem and Lemma~\ref{lem:cocycle}.

	\emph{Step 2: the generator.} By \eqref{eq:cocycle} and Lemma~\ref{lem:block}, $(\U_t)$ is a group, and it is strongly continuous on $\Hx$ because $t\mapsto g_t$ is norm-continuous (Lemma~\ref{lem:cocycle}(i)). It is a $C_0$-group of bounded, non-unitary operators, so we cannot use Stone's theorem. Let $G$ be its generator. The dense subspace $\mathcal D\doteq D(\Upsilon)\oplus D(\Upsilon^{\top})\oplus\mathbb C$ is $\U_t$-invariant. By \eqref{eq:U-block}, the bottom row of $\U_t$ is $(x,y)\mapsto i\langle g_t,x\rangle-i\langle Cg_t,y\rangle$, with $\langle Cg_t,y\rangle=\overline{\langle g_t,Cy\rangle}$. Since $t^{-1}g_t\to\theta$ in $\h_{-1}(\Upsilon)$ (Lemma~\ref{lem:cocycle}(i)), $\U_t\varphi$ is differentiable at $t=0$ for any $\varphi\in \mathcal D$. Therefore, $\mathcal D\subseteq D(G)$ and $G=i\mathbf K$ on $\mathcal D$. Hence $\mathcal D$ is a core for $G$ \cite[Proposition II.1.7]{EN00}. Finally, $\mathbf K$ is closed on $\mathcal D$: its diagonal part is closed on $D(\Upsilon)\oplus D(\Upsilon^{\top})$, and its bottom row is continuous for the graph norm of this diagonal part, since \[|\langle\theta,x\rangle|\leq\|\theta\|_{-1}\|(|\Upsilon|+1)x\|_{\h},\qquad x\in D(\Upsilon),\]
	and similarly $ |\langle C\theta,y\rangle|=|\langle\theta,Cy\rangle|\leq\|\theta\|_{-1}\|(|\Upsilon^{\top}|+\mathbf 1)y\|_{\h}$ for any $y\in D(\Upsilon^{\top})$. Therefore $G=i\mathbf K$ with domain $\mathcal D$.

	(ii) \emph{Step 1: the multiplier is a coboundary.} For any $t\in\mathbb R$, set
	\[b_t\doteq iu_tf_t(\Upsilon)\theta,\qquad u_t\doteq e^{it\Upsilon}.\]
	By Lemma~\ref{lem:cocycle}(i), $b$ is a continuous $1$-cocycle, and $-iu_t^{*}b_t=g_t$. We decompose
	\[ \h = \h'\oplus\h'',\quad \h'\doteq(\Ker\Upsilon)^{\perp},\quad \h''\doteq\Ker\Upsilon,\qquad b=P_{\h'}b+P_{\h''}b\doteq b' + b'',\]
	where $P_{\mathfrak k}$ denotes the orthogonal projection onto a closed subspace $\mathfrak k\subseteq\h$. Since $u_s$ commutes with $P_{\h''}$ and $P_{\h'}$, the multiplier $\mathsf m(s,t)=e^{i\operatorname{Im}(b_s|u_sb_t)}$ of Step 1 of (i), now built from this $b$, factorizes as $\mathsf m=\mathsf m'\mathsf m''$, with $\mathsf m'(s,t)=e^{i\operatorname{Im}(b'_s|u_sb'_t)}$ and $\mathsf m''(s,t)=e^{i\operatorname{Im}(b''_s|b''_t)}$. By \cite[Sect.~3.3, Case~1 (Lemma~3.3) and Case~3]{Gui72}, applied to $u|_{\h'}$, which has no non-zero invariant vector, $\mathsf m'=\delta c$ for some continuous $c$. On $\h''$, $b''$ is a continuous additive map $\mathbb R\to\h''$ (\cite[Sect.~3.3, Case~2]{Gui72}), hence $b''_t=t\,b''_1$, so that $\operatorname{Im}(b''_s|b''_t)=st\operatorname{Im}\|b''_1\|^{2}=0$, i.e., $\mathsf m''=1$. Thus $\mathsf m=\delta c$, and $c$ is unique up to a continuous character of $\mathbb R$, i.e., up to $t\mapsto e^{iEt}$, $E\in\mathbb R$. Note that $c_0=1$, since $\mathsf m(0,0)=1$.

	\emph{Step 2: the group and its generator.} By \cite[Thm.~2.1]{Gui72}, $t\mapsto U_{u_t,b_t,c_t}$ is then a strongly continuous unitary group, thus, it has a self-adjoint generator by Stone's theorem. We denote the generator by $H$. By Lemma~\ref{lem:type-S}, $H$ is an affine quadratic Hamiltonian with data $(\Upsilon,\theta)$, and \eqref{eq:implementation} holds with $\gamma$ such that $c_t=e^{i\gamma_t}$ for any $t\in\mathbb R$. Such a continuous $\gamma$ exists since $\mathbb R$ is simply connected.

	\emph{Step 3: uniqueness up to a constant.} If $\tilde H$ is another affine quadratic Hamiltonian with data $(\Upsilon,\theta)$, then, by Lemma~\ref{lem:type-S}, \[e^{it\tilde H}=\lambda_te^{itH},\quad \lambda_t\in\mathbb T,\qquad t\in\mathbb R.\]
	This in turn implies that $t\mapsto\lambda_t$ is a continuous character, $\lambda_t=e^{iEt}$, for some $E\in\mathbb R$. Consequently, $\tilde H=H+E$.

	\emph{Step 4: comparison with \cite{Der03}.} Let $\Upsilon\geq0$ with $\Ker\Upsilon=\{0\}$. For all $t\in\mathbb R$, $\Gamma(e^{it\Upsilon})\piF(W(\nu_t))$ corresponds to the unitary $V(t)$ of \cite[Sect.~3.3]{Der03} (see Appendix~\ref{app:dico-Der}). Since $\mathcal B(\Fb)'=\mathbb C\mathbf 1$, a self-adjoint operator $H'$ is a Hamiltonian of the dynamics $\beta^t=\mathrm{Ad}\,V(t)$ in the sense of \cite[(3.19)]{Der03} if and only if $e^{itH'}\in\mathbb T\,V(t)$ for all $t\in\mathbb R$. By Corollary~\ref{cor:implementation} and Lemma~\ref{lem:type-S}, this is equivalent to saying that $H'$ is an $N$-diagonal affine quadratic Hamiltonian with data $(\Upsilon,\theta)$. These are, by definition, the van Hove Hamiltonians of \cite[Sect.~3.4]{Der03}.

	(iii) \emph{Step 1: the phase and the group.} Let
	\[V_t\doteq e^{i\gamma_t}\Gamma(u_t)\piF(W(\nu_t)),\qquad t\in\mathbb R.\]
	By \eqref{eq:weyl}, \eqref{eq:bogotransfo}, \eqref{eq:cocycle} and $\sigma(U_s^{-1}\nu_t,\nu_s)=2\operatorname{Im}\langle u_s^{*}g_t,g_s\rangle_{\h}$, for any $t,s\in\mathbb R$, one has
	\[
	V_tV_s=e^{i(\gamma_t+\gamma_s-\operatorname{Im}\langle u_s^{*}g_t,g_s\rangle_{\h}-\gamma_{t+s})}\,V_{t+s}.
	\]
	Hence $(V_t)$ is a group as soon as the multiplier identity $\gamma_{t+s}-\gamma_t-\gamma_s=-\operatorname{Im}\langle u_s^{*}g_t,g_s\rangle_{\h}$ holds. This is the translation of \cite[(3.3)]{Gui72}. It holds: since $\theta\in\h$, $s\mapsto g_s$ is $C^{1}$ in $\h$, with derivative $s\mapsto u_s^{*}\theta$, and $g_s\in D(\Upsilon)$. Using  $\Upsilon g_s=i(u_s^{*}-\mathbf 1)\theta$, we compute that
	\[\partial_s\big(\gamma_{t+s}-\gamma_t-\gamma_s\big)=-\operatorname{Im}\langle u_s^{*}g_t,\theta\rangle_{\h} = -\partial_s\operatorname{Im}\langle u_s^{*}g_t,g_s\rangle_{\h},\qquad s\in\mathbb R.\]
	The functions $\gamma_{t+s}-\gamma_t-\gamma_s$ and $-\operatorname{Im}\langle u_s^{*}g_t,g_s\rangle_{\h}$ have the same derivative and vanish at $s=0$, so they coincide: the multiplier identity holds, and $(V_t)_{t\in\mathbb R}$ is a one-parameter unitary group, which is strongly continuous since $\gamma$, $t\mapsto g_t$ and $t\mapsto\Gamma(u_t)$ are continuous. By Stone's theorem and Lemma~\ref{lem:type-S}, its generator $H$ is an $N$-diagonal affine quadratic Hamiltonian with data $(\Upsilon,\theta)$, i.e., one of the operators $H_{\Upsilon,\theta}$ of (ii), and we fix this choice.

	\emph{Step 2: the vacuum lies in $D(H)$.} Since, for all $t\in\mathbb R$, $|\gamma_t|\leq t^{2}\|\theta\|^{2}/2$ and $\|g_t\|\leq|t|\|\theta\|$, and since $\langle\Omega,\piF(W((g,Cg)))\Omega\rangle=e^{-\|g\|^{2}/2}$ by \eqref{eq:vac-weyl}, we deduce that
	\[
	\langle\Omega,e^{itH}\Omega\rangle=e^{i\gamma_t}e^{-\|g_t\|^{2}/2}=1+O(t^{2}).
	\]
	This yields $\|(e^{itH}-\mathbf 1)\Omega\|^{2}=2-2\operatorname{Re}\langle\Omega,e^{itH}\Omega\rangle=O(t^{2})$. By using Fatou's lemma in the spectral representation of $H$, we obtain that $\Omega\in D(H)$, and then \[\langle\Omega,H\Omega\rangle=-i\partial_t\langle\Omega,e^{itH}\Omega\rangle|_{t=0}=0.\]
	Hence, since $\theta\in\h$, $H$ is the van Hove Hamiltonian of the first kind $H_{\mathrm I}$ of \cite[Sect.~3.4]{Der03}, by \cite[Thm.~3.2(2)]{Der03}.

	\emph{Step 3: $H$ is the closure of $\mathrm d\Gamma(\Upsilon)+a(\theta)+a^{*}(\theta)$.}
	By \cite[Prop.~3.11 and its proof]{Der03}, $A\doteq\mathrm d\Gamma(\Upsilon)+a(\theta)+a^{*}(\theta)$ is essentially self-adjoint on $D(\mathrm d\Gamma(\mathbf 1+\Upsilon))$. Its closure is $H$. Indeed, for $f\in D(\Upsilon)$, \eqref{eq:Weyl-exp} and $\Gamma(u)\epsilon(x)=\epsilon(ux)$ give $e^{itH}\epsilon(f)=e^{i\gamma_t-\|g_t\|^{2}/2-\langle ig_t,f\rangle}\,\epsilon(u_t(f+ig_t))$. Since $\gamma$ and $t\mapsto g_t$ are $C^{1}$ with $\gamma_0=0$, $\partial_t\gamma_t|_{t=0}=0$, $g_0=0$ and $\partial_tg_t|_{t=0}=\theta$, and since $\epsilon$ is Fréchet differentiable, with derivative $k\mapsto a^{*}(k)\epsilon(x)$ at $x\in\h$, this map is differentiable at $t=0$, with derivative $i\big(a^{*}(\Upsilon f)+\langle\theta,f\rangle+a^{*}(\theta)\big)\epsilon(f)=iA\epsilon(f)$, because $\mathrm d\Gamma(\Upsilon)\epsilon(f)=a^{*}(\Upsilon f)\epsilon(f)$ and $a(\theta)\epsilon(f)=\langle\theta,f\rangle\epsilon(f)$. By Stone's theorem, $H\supseteq A|_{\mathcal E}$, where $\mathcal E\doteq\mathrm{span}\{\epsilon(f):f\in D(\Upsilon)\}$. The space $\mathcal E$ is a core for $\mathrm d\Gamma(\mathbf 1+\Upsilon)$, since it is dense, contained in $D(\mathrm d\Gamma(\mathbf 1+\Upsilon))$ and invariant under $\Gamma(e^{is(\mathbf 1+\Upsilon)})$, $s\in\mathbb R$ \cite[Proposition II.1.7]{EN00}, and $A$ is $\mathrm d\Gamma(\mathbf 1+\Upsilon)$-bounded, since $a(\theta)$ and $a^{*}(\theta)$ are $\mathrm d\Gamma(\mathbf 1)^{1/2}$-bounded. Hence the closure of $A|_{\mathcal E}$ contains $A|_{D(\mathrm d\Gamma(\mathbf 1+\Upsilon))}$, so that it is self-adjoint; as a self-adjoint restriction of $H$, it equals $H$.
\end{proof}

 Theorem~\ref{thm:affine} is the present technical section version of Theorem~\ref{thm:intro-vH}(i)--(ii) and the template for Section~\ref{sec:spin} (Remark~\ref{rem:sd-cocycle}); part~(iii) is used for Corollary~\ref{cor:sd-uniqueness}.

\begin{remark}[Known results and cohomology]
	\label{rem:vanhove}
	Theorem~\ref{thm:affine} is a reformulation of known results. More precisely, (i) and the existence part of (ii) are the case of the group $G=\mathbb R$ in Guichardet's theory of representations of type~(S) \cite{Gui72}, and (iii) is in \cite{Der03}. What the self-dual formulation adds is the description through the group $\ASp_0(\Hs)$ and the generator $\mathbf K$, parallel to the fermionic theory of \cite[Section~3]{Quadratic}. It also gives a cohomological reading of the infrared problem. By Lemma~\ref{lem:cocycle} and \eqref{eq:cocycle-std}, $Z^{1}(\mathbb R,e^{it\Upsilon})\cong\h_{-1}(\Upsilon)$ and $B^{1}\cong\Upsilon\h$, so that
	\[
	H^{1}\big(\mathbb R,e^{it\Upsilon}\big)\cong\h_{-1}(\Upsilon)/\Upsilon\h .
	\]
	For $\Upsilon\geq0$ with $\Ker\Upsilon=\{0\}$, the class of $\theta$ vanishes if and only if $H_{\Upsilon,\theta}$ is unitarily equivalent to $\mathrm d\Gamma(\Upsilon)+E$ by a Weyl operator, if and only if it has a ground state (Proposition~\ref{prop:hierarchy}). When $\theta\notin\h+\Upsilon^{1/2}\h$ (Remark~\ref{rem:hierarchy}), $\Omega$ is not even in the form domain, the additive constant is ``infinite'', and $H_{\Upsilon,\theta}$ is a renormalized operator, cf.\ \cite[Thm.~3.2(2) and Sect.~3.4; Sect.~1.1(3) and Thm.~3.4(3) in the massive case]{Der03}.
\end{remark}

\Needspace{10\baselineskip}
\begin{proposition}[Generator conditions]
	\label{prop:hierarchy}
	Let $\Upsilon\geq0$ with $\mathrm{Ker}\Upsilon=\{0\}$, let $\theta\in\h_{-1}(\Upsilon)$, and let $H=H_{\Upsilon,\theta}$.
	\begin{enumerate}
		\item[(i)] $\Omega\in D(H)$ iff $\theta\in\h$.
		\item[(ii)] If $\Upsilon^{-1}\theta\in\h$, then, with $\rho\doteq(-i\Upsilon^{-1}\theta,\,C(-i\Upsilon^{-1}\theta))\in\Hs_{\Ar}$,
		\[
		H=\piF(W(\rho))^{*}\,\mathrm d\Gamma(\Upsilon)\,\piF(W(\rho))+E\mathbf 1\qquad\text{for some }E\in\mathbb R .
		\]
		\item[(iii)] $H$ has a ground state iff $\Upsilon^{-1}\theta\in\h$.
	\end{enumerate}
\end{proposition}

\begin{proof}
	(i) ``$\Leftarrow$'' is shown in the proof of Theorem~\ref{thm:affine}(iii). For ``$\Rightarrow$'', note that $\Gamma(e^{it\Upsilon})\Omega=\Omega$, so \eqref{eq:implementation} and \eqref{eq:vac-weyl} give
	\[
	\|(e^{itH}-\mathbf 1)\Omega\|^{2}=2-2\operatorname{Re}\big(e^{i\gamma_t}e^{-\|g_t\|^{2}/2}\big)\geq2\big(1-e^{-\|g_t\|^{2}/2}\big).
	\]
	If $\Omega\in D(H)$, the left-hand side is $O(t^{2})$, hence $\|g_t\|=O(|t|)$. Since $t^{-1}g_t\to\theta$ in $\h_{-1}(\Upsilon)$ (Lemma~\ref{lem:cocycle}(i)), weak compactness of balls in $\h$ gives $\theta\in\h$.

	(ii) Write $r\doteq-i\Upsilon^{-1}\theta$, so that $\rho=(r,Cr)$. Then $g_t=r-u_t^{*}r$ (Lemma~\ref{lem:cocycle}(iii)), hence $\nu_t=\rho-U_{-t}\rho$. By \eqref{eq:weyl} and \eqref{eq:bogotransfo}, $\piF(W(\rho))^{*}\,\Gamma(e^{it\Upsilon})\,\piF(W(\rho))$ coincides, up to a phase, with $\Gamma(e^{it\Upsilon})\piF(W(\nu_t))$, hence with $e^{itH}$, by \eqref{eq:implementation}. Two strongly continuous unitary groups that agree up to phases differ by a character $e^{iEt}$, which gives the formula. This is also \cite[Thm.~3.2(4) and (3.20)]{Der03}, where the dressing operator is $\piF(W(\rho))^{*}$.

	(iii) This is \cite[Thm.~3.2(4)]{Der03}. Its proof is in \cite[Prop.~3.13]{Der03}. 
\end{proof}

Parts~(i) and~(iii) are Theorem~\ref{thm:intro-vH}(iii); the formula $\nu_t=\rho-U_{-t}\rho$ of~(ii) is used again in Remark~\ref{rem:sd-asymptotic}.

\begin{remark}[Form domain and lower bound]
	\label{rem:hierarchy}
	Part (i) is the bosonic affine counterpart of \cite[Theorem~3.9]{Quadratic} (Shale--Stinespring-like condition for generators). Its form-domain version, namely $\Omega\in D(|H|^{1/2})$ if and only if $\theta\in\h+\Upsilon^{1/2}\h$, is \cite[Thm.~3.2(2)]{Der03}. Moreover, $H$ is bounded from below if and only if $\theta\in\Upsilon^{1/2}\h+\Upsilon\h$ \cite[Thm.~3.2(3)]{Der03}.
\end{remark}

\section{Spin--Boson Hamiltonians as Spin-Dressed Implementations}
\label{sec:spin}

For a field coupled to a spin (we abuse the language here by calling spin any $n$-level subsystem, $n\in\mathbb N + 1$), we need to replace $\mathbb C$ by $\mathbb C^{n}$, $n\geq2$, the group $\ASp_0(\Hs)$ of Section~\ref{sec:affine} is no longer available. We replace it by a dynamical definition, that of a \emph{spin-dressed implementation} of a Bogoliubov flow. The structure theory (Theorem~\ref{thm:sd-structure}) and the characterization in Fock space for regular form factors (Theorem~\ref{thm:sd-fock}) still extend to $n\geq2$ those of Section~\ref{sec:affine}.\\

We first rewrite, in the self-dual notation, the formal computation of Subsection~\ref{subsec:main-sd} which motivates our definition. Recall that $\Phi\doteq\Phi_{\piF}$ is the Segal field of the Fock representation, and that \eqref{eq:sd-scalar-fields} is the scalar case. Let $M_n^{\mathrm{sa}}\subset\mathcal B(\mathbb C^{n})$ be the real vector space of self-adjoint $n\times n$ matrices. For a self-adjoint operator $\Upsilon$ on $\h$ and $m\in\mathbb N$, the spin--boson Hamiltonian with spin Hamiltonian $E$, coupling matrices $\tau_l$ and form factors $\vartheta_l$ reads, formally,
\begin{equation}
	H=E\otimes\mathbf 1+\mathbf 1\otimes\mathrm d\Gamma(\Upsilon)+\sum_{l=1}^{m}\tau_l\otimes\Phi(\vartheta_l),\qquad E,\tau_l\in M_n^{\mathrm{sa}},\quad \vartheta_l\in\Hs_{\Ar}.
	\label{eq:sd-formal-H}
\end{equation}
Let $K\doteq\Upsilon\oplus(-\Upsilon^{\top})$ be the diagonal part of $\mathbf K$ in \eqref{eq:K-block}, so that $e^{itK}=e^{it\Upsilon}\oplus e^{-it\Upsilon^{\top}}$ is the linear part of $\U_t$ in \eqref{eq:affine-data} (the free flow). By Remark~\ref{rem:selfdual-ccr}, $[\Phi(\psi),\Phi(\varphi)]=i\sigma(\psi,\varphi)\mathbf 1$, and, by \eqref{eq:bogotransfo}, $\Gamma(e^{it\Upsilon})\Phi(\psi)\Gamma(e^{it\Upsilon})^{*}=\Phi(e^{itK}\psi)$. Hence, formally,
\begin{equation}
	i\big[H,\mathbf 1\otimes\Phi(\psi)\big]=\mathbf 1\otimes\Phi(iK\psi)+\Theta(\psi)\otimes\mathbf 1,\qquad \Theta(\psi)\doteq\sum_{l=1}^{m}\sigma(\psi,\vartheta_l)\,\tau_l .
	\label{eq:sd-formal-deriv}
\end{equation}
The derivation $i[H,\cdot\,]$ therefore maps fields to fields, up to the matrix $\Theta(\psi)$, which is the rate at which the field is displaced. Integrating \eqref{eq:sd-formal-deriv} in time gives, still formally,
\begin{equation}
	e^{itH}\big(\mathbf 1\otimes\Phi(\psi)\big)e^{-itH}=\mathbf 1\otimes\Phi(e^{itK}\psi)+\int_0^{t}e^{isH}\big(\Theta(e^{i(t-s)K}\psi)\otimes\mathbf 1\big)e^{-isH}\,\mathrm ds .
	\label{eq:sd-formal-langevin}
\end{equation}
\newline

This leads us to a new definition. For $n=1$, \eqref{eq:sd-formal-langevin} is \eqref{eq:sd-scalar-fields}. For $n\geq2$, the displacement (the integral term) is a bounded operator built from the spin observables evolved by the interacting dynamics, and in general it does not commute with the fields. Composing two time steps, the displacement of the step applied first is twisted by the dynamics of the second one \eqref{eq:sd-cocycle}. Hence there is no analogue of $\ASp_0(\Hs)$ (Remarks~\ref{rem:sd-cocycle} and~\ref{rem:sd-nogo}). We keep the structure of \eqref{eq:sd-formal-langevin} as a definition (Definition~\ref{def:sd-impl}): $H$ is a spin-dressed implementation of the Bogoliubov flow if it implements the flow up to bounded displacements whose derivative at $t=0$ is a spin observable $\Theta(\psi)\otimes\mathbf 1$.

The displacements are required to be small as $t\to0$ in the norm $\|\Lambda_L\psi\|$ depending on a self-adjoint operator $L$ on $\h$, the regularity scale (Subsection~\ref{subsec:sd-def}). At scale $L=\Upsilon$, the definition captures all van Hove Hamiltonians, renormalized ones included (Corollary~\ref{cor:sd-scalar-scale}). Subsection~\ref{subsec:sd-displaced} introduces the representations in which the definition is phrased, \ref{subsec:sd-def} the structure theory, \ref{subsec:sd-fock} the Fock-space characterization, and \ref{subsec:sd-renorm} singular commuting couplings at the scale $\Upsilon$ (uniqueness there is open, Remark~\ref{rem:sd-open}).

\subsection{Displaced Representations}
\label{subsec:sd-displaced}

Conjugating a representation of $\Wa$ by $e^{itH}$, for any $t\in\mathbb R$ gives a new representation of $\Wa$ whose fields, by \eqref{eq:sd-formal-langevin}, are the original fields modified by a Bogoliubov transformation plus a bounded displacement. If the displacement is not a scalar, the canonical commutation relations constrain it. The admissibility we define below encodes this.

\begin{definition}[Displaced representations]
	\label{def:sd-displaced}
	Let $\varrho$ be a regular representation of $\Wa$ on a Hilbert space $\Kc$, and let $\Delta:\Hs_{\Ar}\to\mathcal B(\Kc)$ be a bounded real-linear map with self-adjoint values. We say that $\Delta$ is \emph{admissible} for $\varrho$ if the unitaries
	\begin{equation}
		V_\Delta(\psi)\doteq e^{i(\Phi_\varrho(\psi)+\Delta(\psi))},\qquad\psi\in\Hs_{\Ar},
		\label{eq:sd-V}
	\end{equation}
	satisfy \eqref{eq:weyl}. In this case, the representation $\varrho^{\Delta}$ of $\Wa$ such that $\varrho^{\Delta}(W(\psi))=V_\Delta(\psi)$ for all $\psi\in\Hs_{\Ar}$ is called the \emph{displacement of $\varrho$ by $\Delta$}.
\end{definition}

For $\psi\in\Hs_\Ar$, $\Phi_\varrho(\psi)+\Delta(\psi)$ is symmetric by definition and is the bounded perturbation of a self-adjoint operator on $D(\Phi_\varrho(\psi))$, thus it is self-adjoint. Since $\Phi_\varrho(-\psi)=-\Phi_\varrho(\psi)$, one has $V_\Delta(\psi)^{*}=V_\Delta(-\psi)$, so that $\varrho^{\Delta}$ exists by universality of $\Wa$, and it is unique. It is regular and, by Stone's theorem,
\begin{equation}
	\Phi_{\varrho^{\Delta}}(\psi)=\Phi_\varrho(\psi)+\Delta(\psi),\qquad\psi\in\Hs_{\Ar},
	\label{eq:sd-displaced-field}
\end{equation}
since $\Phi_\varrho(s\psi)=s\Phi_\varrho(\psi)$ and $\Delta(s\psi)=s\Delta(\psi)$ for $s\in\mathbb R$. Formally, admissibility translates as the following relations prescribed by the CCR:
\begin{equation}
	[\Phi_\varrho(\psi),\Delta(\varphi)]+[\Delta(\psi),\Phi_\varrho(\varphi)]+[\Delta(\psi),\Delta(\varphi)]=0,\qquad\psi,\varphi\in\Hs_{\Ar}.
	\label{eq:sd-compensation}
\end{equation}
That is, the non-commutativity of the displacements must be compensated by their defect of commutativity with the fields. This identity is formal while admissibility is the rigorous condition. For displacements commuting with the fields it reduces to $[\Delta(\psi),\Delta(\varphi)]=0$, see Lemma~\ref{lem:sd-displaced}(ii) below.

\begin{lemma}[Basic properties of displaced representations]
	\label{lem:sd-displaced}
	Let $\varrho$ be a regular representation of $\Wa$ on a Hilbert space $\Kc$.
	\begin{enumerate}
		\item[(i)] For every $\nu\in\Hs_{\Ar}$, $\Delta=\sigma(\cdot,\nu)\mathbf 1$ is admissible for $\varrho$, and $\varrho^{\Delta}=\varrho\circ\chi_{\U_{(\mathbf 1,\nu)}}$.
		\item[(ii)] Let $\Delta$ take values in $\varrho(\Wa)'$. Then $\Delta$ is admissible for $\varrho$ if and only if $[\Delta(\psi),\Delta(\varphi)]=0$ for all $\psi,\varphi\in\Hs_{\Ar}$. In this case, $\varrho^{\Delta}(W(\psi))=\varrho(W(\psi))\,e^{i\Delta(\psi)}$, and $\psi\mapsto e^{i\Delta(\psi)}$ is a norm-continuous unitary representation of the additive group $\Hs_{\Ar}$.
		\item[(iii)] (Rigidity.) Let $\psi_1,\psi_2\in\Hs_{\Ar}$ and $B,B'\in\mathcal B(\Kc)$ be such that $\Phi_\varrho(\psi_1)+B=\Phi_\varrho(\psi_2)+B'$. Then $\psi_1=\psi_2$ and $B=B'$.
	\end{enumerate}
\end{lemma}

Part (i) contains the scalar displacements of Section~\ref{sec:affine}. It is used for $n=1$ in Corollary~\ref{cor:sd-scalar}. Part (ii) applies to the spin-valued displacements of Remarks~\ref{rem:sd-nogo} and~\ref{rem:sd-dressing}, which commute with the fields. Part (iii) is rigidity: it makes the data in (D1) unique (see Theorem~\ref{thm:sd-structure}(i)).

\begin{proof}
	(i) By Lemma~\ref{lem:field-cov} with $U=\mathbf 1$, \[\Phi_{\varrho\circ\chi_{\U_{(\mathbf 1,\nu)}}}(\psi)=\Phi_\varrho(\psi)+\sigma(\psi,\nu)\mathbf 1,\qquad \psi\in\Hs_{\Ar}.\]
	 Hence $V_\Delta(\psi)=\varrho\circ\chi_{\U_{(\mathbf 1,\nu)}}(W(\psi))$, and these unitaries satisfy \eqref{eq:weyl}.

	(ii) Let $\psi\in\Hs_\Ar$. Since $\Delta(\psi)$ commutes with the unitary group $s\mapsto\varrho(W(s\psi))=e^{is\Phi_\varrho(\psi)}$, the maps $s\mapsto\varrho(W(s\psi))e^{is\Delta(\psi)}$ and $s\mapsto e^{is\Delta(\psi)}\varrho(W(s\psi))$ are strongly continuous unitary groups, whose generators extend the self-adjoint operator $\Phi_\varrho(\psi)+\Delta(\psi)$, hence are equal to it. Thus $V_\Delta(\psi)=\varrho(W(\psi))e^{i\Delta(\psi)}=e^{i\Delta(\psi)}\varrho(W(\psi))$.

	Next, \eqref{eq:weyl} for $V_\Delta$ holds if and only if the values of $\Delta$ commute. If $[\Delta(\psi),\Delta(\varphi)]=0$ for all $\psi,\varphi\in\Hs_\Ar$, then $e^{i\Delta(\psi)}e^{i\Delta(\varphi)}=e^{i\Delta(\psi+\varphi)}$, and \eqref{eq:weyl} holds for $V_\Delta$. Conversely, \eqref{eq:weyl} implies $W(s\psi)W(r\varphi)W(s\psi)^{*}=e^{-irs\sigma(\psi,\varphi)}W(r\varphi)$ for $r,s\in\mathbb R$ and $\psi,\varphi\in\Hs_\Ar$. If \eqref{eq:weyl} holds for $V_\Delta$ then the previous equality also holds for $V_\Delta$ and this in turn implies that
	\[
	e^{is\Delta(\psi)}e^{ir\Delta(\varphi)}e^{-is\Delta(\psi)}=e^{ir\Delta(\varphi)},\qquad r,s\in\mathbb R,
	\]
	since the factors $e^{i\Delta(\cdot)}$ commute with $\varrho(\Wa)$. Differentiating in both variables at $r=s=0$ gives $[\Delta(\psi),\Delta(\varphi)]=0$.

	Finally, the values of $\Delta$ commute, so $e^{i\Delta(\psi)}e^{i\Delta(\varphi)}=e^{i\Delta(\psi+\varphi)}$, and this gives a unitary representation of $\Hs_\Ar$. We estimate \[\|e^{i\Delta(\psi)}-e^{i\Delta(\varphi)}\|\leq\|\Delta(\psi-\varphi)\|,\qquad \psi,\varphi\in\Hs_\Ar.\]
    Since $\Delta$ is bounded, this gives the norm-continuity of $\psi\mapsto e^{i\Delta(\psi)}$.

	(iii) Let $X\doteq\Phi_\varrho(\psi_1)$ and $Y\doteq\Phi_\varrho(\psi_2)$. The assumption means that $\mathcal D\doteq D(X)=D(Y)$ and $(X-Y)\zeta=(B'-B)\zeta$ for $\zeta\in\mathcal D$. For $\varphi\in\Hs_{\Ar}$ and $s\in\mathbb R$, let $V_s\doteq\varrho(W(s\varphi))$. By \eqref{eq:segal-weyl} and having in mind that $V_s^{*}\mathcal D=\mathcal D$, we compute that for any $s\in\mathbb R$,
	\[
 V_s\left(X-Y\right)V^*_s\zeta =	(B'-B)\zeta-s\,\sigma(\varphi,\psi_1-\psi_2)\zeta=V_s(B'-B)V_s^{*}\zeta,\qquad\zeta\in\mathcal D .
	\]
	Since $\mathcal D$ is dense, $|s\,\sigma(\varphi,\psi_1-\psi_2)|\leq2\|B'-B\|$ for all $s\in\mathbb R$, i.e., $\sigma(\varphi,\psi_1-\psi_2)=0$ for all $\varphi\in\Hs_{\Ar}$. Thus $\psi_1=\psi_2$ by non-degeneracy of $\sigma$, and then $B=B'$.
\end{proof}

\subsection{Spin-Dressed Implementations}
\label{subsec:sd-def}

Let $n\in\mathbb N$. For a regular representation $\pi$ of $\Wa$ on $\Kc_\pi$, the representation $\pi^{(n)}\doteq\mathbf 1\otimes\pi$ of $\Wa$ on $\mathbb C^{n}\otimes\Kc_\pi$ is regular, with fields 
\begin{equation}\label{eq:extendedfields}
\Phi_{\pi^{(n)}}(\psi)=\mathbf 1\otimes\Phi_\pi(\psi).
\end{equation}
Its commutant is $\pi^{(n)}(\Wa)'=M_n\otimes\pi(\Wa)'$, which is $M_n\otimes\mathbb C\mathbf 1$ if $\pi$ is irreducible. For $U\in\Sp(\Hs)$, we write $\chi_U\doteq\chi_{\U_{(U,0)}}$, so that $\Phi_{\pi^{(n)}\circ\chi_U}(\psi)=\mathbf 1\otimes\Phi_\pi(U\psi)$ by Lemma~\ref{lem:field-cov}.

In Definition~\ref{def:affine-quad}, $\mathrm{Ad}\,e^{itH}$ implements the affine automorphism $\chi_{\U_t}$, that is, the Bogoliubov part $\chi_{U_t}$ followed by the scalar displacement of \eqref{eq:sd-scalar-fields}. Definition~\ref{def:sd-impl} below keeps $\chi_{U_t}$ and replaces the scalar by a bounded operator-valued displacement $\Delta_t$. The group property of $(U_t)$ and a cocycle identity for $(\Delta_t)$ are then consequences (Theorem~\ref{thm:sd-structure}).

\paragraph{Regularity scales.}
Let $L$ be a self-adjoint operator on $\h$ that we call a \emph{regularity scale}, and let $L^{\top}\doteq CLC$. We set
\[
\Lambda_L\doteq\big(|L|+\mathbf 1\big)\oplus\big(|L^{\top}|+\mathbf 1\big),\qquad \mathcal V_L\doteq\Hs_{\Ar}\cap D(\Lambda_L)=\{(f,Cf):f\in D(L)\}.
\]
Since $C(|L|+\mathbf 1)=(|L^{\top}|+\mathbf 1)C$, the self-adjoint operator $\Lambda_L\geq\mathbf 1$ commutes with $S$ and $\Ar$. Hence $\mathcal V_L$ is dense in $\Hs_{\Ar}$, and $\|\Lambda_L(f,Cf)\|_{\Hs}=\sqrt2\,\|(|L|+\mathbf 1)f\|_{\h}$. The main cases of interest are $L=0$, for which $\Lambda_0=\mathbf 1$ and $\mathcal V_0=\Hs_{\Ar}$, and $L=\Upsilon$, the one-particle Hamiltonian, for which $\Lambda_\Upsilon=|K|+\mathbf 1$. The scale enters through the weighted norm $\|\Lambda_L\psi\|$ in the bound of (D2$_L$) below, it discriminates which displacements are allowed. A scale with larger $\|\Lambda_L\psi\|$ weakens this bound and so accepts more singular coupling maps. At the scale $L=\Upsilon$ of the free flow, the renormalized van Hove Hamiltonians appear (Corollary~\ref{cor:sd-scalar-scale}). The symplectic form extends to the extrapolation space $\h_{-1}(L)$ of Subsection~\ref{subsec:affine-ham} (defined with $L$ in place of $\Upsilon$).

For $w\in\h_{-1}(L)$, let $Cw\in\h_{-1}(L^{\top})$ be defined as in Theorem~\ref{thm:affine}(i), and, for $\psi=(f,Cf)\in\mathcal V_L$, set
\begin{equation}
	\sigma\big(\psi,(w,Cw)\big)\doteq2\,\mathrm{Im}\langle f,w\rangle,\qquad \langle f,w\rangle\doteq\overline{\langle w,f\rangle}.
	\label{eq:sd-sigma-ext}
\end{equation}
Here $\langle f,w\rangle$ is the extension by continuity of the scalar product of $\h$: for $f\in D(L)$ and $w\in\h_{-1}(L)$, $\langle f,w\rangle=\langle(|L|+\mathbf 1)f,J_Lw\rangle_{\h}$, where $J_L:\h_{-1}(L)\to\h$ is the unitary extension of $(|L|+\mathbf 1)^{-1}$.
Let $\Hs_{-1,\Ar}(L)\doteq\{(w,Cw):w\in\h_{-1}(L)\}\subseteq\h_{-1}(L)\oplus\h_{-1}(L^{\top})$. Thus \eqref{eq:sd-sigma-ext} is the duality pairing $\mathcal V_L\times\Hs_{-1,\Ar}(L)\to\mathbb R$ induced by $\sigma$. For $w\in\h$, it is the symplectic form of $\Hs_{\Ar}$. In general, 
\[|\sigma(\psi,(w,Cw))|\leq\sqrt2\,\|w\|_{-1}\|\Lambda_L\psi\|,\]
 and the map $w\mapsto\sigma(\cdot,(w,Cw))$ is injective.

\begin{definition}[Spin-dressed implementations]
	\label{def:sd-impl}
	Let $\pi$ be a regular representation of $\Wa$ on $\Kc_\pi$, let $(U_t)_{t\in\mathbb R}\subseteq\Sp(\Hs)$, let $L$ be a regularity scale, and let $\Theta:\mathcal V_L\to M_n^{\mathrm{sa}}$ be a map. A self-adjoint operator $H$ on $\mathbb C^{n}\otimes\Kc_\pi$ \emph{$\Theta$-implements $(U_t)_{t\in\mathbb R}$ in $\pi$ at scale $L$} if, for every $t\in\mathbb R$, there is a bounded real-linear map $\Delta_t:\Hs_{\Ar}\to\mathcal B(\mathbb C^{n}\otimes\Kc_\pi)$ with self-adjoint values such that:
	\begin{enumerate}
		\item[(D1)] $\Delta_t$ is admissible for $\pi^{(n)}\circ\chi_{U_t}$, and
		\[
		e^{itH}\,\pi^{(n)}(A)\,e^{-itH}=\big(\pi^{(n)}\circ\chi_{U_t}\big)^{\Delta_t}(A),\qquad A\in\Wa .
		\]
		\item[(D2$_L$)] There is $C_\Delta<\infty$ such that $\|\Delta_t(\psi)\|\leq C_\Delta|t|\,\|\Lambda_L\psi\|$ for all $\psi\in\mathcal V_L$ and $|t|\leq1$, and, for every $\psi\in\mathcal V_L$, $t^{-1}\Delta_t(\psi)\to\Theta(\psi)\otimes\mathbf 1$ in the strong topology as $t\to0$.
	\end{enumerate}
	The maps $\Delta_t$ are called the \emph{displacements} of $H$, and $\Theta$ its \emph{coupling map}. At the scale $L=0$, (D2$_L$) reads
	\begin{enumerate}
		\item[(D2$_0$)] $\sup_{0<|t|\leq1}|t|^{-1}\|\Delta_t\|<\infty$, and, for every $\psi\in\Hs_{\Ar}$, $t^{-1}\Delta_t(\psi)\to\Theta(\psi)\otimes\mathbf 1$ in the strong topology as $t\to0$.
	\end{enumerate}
	In this case, we simply say that $H$ \emph{$\Theta$-implements $(U_t)_{t\in\mathbb R}$ in $\pi$}.
\end{definition}

By \eqref{eq:sd-displaced-field}, (D1) is equivalent to the field identity
\begin{equation}
	e^{itH}\big(\mathbf 1\otimes\Phi_\pi(\psi)\big)e^{-itH}=\mathbf 1\otimes\Phi_\pi(U_t\psi)+\Delta_t(\psi),\qquad\psi\in\Hs_{\Ar},
	\label{eq:sd-D1}
\end{equation}
as self-adjoint operators, or, in Weyl form, to
\begin{equation}
	e^{itH}\,\pi^{(n)}(W(\psi))\,e^{-itH}=e^{i(\mathbf 1\otimes\Phi_\pi(U_t\psi)+\Delta_t(\psi))},\qquad\psi\in\Hs_{\Ar}.
	\label{eq:sd-D1-weyl}
\end{equation}
Indeed, $\mathrm{Ad}\,e^{itH}\circ\pi^{(n)}$ is a regular representation of $\Wa$ whose fields are the left-hand side of \eqref{eq:sd-D1}, so that \eqref{eq:sd-D1} implies that $\Delta_t$ is admissible.

Thus, $\mathrm{Ad}\,e^{itH}$ implements the Bogoliubov automorphism $\chi_{U_t}$ up to a bounded displacement of the fields, whose generator $\Theta$ is spin-valued. As in Definition~\ref{def:affine-quad}, neither the group property nor the continuity of $t\mapsto U_t$ is assumed. The displacements are required neither to commute with the fields, nor to take values in $M_n\otimes\mathbf 1$. 

At every scale, each $\Delta_t$ is a bounded map on all of $\Hs_{\Ar}$. The scale is used only to measure the rate at which the displacements vanish as $t\to0$. Scales can be compared: if $D(L)\subseteq D(L')$ and $\|\Lambda_{L'}\psi\|\leq c\|\Lambda_L\psi\|$ for $\psi\in\mathcal V_L$, then a $\Theta$-implementation at scale $L'$ is a $\Theta|_{\mathcal V_L}$-implementation at scale $L$. Since $\|\Lambda_L\psi\|\geq\|\psi\|$, a $\Theta$-implementation at the scale $0$ is thus a $\Theta|_{\mathcal V_L}$-implementation at every scale $L$.

\Needspace{10\baselineskip}
\begin{theorem}[Structure of spin-dressed implementations]
	\label{thm:sd-structure}
	Let $\pi$, $\Theta$, $L$ and $U$ be as in Definition \ref{def:sd-impl}. Let $H$ $\Theta$-implement $(U_t)_{t\in\mathbb R}$ in $\pi$ at scale $L$, with displacements $(\Delta_t)_{t\in\mathbb R}$, and let $\alpha_t\doteq\mathrm{Ad}\,e^{itH}$ on $\mathcal B(\mathbb C^{n}\otimes\Kc_\pi)$, $t\in\mathbb R$.
	\begin{enumerate}
		\item[(i)] For every $t\in\mathbb R$, $U_t$ and $\Delta_t$ are uniquely determined by $H$ and $\pi$. Moreover, $U_0=\mathbf 1$, $\Delta_0=0$, and $U_{t+s}=U_tU_s$ for all $s,t\in\mathbb R$. If $\pi=\piF$, and if either $L=0$ or $\sup_{|t|\leq1}\|U_t\|<\infty$, then $t\mapsto U_t$ is strongly continuous.
		\item[(ii)] The displacements satisfy the cocycle identity
		\begin{equation}
			\Delta_{t+s}(\psi)=\Delta_t(U_s\psi)+\alpha_t\big(\Delta_s(\psi)\big),\qquad s,t\in\mathbb R,\ \psi\in\Hs_{\Ar}.
			\label{eq:sd-cocycle}
		\end{equation}
		\item[(iii)] $\Theta$ is real-linear, and $\|\Theta(\psi)\|\leq C_\Delta\|\Lambda_L\psi\|$ for $\psi\in\mathcal V_L$, with $C_\Delta$ as in (D2$_L$). For every basis $(\tau_l)_{l=1}^{n^{2}}$ of $M_n^{\mathrm{sa}}$, there are unique $w_l\in\h_{-1}(L)$, $l=1,\dots,n^{2}$, such that, upon $\vartheta_l\doteq(w_l,Cw_l)\in\Hs_{-1,\Ar}(L)$ and the notation \eqref{eq:sd-sigma-ext},
		\begin{equation}
			\Theta(\psi)=\sum_{l=1}^{n^{2}}\sigma(\psi,\vartheta_l)\,\tau_l,\qquad\psi\in\mathcal V_L.
			\label{eq:sd-tensor}
		\end{equation}
		For $L=0$, the $\vartheta_l$ are vectors of $\Hs_{\Ar}$.
		\item[(iv)] If $t\mapsto U_t$ is strongly continuous and $U_t$ commutes with $\Lambda_L$ for every $t\in\mathbb R$ (the commutation is automatic if $L=0$), then, for all $t\in\mathbb R$ and $\psi\in\mathcal V_L$,
		\begin{equation}
			\Delta_t(\psi)=\int_0^{t}\alpha_s\big(\Theta(U_{t-s}\psi)\otimes\mathbf 1\big)\,\mathrm ds ,
			\label{eq:sd-langevin}
		\end{equation}
		where the integral converges in the strong operator topology. In particular, for any $t\in\mathbb R$, \[\|\Delta_t(\psi)\|\leq|t|\,\|\Theta\|_L\sup_{|s|\leq|t|}\|U_s\Lambda_L\psi\|,\qquad \|\Theta\|_L\doteq\sup\big\{\|\Theta(\varphi)\|:\varphi\in\mathcal V_L,\ \|\Lambda_L\varphi\|\leq1\big\},\] and, for every $\psi\in\Hs_{\Ar}$, $\Delta_t(\psi)$ belongs to the von Neumann algebra generated by $\{\alpha_s(M_n\otimes\mathbf 1): s\in[0,t]\}$, where $[0,t]\doteq[t,0]$ if $t<0$.
	\end{enumerate}
\end{theorem}

Part (iv), together with \eqref{eq:sd-D1}, is the rigorous form of \eqref{eq:sd-formal-langevin}. In physical terms, it is the generalized quantum Langevin equation of the field: the field evolves following the kinetic part of the model, plus a displacement generated by the spin memory. Part (iii) reads the form factors, that appear in the formal Hamiltonian \eqref{eq:sd-formal-H} as the vectors $\vartheta_l$, from the dynamics.

Part (i) also shows that the data of Definition~\ref{def:intro-sd} are determined by $H$. Indeed, let $H$ $\Theta$-implement $(e^{itK})_{t\in\mathbb R}$ in $\pi$ at a scale $L$, with $K=\Upsilon\oplus(-\Upsilon^{\top})$, and $\Theta'$-implement $(e^{itK'})_{t\in\mathbb R}$ in $\pi$ at a scale $L'$, with $K'=\Upsilon'\oplus(-\Upsilon'^{\top})$ and displacements $\Delta'_t$. By (i), which only uses (D1), $e^{itK'}=e^{itK}$ and $\Delta'_t=\Delta_t$ for all $t\in\mathbb R$. Hence $\Upsilon'=\Upsilon$, by the uniqueness in Stone's theorem, and $\Theta'=\Theta$ on $\mathcal V_L\cap\mathcal V_{L'}$, by (D2$_L$) and (D2$_{L'}$). This is the first assertion of Theorem~\ref{thm:intro-structure}(i).

\begin{proof}
	(i) Let $(U_t,\Delta_t)$ and $(U_t',\Delta_t')$ satisfy (D1) for some $t\in\mathbb R$. Applying \eqref{eq:sd-D1}, \[\Phi_{\pi^{(n)}}(U_t\psi)+\Delta_t(\psi)=\Phi_{\pi^{(n)}}(U'_t\psi)+\Delta'_t(\psi),\qquad \psi\in\Hs_{\Ar}.\] Lemma~\ref{lem:sd-displaced}(iii) with $\varrho=\pi^{(n)}$ gives $U_t\psi=U'_t\psi$ and $\Delta_t(\psi)=\Delta'_t(\psi)$. Since $\Hs=\Hs_{\Ar}+i\Hs_{\Ar}$ and $U_t,U'_t$ are linear, $U_t=U'_t$. At $t=0$, the pair $(\mathbf 1,0)$ satisfies (D1), hence $U_0=\mathbf 1$ and $\Delta_0=0$. Next, for the group property, let $t,s\in\mathbb R$ and observe that $\alpha_{t+s}=\alpha_t\circ\alpha_s$. Apply \eqref{eq:sd-D1} sequentially at time $s\in\mathbb R$, and then at time $t\in\mathbb R$ to $U_s\psi$ and apply \eqref{eq:sd-D1} directly at time $t+s$ to  obtain
	\[
\mathbf 1\otimes\Phi_\pi(U_{t+s}\psi)+\Delta_{t+s}(\psi)=\mathbf 1\otimes\Phi_\pi(U_tU_s\psi)+\Delta_t(U_s\psi)+\alpha_t\big(\Delta_s(\psi)\big).
	\]
 Lemma~\ref{lem:sd-displaced}(iii) yields $U_{t+s}=U_tU_s$, together with \eqref{eq:sd-cocycle} which already proves (ii).

	It remains to show the strong continuity. Fix $\pi=\piF$ and let $\psi\in\mathcal V_L$. By Lemma~\ref{lem:sd-bounded}(i) (Appendix~\ref{app:bounded}) and \eqref{eq:sd-D1-weyl},
	\[
	\big\|\alpha_t\big(\piFn(W(\psi))\big)-\piFn(W(U_t\psi))\big\|\leq\|\Delta_t(\psi)\|,\quad t\in\mathbb R
	\]
	which tends to $0$ as $t\to0$ by (D2$_L$). Since $\alpha_t(B)\to B$ strongly for every bounded operator $B$, it follows that $\piF(W(U_t\psi))\to\piF(W(\psi))$ strongly. In particular,
	\[
	\big\langle\Omega,\piF(W(\psi))^{*}\,\piF(W(U_t\psi))\,\Omega\big\rangle\mathop \rightarrow_{t\to 0}1.
	\]
	By \eqref{eq:weyl}, $W(\psi)^{*}W(U_t\psi)=e^{\frac i2\sigma(\psi,U_t\psi)}W(U_t\psi-\psi)$ for all $t\in\mathbb R$. Moreover, $|\langle\Omega,\piF(W(\varphi))\Omega\rangle|=e^{-\|\varphi\|^{2}/4}$ for $\varphi\in\Hs_{\Ar}$, by \eqref{eq:vac-weyl}, since $\|(g,Cg)\|^{2}=2\|g\|^{2}$. 
Hence $U_t\psi\to\psi$ for every $\psi\in\mathcal V_L$. If $L=0$, then $\mathcal V_L=\Hs_{\Ar}$. Otherwise, $U_t\psi\to\psi$ for every $\psi\in\Hs_{\Ar}$ follows from the density of $\mathcal V_L$ in $\Hs_{\Ar}$ and the bound $\sup_{|t|\leq1}\|U_t\|<\infty$. As $U_t$ commutes with $\Ar$, $U_t\to\mathbf 1$ strongly on $\Hs$, and the group property gives strong continuity everywhere.

	(iii) By (D2$_L$), $\Theta$ is a pointwise limit of real-linear maps on $\mathcal V_L$, hence real-linear, and \[\|\Theta(\psi)\|\leq C_\Delta\|\Lambda_L\psi\|,\qquad \psi\in\mathcal V_L.\]
	 Let $(\tau_l)_{l=1}^{n^{2}}$ be a basis of $M_n^{\mathrm{sa}}$, and expand $\Theta(\psi)=\sum_l\lambda_l(\psi)\tau_l$. The coordinates $\lambda_l$ are real-linear functionals on $\mathcal V_L$ such that $|\lambda_l(\psi)|\leq c\|\Lambda_L\psi\|$ for some $c<\infty$, since all norms on $M_n^{\mathrm{sa}}$ are equivalent. Since $|L|+\mathbf 1$ maps $D(L)$ bijectively onto $\h$, the map \[\mu_l(g)\doteq\lambda_l\big(((|L|+\mathbf 1)^{-1}g,C(|L|+\mathbf 1)^{-1}g)\big),\qquad g\in\h\] 
	 is a bounded real-linear functional on the real Hilbert space $(\h,\mathrm{Re}\langle\cdot,\cdot\rangle_{\h})$. By the Riesz theorem, there is a unique $\eta_l\in\h$ such that $\mu_l(g)=\mathrm{Re}\langle g,\eta_l\rangle_{\h}$ for all $g\in\h$; with $\xi_l\doteq i\eta_l/2$, this reads $\mu_l(g)=2\,\mathrm{Im}\langle g,\xi_l\rangle_{\h}$. Let $J_L:\h_{-1}(L)\to\h$ be the unitary extension of $(|L|+\mathbf 1)^{-1}$, and $w_l\doteq J_L^{-1}\xi_l\in\h_{-1}(L)$. For $f\in D(L)$, $\langle f,w_l\rangle=\overline{\langle J_Lw_l,(|L|+\mathbf 1)f\rangle}=\langle(|L|+\mathbf 1)f,\xi_l\rangle_{\h}$, thus
	\[
	\lambda_l\big((f,Cf)\big)=\mu_l\big((|L|+\mathbf 1)f\big)=2\,\mathrm{Im}\langle f,w_l\rangle=\sigma\big((f,Cf),\vartheta_l\big).
	\]
	Uniqueness of $w_l$ follows from the injectivity of $w\mapsto\sigma(\cdot,(w,Cw))$, see after \eqref{eq:sd-sigma-ext}. For $L=0$, $J_0=\mathbf 1$, so that $w_l\in\h$ and $\vartheta_l\in\Hs_{\Ar}$.

	(iv) Fix $t\in\mathbb R$ and $\psi\in\mathcal V_L$, and let 
	\[Z(s)\doteq\Delta_s(U_{t-s}\psi),\qquad s\in[0,t].\] Since $U_r$ commutes with $\Lambda_L$ and $\Ar$ for every $r\in\mathbb R$ , $U_r\psi\in\mathcal V_L$ and \[\|\Lambda_L(U_r\psi-U_{r'}\psi)\|=\|(U_r-U_{r'})\Lambda_L\psi\|\mathop\rightarrow_{r\to r'}0.\]
	 Let $s$ and $s+h$ lie between $0$ and $t$, with $0<|h|\leq1$. If $h>0$, \eqref{eq:sd-cocycle} applied to the times $(s,h)$ and to the vector $U_{t-s-h}\psi$ gives $Z(s+h)=Z(s)+\alpha_s\big(\Delta_h(U_{t-s-h}\psi)\big)$, hence
	\[
	h^{-1}\big(Z(s+h)-Z(s)\big)=\alpha_s\Big(h^{-1}\Delta_h\big(U_{t-s-h}\psi-U_{t-s}\psi\big)\Big)+\alpha_s\Big(h^{-1}\Delta_h\big(U_{t-s}\psi\big)\Big).
	\]
	By (D2$_L$), the first term has norm at most $C_\Delta\|\Lambda_L(U_{t-s-h}\psi-U_{t-s}\psi)\|\to0$, while the second one converges strongly to $\alpha_s(\Theta(U_{t-s}\psi)\otimes\mathbf 1)$ as $h\to0^{+}$. If $h<0$, \eqref{eq:sd-cocycle} applied to the times $(s+h,-h)$ and to the vector $U_{t-s}\psi$ gives $Z(s)=Z(s+h)+\alpha_{s+h}\big(\Delta_{-h}(U_{t-s}\psi)\big)$, hence
	\[
	h^{-1}\big(Z(s+h)-Z(s)\big)=e^{i(s+h)H}\,\Big((-h)^{-1}\Delta_{-h}\big(U_{t-s}\psi\big)\Big)\,e^{-i(s+h)H}.
	\]
The map $h\mapsto(-h)^{-1}\Delta_{-h}(U_{t-s}\psi)$ is uniformly bounded and converges strongly to $\Theta(U_{t-s}\psi)\otimes\mathbf 1$ as $h\to0^{-}$, by (D2$_L$), and $e^{\mp i(s+h)H}\to e^{\mp isH}$ strongly. Hence, in the strong sense,
\[\partial_sZ(s) = \alpha_s\left(\Theta\left(U_{t-s}\psi\right)\otimes \mathbf 1\right),\qquad s\in[0,t].\]
 This derivative is continuous in $s$, since $s\mapsto\Theta(U_{t-s}\psi)$ is norm-continuous by (iii) and $s\mapsto\alpha_s(B)$ is strongly continuous for every bounded $B$. Since $Z(0)=\Delta_0(U_t\psi)=0$ and $Z(t)=\Delta_t(\psi)$, this gives \eqref{eq:sd-langevin}. The bound follows from $\|\Theta(U_{t-s}\psi)\|\leq\|\Theta\|_L\|U_{t-s}\Lambda_L\psi\|$. For $\psi\in\mathcal V_L$, the integrand of \eqref{eq:sd-langevin} lies in the von Neumann algebra of the statement, which is strongly closed and thus contains the limits of the Riemann sums. For $\psi\in\Hs_{\Ar}$, choose $\psi_k\in\mathcal V_L$ with $\psi_k\to\psi$, since $\Delta_t$ is bounded, $\Delta_t(\psi_k)\to\Delta_t(\psi)$ in norm.
\end{proof}

For $n=1$, Definition~\ref{def:sd-impl} must reproduce the affine quadratic Hamiltonians of Subsection~\ref{subsec:affine-ham}. This is achieved by Corollary~\ref{cor:sd-scalar} at the scale $0$, and by Corollary~\ref{cor:sd-scalar-scale} at larger scales.

\begin{corollary}[Scalar case]
	\label{cor:sd-scalar}
	Let $n=1$ and $\pi=\piF$, and let $(U_t)_{t\in\mathbb R}\subseteq\Sp(\Hs)$ commute with $S$. Let $H$ be a self-adjoint operator acting on $\Fb$. The following two assertions are equivalent:
	\begin{enumerate}
		\item $H$ $\Theta$-implements $(U_t)_{t\in\mathbb R}$ in $\piF$.
		\item $H$ is an $N$-diagonal affine quadratic Hamiltonian with data $(\Upsilon,\theta)$ (see Theorem~\ref{thm:affine}(i)) such that $\theta\in\h$, $U_t=e^{it\Upsilon}\oplus e^{-it\Upsilon^{\top}}$ for all $t\in\mathbb R$, and $\Theta=\sigma(\cdot,(\theta,C\theta))$.
	\end{enumerate}  
	 In this case, $\Delta_t(\psi)=\sigma(\psi,\nu_t)\mathbf 1$ for any $t\in\mathbb R$ and any $\psi\in\Hs_\Ar$ with $\nu_t$ as in \eqref{eq:affine-data}, and \eqref{eq:sd-cocycle} reduces to \eqref{eq:cocycle}. In particular, if $\Upsilon\geq0$ and $\Ker\Upsilon=\{0\}$, an $N$-diagonal affine quadratic Hamiltonian $H$ with data $(\Upsilon,\theta)$ $\Theta$-implements $(e^{it\Upsilon}\oplus e^{-it\Upsilon^{\top}})_{t\in\mathbb R}$ in $\piF$ for some coupling map $\Theta$ if and only if $\Omega\in D(H)$.
\end{corollary}

\begin{proof}
	Let $H$ $\Theta$-implement $(U_t)$. By Theorem~\ref{thm:sd-structure}(iii) with $n=1$, $\Theta=\sigma(\cdot,\vartheta)$ for a unique $\vartheta\in\Hs_{\Ar}$. By Theorem~\ref{thm:sd-structure}(i), $t\mapsto U_t$ is strongly continuous. By \eqref{eq:sd-langevin}, and since $\sigma(U\psi,\vartheta)=\sigma(\psi,U^{-1}\vartheta)$ for $U\in\Sp(\Hs)$,
	\begin{equation}\label{eq:vtappears}
	\Delta_t(\psi)=\int_0^{t}\sigma(U_{t-s}\psi,\vartheta)\,\mathrm ds\,\mathbf 1=\sigma(\psi,\nu_t)\mathbf 1,\qquad \nu_t\doteq\int_0^{t}U_{-r}\vartheta\,\mathrm dr\in\Hs_{\Ar}.
	\end{equation}
	By Lemma~\ref{lem:sd-displaced}(i) and Lemma~\ref{lem:block}, (D1) becomes \begin{equation}\label{eq:D1asrepres}
		\mathrm{Ad}\,e^{itH}\circ\piF=\piF\circ\chi_{U_t}\circ\chi_{\U_{(\mathbf 1,\nu_t)}}=\piF\circ\chi_{\U_{(U_t,\nu_t)}},\qquad t\in\mathbb R,
	\end{equation}
	 with $\U_{(U_t,\nu_t)}\in\ASp_0(\Hs)$. So $H$ is an $N$-diagonal affine quadratic Hamiltonian. Let $(\Upsilon,\theta)$ be its data. By Theorem~\ref{thm:affine}(i), the injectivity in Proposition~\ref{prop:affine-aut} and the faithfulness of $\piF$, $U_t=e^{it\Upsilon}\oplus e^{-it\Upsilon^{\top}}$ and $\nu_t=(g_t,Cg_t)$ with $g_t=f_t(\Upsilon)\theta$. Since $r\mapsto U_{-r}\vartheta$ is norm-continuous, $t^{-1}\nu_t\to\vartheta$ in $\Hs$ as $t\to0$. Writing $\vartheta=(\theta',C\theta')$ ($\theta'$ is the form factor of $\Theta$, not the datum $\theta$ of $H$), we get $t^{-1}g_t\to\theta'$ in $\h$, hence in $\h_{-1}(\Upsilon)$, and Lemma~\ref{lem:cocycle}(i) gives $\theta=\theta'\in\h$, i.e., $\vartheta=(\theta,C\theta)$.

	Conversely, let $H$ be an $N$-diagonal affine quadratic Hamiltonian with data $(\Upsilon,\theta)$ and $\theta\in\h$. By Lemma~\ref{lem:field-cov}, \eqref{eq:sd-D1} holds with $\Delta_t(\psi)\doteq\sigma(\psi,\nu_t)\mathbf 1$ for any $t\in\mathbb R$, and $\|\Delta_t\|=\sqrt2\|f_t(\Upsilon)\theta\|\leq\sqrt2|t|\,\|\theta\|$, since $|f_t(x)|\leq|t|$. Moreover, $t^{-1}f_t(\Upsilon)\theta\to\theta$ in $\h$, by dominated convergence, since $|t^{-1}f_t(x)|\leq1$ and $t^{-1}f_t(x)\to1$. Hence $t^{-1}\Delta_t(\psi)\to\sigma(\psi,(\theta,C\theta))$. The cocycle identity \eqref{eq:sd-cocycle} reads \[\sigma(\psi,\nu_{t+s})=\sigma(\psi,U_s^{-1}\nu_t+\nu_s),\qquad t,s\in\mathbb R,\]
	 which is \eqref{eq:cocycle}. The last assertion follows from Proposition~\ref{prop:hierarchy}(i), since the additive constant does not affect $D(H)$.
\end{proof}

\begin{remark}[Cocycle structure]
	\label{rem:sd-cocycle}
	Let $\mathsf c_t\doteq\Delta_t\circ U_{-t}$. Then \eqref{eq:sd-cocycle} reads $\mathsf c_{t+s}=\mathsf c_t+t\triangleright\mathsf c_s$, where
	\[
	(t\triangleright \mathcal L)(\psi)\doteq\alpha_t\big(\mathcal L(U_{-t}\psi)\big),\qquad\psi\in\Hs_{\Ar},
	\]
	defines an action of $\mathbb R$ on the real-linear maps $\mathcal L:\Hs_{\Ar}\to\mathcal B(\mathbb C^{n}\otimes\Kc_\pi)$. Thus $\mathsf c$ is a $1$-cocycle for this action (see \eqref{eq:cocycle-std} and Appendix~\ref{app:cohomology}), as in Lemma~\ref{lem:cocycle}. For $n=1$ (Corollary~\ref{cor:sd-scalar}), $\mathsf c_t=\sigma(\cdot,\mu_t)\mathbf 1$ with $\mu_t\doteq U_t\nu_t$, and the cocycle identity becomes $\mu_{t+s}=\mu_t+U_t\mu_s$. This is the composition law of the semidirect product $\Hs_{\Ar}\rtimes\Sp(\Hs)\cong\ASp(\Hs)$ of Section~\ref{sec:affine}. In general, however, the module in which $\mathsf c$ takes its values is acted upon by the dynamics $\alpha_t$ itself. The pairs $(U_t,\mathsf c_t)$ therefore do not form a semidirect product with a group given in advance. In other words, it is hopeless to find a ``group of spin-dressed transformations''.
\end{remark}

\begin{corollary}[Scalar case at a scale $F(\Upsilon)$]
	\label{cor:sd-scalar-scale}
	Let $n=1$ and $\pi=\piF$. Let $\Upsilon$ be a self-adjoint operator on $\h$, let $U_t\doteq e^{itK}=e^{it\Upsilon}\oplus e^{-it\Upsilon^{\top}}$, $t\in\mathbb R$, and let $L\doteq F(\Upsilon)$ for a real Borel function $F$ such that $|F(x)|\leq c(1+|x|)$ for some $c<\infty$. Then $\h_{-1}(L)$ embeds continuously and injectively into $\h_{-1}(\Upsilon)$, and, for a self-adjoint operator $H$ on $\Fb$, the following assertions are equivalent:
	\begin{enumerate}
		\item $H$ $\Theta$-implements $(U_t)_{t\in\mathbb R}$ in $\piF$ at scale $L$, for some map $\Theta:\mathcal V_L\to\mathbb R$.
		\item $H$ is an $N$-diagonal affine quadratic Hamiltonian with data $(\Upsilon,\theta)$ such that $\theta\in\h_{-1}(L)$.
	\end{enumerate}
	In this case, $\Theta=\sigma(\cdot,(\theta,C\theta))$, and $\Delta_t(\psi)=\sigma(\psi,\nu_t)\mathbf 1$ for all $t\in\mathbb R$ and $\psi\in\Hs_{\Ar}$, with $\nu_t$ as in \eqref{eq:affine-data}. In particular, at the scale $L=\Upsilon$, the $\Theta$-implementations of $(U_t)_{t\in\mathbb R}$ in $\piF$ are exactly the $N$-diagonal affine quadratic Hamiltonians with one-particle Hamiltonian $\Upsilon$ and \emph{arbitrary} $\theta\in\h_{-1}(\Upsilon)$, that is, for $\Upsilon\geq0$ with $\Ker\Upsilon=\{0\}$, all the van Hove Hamiltonians of \cite{Der03} with $h=\Upsilon$ (Theorem~\ref{thm:affine}(ii)).
\end{corollary}

\begin{proof}
	\emph{Embedding.} Let $G(x)\doteq(|F(x)|+1)/(|x|+1)$, so that $0<G\leq c+1$. For $\theta\in\h$, \[\|(|\Upsilon|+\mathbf 1)^{-1}\theta\|=\|G(\Upsilon)(|L|+\mathbf 1)^{-1}\theta\|\leq(c+1)\|(|L|+\mathbf 1)^{-1}\theta\|.\]
	Hence the identity of $\h$ extends to a continuous map $\iota:\h_{-1}(L)\to\h_{-1}(\Upsilon)$, such that $J_\Upsilon\iota=G(\Upsilon)J_L$, and it is injective since $G>0$. Moreover, $D(\Upsilon)\subseteq D(L)$ and, for $\theta\in\h_{-1}(L)$ and $f\in D(\Upsilon)$, \[\langle\iota\theta,f\rangle=\langle G(\Upsilon)J_L\theta,(|\Upsilon|+\mathbf 1)f\rangle=\langle J_L\theta,(|L|+\mathbf 1)f\rangle=\langle\theta,f\rangle .\] We identify $\theta$ with $\iota\theta$. Likewise, $f_t(\Upsilon)$ extends to a bounded operator $\h_{-1}(L)\to\h$, namely $f_t(\Upsilon)\theta=[f_t(\Upsilon)(|L|+\mathbf 1)]J_L\theta$, since $|f_t(x)|(|F(x)|+1)\leq(c+1)(|t|+2)$, and this extension is compatible with $\iota$.

	\emph{(1)$\Rightarrow$(2).} As in the proof of Corollary~\ref{cor:sd-scalar}, by Theorem~\ref{thm:sd-structure}(iii), $\Theta=\sigma(\cdot,\vartheta)$ on $\mathcal V_L$, with $\vartheta=(\theta',C\theta')$ and $\theta'\in\h_{-1}(L)$. The hypotheses of Theorem~\ref{thm:sd-structure}(iv) hold: the $U_t$ are unitary, strongly continuous, and commute with $\Lambda_L=(|F(\Upsilon)|+\mathbf 1)\oplus(|F(\Upsilon^{\top})|+\mathbf 1)$. Hence \[\Delta_t(\psi)=\int_0^{t}\sigma(U_{t-s}\psi,\vartheta)\,\mathrm ds\,\mathbf 1,\qquad t\in\mathbb R,\quad \psi\in\mathcal V_L.\]
	 For any $t\in\mathbb R$, since $\Delta_t$ is bounded and $\mathcal V_L$ is dense, $\Delta_t(\psi)\in\mathbb R\mathbf 1$ for every $\psi\in\Hs_{\Ar}$, and the Riesz argument of the proof of Theorem~\ref{thm:sd-structure}(iii) with $L=0$ gives a unique $\nu_t\in\Hs_{\Ar}$ such that $\Delta_t=\sigma(\cdot,\nu_t)\mathbf 1$. As in the proof of Corollary~\ref{cor:sd-scalar}, (D1) then reads $\mathrm{Ad}\,e^{itH}\circ\piF=\piF\circ\chi_{\U_{(U_t,\nu_t)}}$, so that $H$ is an $N$-diagonal affine quadratic Hamiltonian. By Theorem~\ref{thm:affine}(i), the injectivity in Proposition~\ref{prop:affine-aut} and the uniqueness in Stone's theorem, its data are $(\Upsilon,\theta)$ for some $\theta\in\h_{-1}(\Upsilon)$, and $\nu_t=(g_t,Cg_t)$ with $g_t=f_t(\Upsilon)\theta$. Let $f\in D(\Upsilon)$ and $\psi\doteq(f,Cf)$. By Lemma~\ref{lem:cocycle}(i), $t^{-1}g_t\to\theta$ in $\h_{-1}(\Upsilon)$, so that $t^{-1}\sigma(\psi,\nu_t)\to2\,\mathrm{Im}\langle f,\theta\rangle$, whereas $t^{-1}\sigma(\psi,\nu_t)\to2\,\mathrm{Im}\langle f,\theta'\rangle$ by (D2$_L$) where $\theta'$ is given by \eqref{eq:sd-tensor}. Replacing $f$ by $if$, we get $\langle\theta-\theta',f\rangle=0$ for all $f\in D(\Upsilon)$, hence $\theta=\theta'\in\h_{-1}(L)$.

	\emph{(2)$\Rightarrow$(1).} Let $\theta\in\h_{-1}(L)$, and let $H$ have data $(\Upsilon,\theta)$. As in the proof of Corollary~\ref{cor:sd-scalar}, Lemma~\ref{lem:field-cov} gives \eqref{eq:sd-D1} with $\Delta_t(\psi)\doteq\sigma(\psi,\nu_t)\mathbf 1$, (recall \eqref{eq:affine-data}) which is bounded on $\Hs_{\Ar}$ since $g_t\doteq f_t(\Upsilon)\theta\in\h$, $t\in\mathbb R$ by Lemma~\ref{lem:cocycle}(i) and $\h_{-1}(L)\subseteq\h_{-1}(\Upsilon)$. For $\psi=(f,Cf)$ with $f\in D(L)$,
	\[
	\langle g_t,f\rangle=\big\langle J_L\theta,\overline{f_t}(\Upsilon)(|L|+\mathbf 1)f\big\rangle_{\h},\qquad t\in\mathbb R.
	\]
	Since $|f_t(x)|\leq|t|$, this gives \[|\sigma(\psi,\nu_t)|\leq2|\langle g_t,f\rangle|\leq\sqrt2\,|t|\,\|\theta\|_{-1}\|\Lambda_L\psi\|,\qquad t\in\mathbb R.\]
	 Moreover, $t^{-1}\langle g_t,f\rangle\to\langle J_L\theta,(|L|+\mathbf 1)f\rangle=\langle\theta,f\rangle$ by dominated convergence, with the same bounds $|t^{-1}f_t(x)|\leq1$ and $t^{-1}f_t(x)\to1$ as in the proof of Corollary~\ref{cor:sd-scalar}. Hence (D2$_L$) holds with $\Theta=\sigma(\cdot,(\theta,C\theta))$. The last assertion is the case $F(x)=x$.
\end{proof}

Thus at the scale $0$ the definition selects the van Hove Hamiltonians whose data satisfy $\theta\in\h$, and at the scale $\Upsilon$ all of them. For $U_t=e^{itK}$, Corollary~\ref{cor:sd-scalar-scale} with $F\equiv0$ (so $L=0$ and $\h_{-1}(L)=\h$) gives the equivalence of Corollary~\ref{cor:sd-scalar} for this family. The latter also allows an arbitrary family $(U_t)_{t\in\mathbb R}\subseteq\Sp(\Hs)$ commuting with $S$, and shows that any such family which is $\Theta$-implemented is of the form $e^{it\Upsilon}\oplus e^{-it\Upsilon^{\top}}$, with $\Upsilon$ read off from $H$.

\begin{remark}[Regularity scales and the infrared/ultraviolet hierarchy]
	\label{rem:sd-scales}
	Let $\Upsilon\geq0$ with $\Ker\Upsilon=\{0\}$, and $H=H_{\Upsilon,\theta}$. By Corollaries~\ref{cor:sd-scalar} and~\ref{cor:sd-scalar-scale}, Proposition~\ref{prop:hierarchy}(i) and Remark~\ref{rem:hierarchy}, the regularity conditions on the vacuum become a choice of norm in the Lipschitz condition (D2$_L$). More precisely,
	\begin{itemize}
		\item at the scale $L=0$: $\theta\in\h$, i.e., $\Omega\in D(H)$
		\item at the scale $L=\Upsilon^{1/2}$: $\theta\in\h_{-1}(\Upsilon^{1/2})=\h+\Upsilon^{1/2}\h$, i.e., $\Omega\in D(|H|^{1/2})$
		\item at the scale $L=\Upsilon$: every $\theta\in\h_{-1}(\Upsilon)$, including the renormalized van Hove Hamiltonians, whose form domain may not contain $\Omega$.
	\end{itemize}
	For $\h=L^{2}(\mathbb R^{d})$ and $\Upsilon$ the multiplication by a dispersion relation $\omega\geq0$, the condition $w\in\h_{-1}(\Upsilon)$ reads $(1+\omega)^{-1}w\in L^{2}(\mathbb R^{d})$. This is the relevant ultraviolet threshold for spin--boson models.

	For normal couplings, renormalized Hamiltonians are constructed by a self-energy renormalization under this condition, together with the infrared condition $\omega^{-1/2}\mathbf 1_{\{\omega\leq\kappa\}}w\in L^{2}(\mathbb R^{d})$ for some $\kappa>0$, and beyond it no cutoff approximation, corrected by bounded counterterms keeping the regularized ground state energies bounded from below, converges in the strong resolvent sense \cite[Theorems~2.5 and~2.8, Remark~2.9]{HLVM25}, see also \cite{ALLVM25}. For the standard spin--boson model with form factor $v$ beyond this threshold, i.e., if $\omega^{-1}v\mathbf 1_{\{\omega>1\}}\notin L^{2}$, and under an infrared condition, the self-energy renormalized cutoff Hamiltonians, conjugated by unitaries independent of the qubit, converge to a limit which is independent of the qubit, in particular of the tunnelling term \cite[Corollary~4.4]{DM20a}.

	On the other hand, by Theorem~\ref{thm:sd-structure}(iii), the form factors of every $\Theta$-implementation at the scale $\Upsilon$ lie in $\h_{-1}(\Upsilon)$, so that this scale does not reach beyond the threshold. Up to the threshold, it contains the renormalized Hamiltonians with commuting couplings, in particular the standard spin--boson model with an arbitrary tunnelling term (Proposition~\ref{prop:sd-commuting} and Example~\ref{ex:sd-SB-ren}). The general case is open (Remark~\ref{rem:sd-open}).
\end{remark}

\paragraph{Asymptotic fields.}
In scattering theory, one compares the interacting evolution with the free one at large times. By \eqref{eq:sd-D1-weyl}, $\alpha_t\big(\pi^{(n)}(W(U_{-t}\psi))\big)=e^{i(\mathbf 1\otimes\Phi_\pi(\psi)+\Delta_t(U_{-t}\psi))}$, which converges if $t\mapsto\Delta_t(U_{-t}\psi)$ does. By \eqref{eq:sd-langevin} the latter is an integral over $[0,t]$ of $\alpha_s\big(\Theta(U_{-s}\psi)\otimes\mathbf 1\big)$, convergent when $\Theta$ is integrable along the flow. Let $\mathbb R_+\doteq[0,\infty)$ and $\mathbb R_-\doteq(-\infty,0]$.

\begin{proposition}[Asymptotic fields]
	\label{prop:sd-asymptotic}
	Let $\pi$, $\Theta$ and $L$ be as in Definition \ref{def:sd-impl}.	Let $H$ $\Theta$-implement $(U_t)_{t\in\mathbb R}$ in $\pi$ at scale $L$, with $t\mapsto U_t$ strongly continuous and $U_t$ commuting with $\Lambda_L$ for every $t\in\mathbb R$, and let $\alpha_t\doteq\mathrm{Ad}\,e^{itH}$. Let
	\[
	\mathcal D_\pm\doteq\Big\{\psi\in\mathcal V_L:\ \int_{\mathbb R_\pm}\big\|\Theta(U_{-s}\psi)\big\|\,\mathrm ds<\infty\Big\}.
	\]
	Then, for every $\psi\in\mathcal D_\pm$, the limits
	\begin{align*}
	W_\pm(\psi)&\doteq\lim_{t\to\pm\infty}\alpha_t\big(\pi^{(n)}(W(U_{-t}\psi))\big),\\
	\Delta_\pm(\psi)&\doteq\lim_{t\to\pm\infty}\Delta_t(U_{-t}\psi)=\int_0^{\pm\infty}\alpha_s\big(\Theta(U_{-s}\psi)\otimes\mathbf 1\big)\,\mathrm ds
	\end{align*}
	exist in norm, and $W_\pm(\psi)=e^{i(\mathbf 1\otimes\Phi_\pi(\psi)+\Delta_\pm(\psi))}$. Moreover:
	\begin{enumerate}
		\item[(i)] $\mathcal D_\pm$ is a real subspace of $\mathcal V_L$, invariant under every $U_r$, $r\in\mathbb R$. The map $\Delta_\pm$ is real-linear on $\mathcal D_\pm$, with self-adjoint values, \[\|\Delta_\pm(\psi)\|\leq\int_{\mathbb R_\pm}\|\Theta(U_{-s}\psi)\|\,\mathrm ds,\] and $\Delta_\pm(\psi)$ belongs to the von Neumann algebra generated by $\{\alpha_s(M_n\otimes\mathbf 1):s\in\mathbb R_\pm\}$.
		\item[(ii)] $W_\pm(\psi)^{*}=W_\pm(-\psi)$, and the CCR hold, i.e., \[W_\pm(\psi)W_\pm(\varphi)=e^{-\frac i2\sigma(\psi,\varphi)}W_\pm(\psi+\varphi),\qquad\psi,\varphi\in\mathcal D_\pm.\]
		\item[(iii)] $\alpha_r\big(W_\pm(\psi)\big)=W_\pm(U_r\psi)$ for all $\psi\in\mathcal D_\pm$ and $r\in\mathbb R$.
	\end{enumerate}
\end{proposition}

\begin{proof}
	We consider $\mathcal D_+$, the case of $\mathcal D_-$ is proven in the same way. Let $\psi\in\mathcal D_+$. For $t\in\mathbb R$, $U_{-t}\psi\in\mathcal V_L$, and Theorem~\ref{thm:sd-structure}(iv), with $U_{t-s}U_{-t}=U_{-s}$, gives
	\[
	\Delta_t(U_{-t}\psi)=\int_0^{t}\alpha_s\big(\Theta(U_{-s}\psi)\otimes\mathbf 1\big)\,\mathrm ds .
	\]
	It follows that for $0\leq t\leq t'$, \[\left\|\Delta_{t'}(U_{-t'}\psi)-\Delta_t(U_{-t}\psi)\right\|\le\int_t^{t'}\|\Theta(U_{-s}\psi)\|\,\mathrm ds.\] 
	As $\psi\in\mathcal D_+$, $(\Delta_t(U_{-t}\psi))_{t\geq0}$ is a Cauchy net in norm, which converges to a bounded self-adjoint operator $\Delta_+(\psi)$ satisfying the bound of (i). It belongs to the von Neumann algebra of (i), since every $\Delta_t(U_{-t}\psi)$, $t\geq0$, does by Theorem~\ref{thm:sd-structure}(iv). By \eqref{eq:sd-D1-weyl} at time $t\in\mathbb R$, applied to $U_{-t}\psi$,
	\[
	\alpha_t\big(\pi^{(n)}(W(U_{-t}\psi))\big)=e^{i(\mathbf 1\otimes\Phi_\pi(\psi)+\Delta_t(U_{-t}\psi))},\qquad t\in\mathbb R
	\]
	and Lemma~\ref{lem:sd-bounded}(i), with $Y=\mathbf 1\otimes\Phi_\pi(\psi)$, gives that the limit defining $W_+(\psi)$ exists in norm, and
	\[W_+(\psi) = e^{i(\mathbf 1\otimes\Phi_\pi(\psi)+\Delta_+(\psi))}.\]

	(i) $\mathcal D_+$ is a real subspace, and $\Delta_+$ is real-linear, since $\Theta$ and the $\Delta_t$ are real-linear. For $r\in\mathbb R$, $U_r\psi\in\mathcal V_L$  and as $s\mapsto\|\Theta(U_{-s}\psi)\|\leq\|\Theta\|_L\|U_{-s}\Lambda_L\psi\|$ is locally bounded,
	\[\int_{\mathbb R_+}\|\Theta(U_{-s}U_r\psi)\|\,\mathrm ds=\int_{-r}^{\infty}\|\Theta(U_{-s}\psi)\|\,\mathrm ds<\infty.\]

	(ii) For every $t$, the map $\psi\mapsto\alpha_t(\pi^{(n)}(W(U_{-t}\psi)))$ satisfies the CCR, since $\alpha_t\circ\pi^{(n)}$ is a representation of $\Wa$ and $\sigma(U_{-t}\psi,U_{-t}\varphi)=\sigma(\psi,\varphi)$ for all $t\in\mathbb R$. They pass to the norm limits.

	(iii) Let $r\in\mathbb R$. $\alpha_r$ is an isometry and $\alpha_r\circ\alpha_t=\alpha_{r+t}$. Therefore, by (i), \[\alpha_r(W_+(\psi))=\lim_{t\to\infty}\alpha_{r+t}\big(\pi^{(n)}(W(U_{-(r+t)}U_r\psi))\big)=W_+(U_r\psi).\]
\end{proof}

\begin{lemma}[Density of $\mathcal D_\pm$]
	\label{lem:sd-density}
	Let $\h=L^{2}(\mathbb R^{d})$, let $\Upsilon$ be the multiplication by a real function $\omega\in C^{\infty}(\mathbb R^{d}\setminus\{0\})$ with $\nabla\omega\neq0$ on $\mathbb R^{d}\setminus\{0\}$ (e.g., $\omega(k)=|k|$ or $\omega(k)=(k^{2}+m^{2})^{1/2}$, the dispersion relations of photons and of massive bosons of mass $m>0$), let $U_t\doteq e^{itK}$, $t\in\mathbb R$, and let $L\doteq F(\Upsilon)$ for a real Borel function $F$ such that $|F(x)|\leq c(1+|x|)$, $c<\infty$. Let $H$ $\Theta$-implement $(U_t)_{t\in\mathbb R}$ in a regular representation $\pi$ at scale $L$, and assume that the $w_l\in\h_{-1}(L)$ in \eqref{eq:sd-tensor} are given by functions in $C^{\infty}(\mathbb R^{d}\setminus\{0\})$. Then $(f,Cf)\in\mathcal D_+\cap\mathcal D_-$ for every $f\in C^{\infty}_c(\mathbb R^{d}\setminus\{0\})$. In particular, $\mathcal D_+\cap\mathcal D_-$ is dense in $\Hs_{\Ar}$, whatever the spin Hamiltonian and the infrared behaviour of the $w_l$.
\end{lemma}

\begin{proof}
	In this representation, $\h_{-1}(L)$ is the space of measurable functions $w$ such that $(|F(\omega)|+1)^{-1}w\in L^{2}(\mathbb R^{d})$, and $\langle f,w\rangle=\int\overline{f}\,w$ for $f\in D(L)$. Let $f\in C^{\infty}_c(\mathbb R^{d}\setminus\{0\})$, with support $\mathcal K$, and $\psi\doteq(f,Cf)$. Since $\omega$ is continuous on the compact set $\mathcal K$, $\omega$ and $F(\omega)$ are bounded on $\mathcal K$, so that $f\in D(L)$, i.e., $\psi\in\mathcal V_L$. By \eqref{eq:sd-tensor} and \eqref{eq:sd-sigma-ext},
	\[
	\Theta(U_{-s}\psi)=\sum_l2\,\mathrm{Im}\langle e^{-is\omega}f,w_l\rangle\,\tau_l,\qquad \langle e^{-is\omega}f,w_l\rangle=\int_{\mathbb R^{d}}e^{is\omega(k)}\,\overline{f(k)}\,w_l(k)\,\mathrm dk,\qquad s\in\mathbb R .
	\]
	Here $\overline f\,w_l\in C^{\infty}_c(\mathbb R^{d}\setminus\{0\})$, and $|\nabla\omega|$ is bounded from below on $\mathcal K$. Since $e^{is\omega}=(is|\nabla\omega|^{2})^{-1}\nabla\omega\cdot\nabla e^{is\omega}$ for $s\neq0$, $N$ integrations by parts give $|\langle e^{-is\omega}f,w_l\rangle|\leq c_N|s|^{-N}$ for every $N\in\mathbb N$ and $|s|\geq1$, while $|\langle e^{-is\omega}f,w_l\rangle|\leq\|\overline f\,w_l\|_{L^{1}}$ for all $s\in\mathbb R$. Hence $s\mapsto\|\Theta(U_{-s}\psi)\|$ is integrable on $\mathbb R$, i.e., $\psi\in\mathcal D_+\cap\mathcal D_-$. The density follows, since $C^{\infty}_c(\mathbb R^{d}\setminus\{0\})$ is dense in $L^{2}(\mathbb R^{d})$.
\end{proof}

\begin{remark}[Asymptotic fields]
	\label{rem:sd-asymptotic}
	By (ii) and (iii), the asymptotic Weyl operators form a representation of the Weyl relations over $(\mathcal D_\pm,\sigma)$ that intertwines the interacting dynamics with the Bogoliubov flow: $\alpha_r\circ W_\pm=W_\pm\circ U_r$. Their fields are the free fields displaced by the operators $\Delta_\pm(\psi)$, which depend on the whole spin history. In the language of Definition~\ref{def:sd-displaced}, the asymptotic representation is the displacement of $\pi^{(n)}$ by $\Delta_\pm$, except that $\Delta_\pm$ is only defined on $\mathcal D_\pm$.
	\begin{itemize}
		\item \emph{The van Hove case.} Let $n=1$, $\pi=\piF$, $U_t=e^{itK}$ with $\Upsilon\geq0$, $\Ker\Upsilon=\{0\}$, and let $H=H_{\Upsilon,\theta}$ with $\Upsilon^{-1}\theta\in\h$, which is a $\Theta$-implementation at the scale $L=\Upsilon$ (Corollary~\ref{cor:sd-scalar-scale}). With $\rho=(-i\Upsilon^{-1}\theta,C(-i\Upsilon^{-1}\theta))$ as in Proposition~\ref{prop:hierarchy}(ii), $\nu_t=\rho-U_{-t}\rho$, so that $\Delta_t(U_{-t}\psi)=\sigma(U_{-t}\psi,\rho)-\sigma(\psi,\rho)$ for every $\psi\in\Hs_{\Ar}$. If the spectrum of $\Upsilon$ is absolutely continuous, then $\sigma(U_{-t}\psi,\rho)\to0$ as $t\to\pm\infty$ by the Riemann--Lebesgue lemma. Hence the limits defining $W_\pm(\psi)$ exist for every $\psi\in\Hs_{\Ar}$, without any integrability condition, and $W_\pm(\psi)=e^{-i\sigma(\psi,\rho)}\piF(W(\psi))=\piF(W(\rho))^{*}\piF(W(\psi))\piF(W(\rho))$. The asymptotic representation is the Fock representation conjugated by the dressing operator, and its vacuum is the ground state $\piF(W(\rho))^{*}\Omega$ of $H$. This recovers \cite[Thm.~4.2(1),(3)]{Der03}, where $\piF(W(\rho))^{*}$ is the dressing operator, see also \cite{FF24}. When $\Upsilon^{-1}\theta\notin\h$ and $\Upsilon$ has absolutely continuous spectrum, the functional $\psi\mapsto\lim_t\Delta_t(U_{-t}\psi)$ is no longer of the form $\sigma(\cdot,\nu)$ with $\nu\in\Hs_{\Ar}$, and the asymptotic representation (on $\{(f,Cf):f\in D(\Upsilon^{-1})\}$) is not unitarily equivalent to the Fock one, see \cite[Thm.~4.2(3)]{Der03}.
		\item \emph{Density of $\mathcal D_\pm$.} For multiplication operators with smooth dispersion relations and form factors, $\mathcal D_+\cap\mathcal D_-$ is dense in $\Hs_{\Ar}$, at every scale $L=F(\Upsilon)$ (Lemma~\ref{lem:sd-density}).
	\end{itemize}
\end{remark}

\paragraph{Comparison of implementations.}
Do two implementations $H$ and $H'$ of the same data differ only by an element of the commutant $\pi^{(n)}(\Wa)'$? In Fock space the commutant is $M_n\otimes\mathbf 1$, so a positive answer would be a uniqueness statement up to a spin Hamiltonian. Lemma~\ref{lem:sd-compare} answers yes in the weak sense, in an arbitrary regular representation and at an arbitrary scale: $H-H'$ commutes with the Weyl operators $\pi^{(n)}(W(\psi))$, $\psi\in\mathcal V_L$, in the sense of forms on $D(H)\cap D(H')$.

\begin{lemma}[Comparison]
	\label{lem:sd-compare}
	Let $H$ and $H'$ both $\Theta$-implement the same family $(U_t)_{t\in\mathbb R}$ in a regular representation $\pi$ at the same scale $L$. Then, for all $\psi\in\mathcal V_L$ and $\eta,\zeta\in D(H)\cap D(H')$,
	\begin{equation}
		\big\langle(H-H')\eta,\ \pi^{(n)}(W(\psi))\,\zeta\big\rangle=\big\langle\eta,\ \pi^{(n)}(W(\psi))\,(H-H')\zeta\big\rangle .
		\label{eq:sd-compare}
	\end{equation}
\end{lemma}

\begin{proof}
 We write $\alpha_t\doteq\mathrm{Ad}\,e^{itH}$ and $\alpha'_t\doteq\mathrm{Ad}\,e^{itH'}$, with displacements $\Delta_t$ and $\Delta_t'$, and write $W_\varphi\doteq\pi^{(n)}(W(\varphi))$ for any $\varphi\in\Hs_\Ar$. Let $\psi\in\mathcal V_L$ and $\eta ,\zeta\in D(H)\cap D(H')$. We introduce the function $q$ defined, for any $t\in\mathbb R$, by
	\[
	q(t)\doteq\big\langle\eta,\big(\alpha_t(W_\psi)-\alpha'_t(W_\psi)\big)\zeta\big\rangle=\big\langle e^{-itH}\eta,W_\psi e^{-itH}\zeta\big\rangle-\big\langle e^{-itH'}\eta,W_\psi e^{-itH'}\zeta\big\rangle .
	\]
	It vanishes at $t=0$, and it is differentiable at $t=0$, with derivative \[q'(0)=i\langle(H-H')\eta,W_\psi\zeta\rangle-i\langle\eta,W_\psi(H-H')\zeta\rangle.\]
	 On the other hand, by \eqref{eq:sd-D1-weyl}, \[\alpha_t(W_{s\psi})=e^{is(\mathbf 1\otimes\Phi_\pi(U_t\psi)+\Delta_t(\psi))},\qquad t,s\in\mathbb R, \quad \psi\in\Hs_\Ar,\]
	 and similarly for $\alpha'_t$, by real-linearity of $\Delta_t$ and $\Delta'_t$. Lemma~\ref{lem:sd-bounded}(i), with $Y=\mathbf 1\otimes\Phi_\pi(U_t\psi)$, therefore gives
	\[
	t^{-1}q(t)=i\int_0^{1}\Big\langle\alpha_t(W_{-s\psi})\eta,\ t^{-1}\big(\Delta_t(\psi)-\Delta'_t(\psi)\big)\,\alpha'_t(W_{(1-s)\psi})\zeta\Big\rangle\,\mathrm ds,\qquad t\in\mathbb R\backslash \{0\} .
	\]
 Recall that $t\mapsto e^{itH}$ and $t\mapsto e^{itH'}$ are strongly continuous groups. By (D2$_L$), the operators $t^{-1}(\Delta_t(\psi)-\Delta'_t(\psi))$ are uniformly bounded for $0<|t|\leq1$ and converge strongly to $\Theta(\psi)\otimes\mathbf 1-\Theta(\psi)\otimes\mathbf 1=0$. Therefore, the integrand tends to $0$ as $t\to 0$ for every $s\in[0,1]$, and it is uniformly bounded, so that $t^{-1}q(t)\to0$. Thus $q'(0)=0$, which is \eqref{eq:sd-compare}.
\end{proof}

 The main obstacle to concluding now that $H-H'$ is affiliated (see \cite[Section 3.1.2]{deNittisbook}) with $\pi^{(n)}(\Wa)'$ is that $D(H)\cap D(H')$ may a priori not  contain a suitable core.

\subsection{Spin--Boson Hamiltonians in Fock Space}
\label{subsec:sd-fock}

This subsection studies the self-adjoint operators $H$ on $\mathbb C^{n}\otimes\Fb$ that $\Theta$-implement the free flow, that is, satisfy (D1)--(D2$_0$) of Definition~\ref{def:sd-impl} (scale $L=0$) in the Fock representation. For form factors in $D(\Upsilon^{-1/2})$ they are exactly the operators $H_{\Upsilon,\Theta}+E\otimes\mathbf 1$ with a unique spin Hamiltonian $E$, and no assumption on the domain of $H$ is made (Theorem~\ref{thm:sd-fock}).

The proof has two stages. First, (D1) and (D2$_0$) force $H\supseteq\widehat H_E$ for a spin Hamiltonian $E$, where $\widehat H_E$ is the formal expression on the span $\mathcal E_\Upsilon$ of the coherent vectors. By the quantum Langevin equation, annihilation operators applied to $e^{-itH}(x\otimes\Omega)$ give displacements whose squares sum to $O(t^{2})$, so the evolved vacuum leaves the vacuum sector at rate $O(|t|)$. A group-law argument then puts the vacuum, and the coherent vectors, into $D(H)$ (Proposition~\ref{prop:sd-vacuum}), where $H$'s action is inferred (Proposition~\ref{prop:sd-universal}). Second, $H_{\Upsilon,\Theta}+E\otimes\mathbf 1$ is self-adjoint by the Kato--Rellich theorem, has the core $\mathcal E_\Upsilon$, and satisfies (D1)--(D2$_0$) by integration of the Heisenberg equation. It is therefore the only self-adjoint extension of $\widehat H_E$.\\

For this subsection, $\pi=\piF$, and $\Upsilon\geq0$ is a self-adjoint operator on $\h$ with $\Ker\Upsilon=\{0\}$. Recall that $\Phi=\Phi_{\piF}$ are the Segal fields \eqref{eq:segal} and $K=\Upsilon\oplus(-\Upsilon^{\top})$, and let
\[
U_t\doteq e^{itK}=e^{it\Upsilon}\oplus e^{-it\Upsilon^{\top}},\qquad t\in\mathbb R ,
\]
so that $iK(f,Cf)=(i\Upsilon f,C(i\Upsilon f))$ for $f\in D(\Upsilon)$. For $w\in\h$, let
\[
\phi(w)\doteq\Phi((w,Cw))=\overline{a(w)+a^{*}(w)}.
\]

The coupling of the spin to the field is encoded by $\tau_1,\dots,\tau_m\in M_n^{\mathrm{sa}}$ and the form factors $w_1,\dots,w_m\in\h$. With $\vartheta_l\doteq(w_l,Cw_l)\in\Hs_{\Ar}$, we define the coupling map $\Theta:\Hs_{\Ar}\to M_n^{\mathrm{sa}}$ and the Hamiltonian $H_{\Upsilon,\Theta}$ on $\mathbb C^{n}\otimes\Fb$ by
\begin{equation}
	\Theta(\psi)\doteq\sum_{l=1}^{m}\sigma(\psi,\vartheta_l)\,\tau_l,\qquad
	H_{\Upsilon,\Theta}\doteq\mathbf 1\otimes\mathrm d\Gamma(\Upsilon)+\sum_{l=1}^{m}\tau_l\otimes\phi(w_l)\quad\text{on }\mathbb C^{n}\otimes D(\mathrm d\Gamma(\Upsilon)).
	\label{eq:sd-HUT}
\end{equation}
The operator $H_{\Upsilon,\Theta}$ is used only when $w_1,\dots,w_m\in D(\Upsilon^{-1/2})$. Elsewhere in this subsection, the standing assumption on the form factors is only $w_1,\dots,w_m\in\h$ (Corollary~\ref{cor:sd-uniqueness}(a) assumes more). It depends on $(\tau_l,w_l)_l$ only through $\Theta$ (see after~\eqref{eq:sd-Hhat}). For $n=1$, $\tau_1=1$ and $\Theta=\sigma(\cdot,(\theta,C\theta))$ with $\theta\in D(\Upsilon^{-1/2})$, it is the van Hove Hamiltonian $H_{\Upsilon,\theta}$ of Theorem~\ref{thm:affine}(iii) with $\langle\Omega,H_{\Upsilon,\theta}\Omega\rangle=0$ (it is self-adjoint by Theorem~\ref{thm:sd-fock}(i)). The subscript of $H_{\Upsilon,\theta}$ is a vector $\theta\in\h_{-1}(\Upsilon)$, that of $H_{\Upsilon,\Theta}$ a coupling map.

With the creation and annihilation notation of \cite[Sect.~2.4]{Der03}, $\phi(g)=g(a^{*})+\bar g(a)$ (the field of \cite[(2.11)]{Der03} carries an extra factor $2^{-1/2}$). For $g\in D(\Upsilon^{-1/2})$ and $\zeta\in D(\mathrm d\Gamma(\Upsilon))$, \cite[Prop.~3.8]{Der03} gives
\begin{equation}
	\|\phi(g)\zeta\|^{2}\leq4\|\Upsilon^{-1/2}g\|^{2}\langle\zeta,\mathrm d\Gamma(\Upsilon)\zeta\rangle+2\|g\|^{2}\|\zeta\|^{2} ,
	\label{eq:sd-relbound}
\end{equation}
so that $\phi(g)$ is $\mathrm d\Gamma(\Upsilon)$-bounded with arbitrarily small relative bound.

\paragraph{Coherent vectors and cores.} Let $\mathcal D_0^{\Upsilon}\subseteq\mathcal D_0$ be the span of $\Omega$ and the vectors $a^{*}(f_1)\cdots a^{*}(f_k)\Omega$, where $k\in\mathbb N$ and $f_1,\dots,f_k\in D(\Upsilon)$. Up to a spin Hamiltonian, (D1)--(D2$_0$) will determine $H$ on the space
\[
\mathcal E_\Upsilon\doteq\mathrm{span}\big\{x\otimes\epsilon(f):\ x\in\mathbb C^{n},\ f\in D(\Upsilon)\big\},
\]
where $\epsilon(f)$ is the exponential vector \eqref{eq:defexpvector}. For $\psi=(f,Cf)$ in $\mathcal V_\Upsilon\doteq\{(f,Cf):f\in D(\Upsilon)\}$ (the space $\mathcal V_L$ of Subsection~\ref{subsec:sd-def} for $L=\Upsilon$) and $g\in D(\Upsilon)$, \eqref{eq:Weyl-exp} shows that $\piF(W(\psi))\epsilon(g)$ is a non-zero multiple of $\epsilon(g+if)$. Hence $\mathcal E_\Upsilon$ is also the span of the coherent vectors $x\otimes\piF(W(\psi))\Omega$, $x\in\mathbb C^{n}$, $\psi\in\mathcal V_\Upsilon$, and it is invariant under $\piFn(W(\psi))$, $\psi\in\mathcal V_\Upsilon$. For $\mathsf h\in\{\Upsilon,\mathbf 1+\Upsilon\}$, $\epsilon(f)\in D(\mathrm d\Gamma(\mathsf h))$ for $f\in D(\Upsilon)$, and $\Gamma(e^{is\mathsf h})\epsilon(f)=\epsilon(e^{is\mathsf h}f)$ by \eqref{eq:exp-facts}. Thus $\mathcal E_\Upsilon$ is contained in $\mathbb C^{n}\otimes D(\mathrm d\Gamma(\mathsf h))$ and invariant under $\mathbf 1\otimes\Gamma(e^{is\mathsf h})$, $s\in\mathbb R$. It is dense, since the exponential vectors are total and depend continuously on their argument \cite[Props.~2.1 and~2.2]{Gui72} and $D(\Upsilon)$ is dense. By \cite[Prop.~II.1.7]{EN00},
\begin{equation}
	\mathcal E_\Upsilon\subseteq\mathbb C^{n}\otimes D(\mathrm d\Gamma(\mathsf h))\ \text{ is a core for }\ \mathbf 1\otimes\mathrm d\Gamma(\mathsf h),\qquad\mathsf h\in\{\Upsilon,\mathbf 1+\Upsilon\} .
	\label{eq:sd-core}
\end{equation}

\paragraph{Particle sectors.} For $p\in\mathbb N_0$, let $P_p$ be the orthogonal projection of $\mathbb C^{n}\otimes\Fb$ onto the $p$-particle sector $\mathbb C^{n}\otimes\h^{\vee p}$, and $v^{(p)}\doteq P_pv$. Thus $P_0=\mathbf 1\otimes|\Omega\rangle\langle\Omega|$ is the vacuum projection, and $P_{-1}\doteq0$. Let $(e_k)_k$ be an orthonormal basis of $\h$. For $v^{(p)}$ in the range of $P_p$,
\begin{equation}
	\sum_k\|(\mathbf 1\otimes a(e_k))v^{(p)}\|^{2}=p\,\|v^{(p)}\|^{2} .
	\label{eq:sd-sectors-sum}
\end{equation}
Moreover, since $a(g)=(a^{*}(g)|_{\mathcal D_0})^{*}$ for $g\in\h$, one has $P_{p-1}(\mathbf 1\otimes a(g))\subseteq(\mathbf 1\otimes a(g))P_p$, i.e.,
\begin{equation}
	v\in D(\mathbf 1\otimes a(g))\ \Longrightarrow\ P_pv\in D(\mathbf 1\otimes a(g)),\quad(\mathbf 1\otimes a(g))P_pv=P_{p-1}(\mathbf 1\otimes a(g))v .
	\label{eq:sd-sectors-shift}
\end{equation}

\paragraph{The formal operator.} Let
\[
c_\Theta\doteq\sum_{l=1}^{m}\|\tau_l\|\,\|w_l\| .
\]
This constant depends on the representation \eqref{eq:sd-HUT} of $\Theta$. Any such representation can be used below. For $E\in M_n^{\mathrm{sa}}$, we restrict the formal expression to $\mathcal E_\Upsilon$:
\begin{equation}
	\widehat H_E\doteq E\otimes\mathbf 1+\mathbf 1\otimes\mathrm d\Gamma(\Upsilon)+\sum_{l=1}^{m}\tau_l\otimes\phi(w_l)\qquad\text{on }\mathcal E_\Upsilon .
	\label{eq:sd-Hhat}
\end{equation}
It is well defined for arbitrary $w_l\in\h$: the exponential vectors $\epsilon(f)$, $f\in D(\Upsilon)$, lie in $D(\mathrm d\Gamma(\Upsilon))$ and in $D(N^{1/2})$, where $N\doteq\mathrm d\Gamma(\mathbf 1)$, and $\phi(w)$ is $N^{1/2}$-bounded for every $w\in\h$ \cite[Sect.~5.2.1]{BR97}. If $w_1,\dots,w_m\in D(\Upsilon^{-1/2})$, then $\widehat H_E=(H_{\Upsilon,\Theta}+E\otimes\mathbf 1)|_{\mathcal E_\Upsilon}$.

Both $H_{\Upsilon,\Theta}$ and $\widehat H_E$ depend on $(\tau_l,w_l)_l$ only through $\Theta$, by real-linearity of $w\mapsto\phi(w)$ on two domains: on $D(N^{1/2})$ for $\widehat H_E$ and $w_l\in\h$, and on $D(\mathrm d\Gamma(\Upsilon))$ for $H_{\Upsilon,\Theta}$ and $w\in D(\Upsilon^{-1/2})$ (by \eqref{eq:sd-relbound} and \cite[Thm.~5]{DerezinskiCCR}). The second is not a case of the first, since $D(\mathrm d\Gamma(\Upsilon))\not\subseteq D(N^{1/2})$ when $0\in\mathrm{spec}(\Upsilon)$. As $\sigma$ is non-degenerate, any two families yielding the same $\Theta$ define the same operator.

The first result shows that (D1), together with the Lipschitz condition (D2$_0$), forces the vacuum, and then the coherent vectors, into the domain of every spin-dressed implementation. It only uses $w_l\in\h$.

\begin{proposition}[Vacuum regularity]
	\label{prop:sd-vacuum}
	Let $H$ be a self-adjoint operator which $\Theta$-implements $(U_t)_{t\in\mathbb R}$ in $\piF$. Then:
	\begin{enumerate}
		\item[(i)] $\|(\mathbf 1-P_0)\,e^{-itH}P_0\|\leq c_\Theta|t|$ for all $t\in\mathbb R$;
		\item[(ii)] $\mathbb C^{n}\otimes\mathbb C\Omega\subseteq D(H)$;
		\item[(iii)] $\mathcal E_\Upsilon\subseteq D(H)$, and, with $W_\psi\doteq\piFn(W(\psi))$,
		\begin{equation}
			HW_\psi\zeta=W_\psi H\zeta+(\Theta(\psi)\otimes\mathbf 1)W_\psi\zeta+(\mathbf 1\otimes\mathrm d\Gamma(\Upsilon))W_\psi\zeta,\qquad \zeta\in\mathbb C^{n}\otimes\mathbb C\Omega,\ \psi\in\mathcal V_\Upsilon .
			\label{eq:sd-coherent}
		\end{equation}
	\end{enumerate}
\end{proposition}

\begin{proof}
	(i) \emph{Step 1: annihilation operators on the evolved vacuum.} Let $\Delta_t$ be the displacements of $H$, $\alpha_t\doteq\mathrm{Ad}\,e^{itH}$, $u_r\doteq e^{ir\Upsilon}$, and fix $x\in\mathbb C^{n}$, $\zeta\doteq x\otimes\Omega$ and $t\in\mathbb R$. Since $\zeta\in D(\mathbf 1\otimes\Phi(\xi))$ for every $\xi\in\Hs_{\Ar}$, \eqref{eq:sd-D1} gives $e^{-itH}\zeta\in D(\mathbf 1\otimes\Phi(\xi))$ and
	\[
	(\mathbf 1\otimes\Phi(\xi))\,e^{-itH}\zeta=e^{-itH}\big(\mathbf 1\otimes\Phi(U_t\xi)+\Delta_t(\xi)\big)\zeta,\qquad\xi\in\Hs_{\Ar}.
	\]
	We will translate this identity in terms of annihilation operators instead of fields. For $g\in\h$, let 
		\begin{equation}\label{eq:defxi1and2}
			\xi_1\doteq(g,Cg)\quad\mbox{and}\quad\xi_2\doteq(ig,C(ig)).
			\end{equation} By \eqref{eq:B-def}, we have a decomposition of $(g,0)$ using \[(g,0)_R=\frac12\xi_1\quad\mbox{ and }\quad (g,0)_I=-\frac12\xi_2,\] so that $B_{\piF}((g,0))=\frac12(\Phi(\xi_1)+i\Phi(\xi_2))$ on $D(\Phi(\xi_1))\cap D(\Phi(\xi_2))$. This operator is contained in the closed annihilation operator $a(g)$. Indeed, let $\eta\in\mathcal D_0$ and $v\in D(\Phi(\xi_1))\cap D(\Phi(\xi_2))$. Since $a^{*}(g)\eta=\frac12(\Phi(\xi_1)-i\Phi(\xi_2))\eta$ and the $\Phi(\xi_j)$ are self-adjoint,
	\[
	\langle a^{*}(g)\eta,v\rangle=\big\langle\eta,\tfrac12(\Phi(\xi_1)+i\Phi(\xi_2))v\big\rangle ,
	\]
	i.e., $v\in D\big((a^{*}(g)|_{\mathcal D_0})^{*}\big)=D(a(g))$ and $a(g)v=\frac12(\Phi(\xi_1)+i\Phi(\xi_2))v$. Since $U_t\xi_1=(u_tg,Cu_tg)$, $U_t\xi_2=(iu_tg,C(iu_tg))$ and $a(u_tg)\Omega=0$, we obtain $e^{-itH}\zeta\in D(\mathbf 1\otimes a(g))$ and
	\begin{equation}
		(\mathbf 1\otimes a(g))\,e^{-itH}\zeta=e^{-itH}\,\beta_t(g)\,\zeta,\qquad \beta_t(g)\doteq\tfrac12\big(\Delta_t(\xi_1)+i\Delta_t(\xi_2)\big).
		\label{eq:sd-annihilation}
	\end{equation}
	\emph{Step 2: the displacement through the Langevin formula.} Since $U$ is strongly continuous, \eqref{eq:sd-langevin} applies. Using $\sigma((g_1,Cg_1),(g_2,Cg_2))=2\,\mathrm{Im}\langle g_1,g_2\rangle$, for any $s\in\mathbb R$, $\langle u_{t-s}ig,w_l\rangle=-i\langle u_{t-s}g,w_l\rangle$ and the definition \eqref{eq:sd-HUT} of $\Theta$, we obtain
	\[
	\beta_t(g)=-i\sum_{l=1}^{m}\int_0^{t}\langle u_{t-s}g,w_l\rangle\,\alpha_s(\tau_l\otimes\mathbf 1)\,\mathrm ds ,\qquad \beta_t(g)\zeta=(\langle g|\otimes\mathbf 1)\,\Xi_t,
	\]
  using the bra-ket notation and where
	\[
	\Xi_t\doteq-i\sum_{l=1}^{m}\int_0^{t}u_{s-t}w_l\otimes\alpha_s(\tau_l\otimes\mathbf 1)\zeta\,\mathrm ds\ \in\ \h\otimes(\mathbb C^{n}\otimes\Fb),\qquad \|\Xi_t\|\leq c_\Theta|t|\,\|\zeta\| .
	\]
	\emph{Step 3: summing over the particle sectors.} For $v\in\bigcap_kD(\mathbf 1\otimes a(e_k))$ and $p\geq1$, \eqref{eq:sd-sectors-shift} gives $P_{p-1}(\mathbf 1\otimes a(e_k))v=(\mathbf 1\otimes a(e_k))v^{(p)}$. Summing over the orthogonal sectors and using \eqref{eq:sd-sectors-sum} and Tonelli's theorem, we estimate
	\[
	\sum_k\|(\mathbf 1\otimes a(e_k))v\|^{2}=\sum_{p\ge 1}p\,\|v^{(p)}\|^{2}\geq\|(\mathbf 1-P_0)v\|^{2}.
	\]
	Applied to $v=e^{-itH}\zeta$, together with \eqref{eq:sd-annihilation} and Parseval's identity in $\h\otimes(\mathbb C^{n}\otimes\Fb)$, this gives
	\[
	\|(\mathbf 1-P_0)e^{-itH}\zeta\|^{2}\leq\sum_k\|\beta_t(e_k)\zeta\|^{2}=\|\Xi_t\|^{2}\leq c_\Theta^{2}\,t^{2}\,\|\zeta\|^{2} .
	\]

	(ii) \emph{Step 1: approximate group law.} Let $Q_t\in M_n$ be defined by \[\langle y,Q_tx\rangle\doteq\langle y\otimes\Omega,e^{-itH}(x\otimes\Omega)\rangle , \qquad x,y\in\mathbb C ^n,\qquad t\in\mathbb R,\]
	 so that $P_0(e^{-itH}-\mathbf 1)(x\otimes\Omega)=((Q_t-\mathbf 1)x)\otimes\Omega$. The map $t\mapsto Q_t$ is continuous, $Q_0=\mathbf 1$ and $\|Q_t\|\leq1$. Since $P_0e^{-itH}(\mathbf 1-P_0)=((\mathbf 1-P_0)e^{itH}P_0)^{*}$, (i) yields
	\begin{equation}
		\|Q_{t+s}-Q_tQ_s\|=\big\|P_0e^{-itH}(\mathbf 1-P_0)e^{-isH}P_0\big\|\leq c_\Theta^{2}|t||s|,\qquad s,t\in\mathbb R .
		\label{eq:sd-approx-group}
	\end{equation}
	\emph{Step 2: Lipschitz bound for $Q_t$.} The key observation of the proof is that the defect in the group law is of order 2: for small $|t|$, $\log Q_{2t}$ and $2\log Q_t$ (that we define below) differ by $O(t^{2})$. The corresponding defect for $t^{-1}\log Q_t$ is of order one. In practice: choose $t_1>0$ such that $\|Q_t-\mathbf 1\|\leq\frac14$ for all $|t|\leq t_1$. For such $t$, we introduce
	\[\mathsf L_t\doteq\log Q_t\doteq-\sum_{j\geq1}j^{-1}(\mathbf 1-Q_t)^{j}.\]
	It satisfies $e^{\mathsf L_t}=Q_t$ and $\|\mathsf L_t\|\leq\log\frac43<\frac12\log2$. Hence $e^{2\mathsf L_t}=Q_t^{2}$. Moreover, $\log(e^{Y})=Y$ for $\|Y\|<\log2$: the composed power series converges absolutely there, and the identity holds for formal power series. Therefore, $\log(e^{2\mathsf L_t})=2\mathsf L_t$. The logarithm is $2$-Lipschitz on $\{X\in M_n:\|X-\mathbf 1\|\leq\frac12\}$, a set which contains $Q_t^{2}$ and $Q_{2t}$ for $|t|\leq t_1/2$. Hence, by \eqref{eq:sd-approx-group},
	\[
	\|\mathsf L_{2t}-2\mathsf L_t\|\leq2\|Q_{2t}-Q_t^{2}\|\leq2c_\Theta^{2}t^{2},\qquad|t|\leq t_1/2 .
	\]
	Thus $\ell(t)\doteq t^{-1}\mathsf L_t$ satisfies $\|\ell(2t)-\ell(t)\|\leq c_\Theta^{2}|t|$ for $0<|t|\leq t_1/2$. Every $t$ with $0<|t|\leq t_1$ reads $t=2^{-d}s$ with $d\in\mathbb N_0$ and $t_1/2<|s|\leq t_1$, and then using a telescopic sum
	\[
	\|\ell(t)-\ell(s)\|\leq\sum_{j=1}^{d}c_\Theta^{2}\,2^{-j}|s|\leq c_\Theta^{2}\,t_1 .
	\]
	Since $\|\ell(s)\|\leq2t_1^{-1}\log\frac43$ for $t_1/2<|s|\leq t_1$, we get \[\|\mathsf L_t\|=\|\mathsf L_t-t\ell(s)+t\ell(s)\|\leq(2t_1^{-1}\log\frac43+c_\Theta^{2}t_1)|t|,\quad |t|\leq t_1,\]  and therefore $\|Q_t-\mathbf 1\|\leq e^{\|\mathsf L_t\|}-1=O(|t|)$. The triangle inequality and (i) yield
	\[
	\big\|(e^{-itH}-\mathbf 1)(x\otimes\Omega)\big\|\leq\|(Q_t-\mathbf 1)x\|+\big\|(\mathbf 1-P_0)e^{-itH}(x\otimes\Omega)\big\|=\mathop O_{|t|\to 0 }(|t|)\|x\|.
	\]
	\emph{Step 3: $x\otimes\Omega\in D(H)$.} We square this last relation and divide by $t^{2}$. Using Fatou's lemma in the spectral representation of $H$ allows us to deduce that $x\otimes\Omega\in D(H)$.

	(iii) Let $x\in\mathbb C^{n}$, $\zeta\doteq x\otimes\Omega$. Let $\psi\in\mathcal V_\Upsilon$, and write $W_\varphi\doteq\piFn(W(\varphi))$ for $\varphi\in\Hs_\Ar$, as in the statement. Set $v\doteq W_\psi\zeta$. Since $e^{itH}W_\psi=\alpha_t(W_\psi)e^{itH}$,
	\[
	t^{-1}(e^{itH}v-v)=\alpha_t(W_\psi)\,t^{-1}(e^{itH}\zeta-\zeta)+t^{-1}\big(\alpha_t(W_\psi)-W_\psi\big)\zeta,\qquad t\in\mathbb R\setminus\{0\} .
	\]
By (ii), $t^{-1}(e^{itH}\zeta-\zeta)\to iH\zeta$ in $\mathbb C^n\otimes \mathcal F_+$, while the unitaries $\alpha_t(W_\psi)$ converge strongly to $W_\psi$. Hence the first term converges to $W_\psi(iH\zeta)$ as $t\to0$. For the second term, let $Y_t\doteq\mathbf 1\otimes\Phi(U_t\psi)$. By \eqref{eq:sd-D1-weyl} and Lemma~\ref{lem:sd-bounded}(i) (Duhamel's formula, $B'=0$), for any $|t|>0$,
	\begin{equation}\label{eq:applduhamel}
	t^{-1}\big(\alpha_t(W_\psi)-W_\psi\big)\zeta=i\int_0^{1}\alpha_t(W_{s\psi})\,t^{-1}\Delta_t(\psi)\,e^{i(1-s)Y_t}\zeta\,\mathrm ds+t^{-1}\big(e^{iY_t}-e^{iY_0}\big)\zeta .
	\end{equation}
First, $e^{i(1-s)Y_t}\zeta=\piFn(W((1-s)U_t\psi))\zeta\to W_{(1-s)\psi}\zeta$, for any $s\in[0,1]$, by strong continuity of $U$ and of $\xi\mapsto\piF(W(\xi))$ (Subsection~\ref{subsec:affine-setting}). Then, $t^{-1}\Delta_t(\psi)\to\Theta(\psi)\otimes\mathbf 1$ strongly, and these operators are uniformly bounded for $0<|t|\leq1$, by (D2$_0$). This in turn implies that
\begin{equation}\label{eq:limit1}
	t^{-1}\Delta_t(\psi)e^{i(1-s)Y_t}\zeta \mathop\rightarrow_{t\to 0} (\Theta(\psi)\otimes\mathbf 1)W_{(1-s)\psi}\zeta, \qquad s\in[0,1].
\end{equation}
Finally, $\alpha_t(W_{s\psi})\to W_{s\psi}$ strongly for all $s\in[0,1]$, and these operators are unitary. This together with \eqref{eq:limit1} and the dominated convergence theorem imply that the integral in \eqref{eq:applduhamel} converges to
	\[
	i\int_0^1W_{s\psi}(\Theta(\psi)\otimes\mathbf 1)W_{(1-s)\psi}\zeta \mathrm ds=i(\Theta(\psi)\otimes\mathbf 1)W_\psi\zeta ,
	\]
	since $W_{s\psi}$ commutes with $\Theta(\psi)\otimes\mathbf 1$ and $W_{s\psi}W_{(1-s)\psi}=W_\psi$. It remains to look at the limit of the last term of \eqref{eq:applduhamel}. By \eqref{eq:bogotransfo} and $\Gamma(u_t)^{*}\Omega=\Omega$, we have $e^{iY_t}\zeta=(\mathbf 1\otimes\Gamma(u_t))\,v=e^{it\,\mathbf 1\otimes\mathrm d\Gamma(\Upsilon)}v$. This is differentiable at $t=0$, since $v\in\mathcal E_\Upsilon\subseteq D(\mathbf 1\otimes\mathrm d\Gamma(\Upsilon))$. Hence all the terms in the right-hand side of \eqref{eq:applduhamel} converge, and $t^{-1}(e^{itH}v-v)$ converges to $iW_\psi H\zeta+i(\Theta(\psi)\otimes\mathbf 1)v+i(\mathbf 1\otimes\mathrm d\Gamma(\Upsilon))v$ as $t\to0$. By Stone's theorem, $v\in D(H)$, and \eqref{eq:sd-coherent} holds. As $\mathcal E_\Upsilon$ is spanned by such vectors $v$, (iii) follows.
\end{proof}

Part (iii) is (D1) in Weyl form, differentiated at $t=0$ on vacuum vectors: $H$ acts on $W_\psi\zeta$ as on $\zeta$, transported by $W_\psi$, plus the derivatives $\Theta(\psi)$ of the displacement and $\mathrm d\Gamma(\Upsilon)$ of the free flow. For $n=1$, (ii) gives a direct proof of the implication ``$\Rightarrow$'' of the last assertion of Corollary~\ref{cor:sd-scalar}, for every form factor in $\h$.

\begin{proposition}[Universality on coherent vectors]
	\label{prop:sd-universal}
	Let $w_1,\dots,w_m\in\h$, and let $H$ be a self-adjoint operator which $\Theta$-implements $(U_t)_{t\in\mathbb R}$ in $\piF$. Then there is a unique $E\in M_n^{\mathrm{sa}}$ such that, identifying $\h$ with the one-particle subspace of $\Fb$,
	\begin{equation}
		H(x\otimes\Omega)=(Ex)\otimes\Omega+\sum_{l=1}^{m}(\tau_lx)\otimes w_l,\qquad x\in\mathbb C^{n}.
		\label{eq:sd-vacuum-image}
	\end{equation}
	Moreover, $H\supseteq\widehat H_E$. In particular, if $\widehat H_E$ is essentially self-adjoint, then $H$ is its closure.
\end{proposition}

Thus every $\Theta$-implementation acts on the coherent vectors by the formal expression \eqref{eq:sd-formal-H}, for some spin Hamiltonian $E$, and uniqueness reduces to the essential self-adjointness of the explicit operators $\widehat H_E$ on $\mathcal E_\Upsilon$. For $w_l\in D(\Upsilon^{-1/2})$, the latter is provided by Theorem~\ref{thm:sd-fock}(i) below. For infrared-singular form factors $w_l\in\h\setminus D(\Upsilon^{-1/2})$, Propositions~\ref{prop:sd-vacuum} and~\ref{prop:sd-universal} hold verbatim, and the essential self-adjointness of $\widehat H_E$ holds for ultraviolet-regular form factors or commuting couplings (Corollary~\ref{cor:sd-uniqueness} below).

\begin{proof}
	Let $x\in\mathbb C^{n}$ and $\zeta\doteq x\otimes\Omega$, which lies in $D(H)$ by Proposition~\ref{prop:sd-vacuum}(ii). Let $g\in\h$, and let $\xi_1,\xi_2$ and $\beta_t(g)$, $t\in\mathbb R$, be defined by \eqref{eq:defxi1and2} and \eqref{eq:sd-annihilation}. By \eqref{eq:sd-annihilation} and $(\mathbf 1\otimes a(g))\zeta=0$,
	\[
	(\mathbf 1\otimes a(g))\,t^{-1}\big(e^{-itH}\zeta-\zeta\big)=e^{-itH}\,t^{-1}\beta_t(g)\,\zeta,\qquad t\neq0 .
	\]
	Taking the limit $t\to0$, $t^{-1}(e^{-itH}\zeta-\zeta)\to-iH\zeta$. By (D2$_0$), the operators $t^{-1}\beta_t(g)$ are uniformly bounded for $0<|t|\leq1$ and converge strongly to $\beta(g)\doteq\frac12(\Theta(\xi_1)+i\Theta(\xi_2))\otimes\mathbf 1$. For all $t\in\mathbb R$, $e^{-itH}$ is unitary and strongly converges to $\mathbf 1$. Therefore, as $\mathbf 1\otimes a(g)$ is closed, $H\zeta\in D(\mathbf 1\otimes a(g))$ and
	\[	(\mathbf 1\otimes a(g))H\zeta =i\beta(g)\zeta = i\left(\frac12(\Theta(\xi_1)+i\Theta(\xi_2))\otimes\mathbf 1\right)\zeta .\] 
	 Now $\Theta(\xi_1)=\sum_l2\,\mathrm{Im}\langle g,w_l\rangle\tau_l$ and $\Theta(\xi_2)=\sum_l2\,\mathrm{Im}\langle ig,w_l\rangle\tau_l=-\sum_l2\,\mathrm{Re}\langle g,w_l\rangle\tau_l$, so that $\beta(g)=-i\sum_l\langle g,w_l\rangle\tau_l\otimes\mathbf 1$ and
	\begin{equation}
		(\mathbf 1\otimes a(g))H\zeta=\sum_{l=1}^{m}\langle g,w_l\rangle\,(\tau_lx)\otimes\Omega,\qquad g\in\h .
		\label{eq:sd-aH}
	\end{equation}
	For $p\geq2$, the right-hand side of \eqref{eq:sd-aH} lies in the vacuum sector, so that \eqref{eq:sd-sectors-shift} gives \[(\mathbf 1\otimes a(e_k))P_pH\zeta=P_{p-1}(\mathbf 1\otimes a(e_k))H\zeta=0,\qquad k\in\mathbb N.\] 
By \eqref{eq:sd-sectors-sum}, this implies that $p\|P_pH\zeta\|^{2}=\sum_k\|(\mathbf 1\otimes a(e_k))P_pH\zeta\|^{2}=0$. For $p=1$, write $P_1H\zeta=\sum_{j=1}^{n}\mathsf e_j\otimes\varphi_j$, with $(\mathsf e_j)$ the canonical basis of $\mathbb C^{n}$ and $\varphi_j\in\h$. Then $(\mathbf 1\otimes a(g))P_1H\zeta=\sum_j\langle g,\varphi_j\rangle \mathsf e_j\otimes\Omega$, and identification in \eqref{eq:sd-aH} gives $\varphi_j=\sum_l\langle \mathsf e_j,\tau_lx\rangle w_l$, i.e., \[P_1H\zeta=\sum_{l=1}^m(\tau_lx)\otimes w_l.\] Finally, $P_0H\zeta=(Ex)\otimes\Omega$ for a linear map $E$ on $\mathbb C^{n}$, and $E$ is self-adjoint, since $\langle y,Ex\rangle=\langle y\otimes\Omega,H(x\otimes\Omega)\rangle=\langle H(y\otimes\Omega),x\otimes\Omega\rangle=\langle Ey,x\rangle$ for $x,y\in\mathbb C^{n}$. This proves \eqref{eq:sd-vacuum-image}, and $E$ is unique, since it is determined by these matrix elements.

	Let $\psi\in\mathcal V_\Upsilon$, with $W_\psi$ as in Proposition~\ref{prop:sd-vacuum}(iii). By \eqref{eq:segal-weyl}, with $\psi$ replaced by $-\psi$, $W_\psi^{*}(\mathbf 1\otimes\Phi(\vartheta_l))W_\psi=\mathbf 1\otimes\Phi(\vartheta_l)+\sigma(\psi,\vartheta_l)\mathbf 1$. Since $\phi(w_l)\Omega=w_l$, this gives $(\mathbf 1\otimes\phi(w_l))W_\psi\zeta=W_\psi(x\otimes w_l)+\sigma(\psi,\vartheta_l)W_\psi\zeta$. As $\tau_l\otimes\mathbf 1$ commutes with $W_\psi$,
	\[
	\widehat H_EW_\psi\zeta=W_\psi\Big((Ex)\otimes\Omega+\sum_{l=1}^{m}(\tau_lx)\otimes w_l\Big)+(\Theta(\psi)\otimes\mathbf 1)W_\psi\zeta+(\mathbf 1\otimes\mathrm d\Gamma(\Upsilon))W_\psi\zeta .
	\]
	By \eqref{eq:sd-vacuum-image} and \eqref{eq:sd-coherent}, the right-hand side equals $HW_\psi\zeta$. Since $\mathcal E_\Upsilon$ is spanned by the vectors $W_\psi(x\otimes\Omega)$, $x\in\mathbb C^{n}$, $\psi\in\mathcal V_\Upsilon$, we obtain $H\supseteq\widehat H_E$. Finally, if $\widehat H_E$ is essentially self-adjoint, its closure is a self-adjoint restriction of the self-adjoint operator $H$, hence equal to $H$.
\end{proof}

\begin{theorem}[Spin-dressed implementations in Fock space]
	\label{thm:sd-fock}
	Let $w_1,\dots,w_m\in D(\Upsilon^{-1/2})$.
	\begin{enumerate}
		\item[(i)] For every $E\in M_n^{\mathrm{sa}}$, the operator $H_{\Upsilon,\Theta}+E\otimes\mathbf 1$ is self-adjoint on $\mathbb C^{n}\otimes D(\mathrm d\Gamma(\Upsilon))$ and bounded from below, $\mathcal E_\Upsilon$ is a core for it, and it $\Theta$-implements $(U_t)_{t\in\mathbb R}$ in $\piF$.
		\item[(ii)] Conversely, every self-adjoint operator $H$ which $\Theta$-implements $(U_t)_{t\in\mathbb R}$ in $\piF$ is of the form $H=H_{\Upsilon,\Theta}+E\otimes\mathbf 1$ for a unique $E\in M_n^{\mathrm{sa}}$.
	\end{enumerate}
\end{theorem}

Thus (D1)--(D2$_0$) fix $H$ up to $E$ without any domain assumption, and the implementations all have the same domain, namely $\mathbb C^{n}\otimes D(\mathrm d\Gamma(\Upsilon))$, and the same core $\mathcal E_\Upsilon$. This is the technical section version of Theorem~\ref{thm:intro-fock}, applied in Example~\ref{ex:sd-SB} and Remark~\ref{rem:sd-nogo}.

 In the situation of Theorem~\ref{thm:sd-fock}, the data $(U,\Theta)$ determine $H$ up to a self-adjoint element of $\piFn(\Wa)'=M_n\otimes\mathbf 1$. In Theorem~\ref{thm:affine}(ii), the ambiguity was $\piF(\Wa)'=\mathbb C\mathbf 1$, i.e., an additive constant, by Schur's lemma applied to the irreducible representation $\piF$. Here $\piFn=\mathbf 1\otimes\piF$ is the direct sum of $n$ copies of $\piF$, hence reducible, and its commutant is no longer reduced to the scalars. The whole spin Hamiltonian $E$ plays the role of the additive constant.

The analogy is however only partial. For the van Hove Hamiltonian, an additive constant changes the unitary group only by the character $t\mapsto e^{iEt}$, cf.\ the phases $c_t$ in Section~\ref{sec:affine}, and leaves the Heisenberg dynamics unchanged. Here, $[E\otimes\mathbf 1,H_{\Upsilon,\Theta}]=\sum_l[E,\tau_l]\otimes\phi(w_l)$. If $[E,\tau_l]=0$ for all $l$, then $e^{it(H_{\Upsilon,\Theta}+E\otimes\mathbf 1)}=(e^{itE}\otimes\mathbf 1)\,e^{itH_{\Upsilon,\Theta}}$, which is the exact analogue of the character ambiguity. In general, however, $E$ modifies the dynamics, including that of the fields through the terms $\alpha_s(\tau_l\otimes\mathbf 1)$ in \eqref{eq:sd-langevin}. The data $(U,\Theta)$ thus do not determine the spin Hamiltonian, which is a parameter of the model (e.g.\ the tunnelling term in the spin--boson model). Conversely, a spin Hamiltonian can be added without changing $(U,\Theta)$ (Lemma~\ref{lem:sd-commutant} and Remark~\ref{rem:sd-commutant} below).\\

 In the fermionic setting of \cite{Quadratic}, the dynamical (à la Bach-Lieb-Solovej \cite{BLS94}) definition of quadratic Hamiltonians  coincides with the operator-theoretic one for compatible quadratic Hamiltonians, provided their off-diagonal coefficient is a Hilbert--Schmidt operator \cite[Corollaries~3.5 and~3.8]{Quadratic}. Under a mild regularity assumption, this Hilbert--Schmidt condition is equivalent to the vacuum lying in the domain \cite[Theorem~3.9 and Corollary~3.11]{Quadratic}. Here the analog vacuum condition is encoded in the Lipschitz condition (D2$_0$). For $n=1$, it is equivalent to $\Omega\in D(H)$ (Corollary~\ref{cor:sd-scalar}), and in general it forces $w_l\in\h$ (Theorem~\ref{thm:sd-structure}(iii)), so that the condition $w_l\in\h$, i.e., $c_\Theta<\infty$, plays the role of the Hilbert--Schmidt condition on the off-diagonal coefficient. Condition (D2$_0$) is thus the analogue of the Shale--Stinespring-like condition of \cite[Theorem~3.9]{Quadratic}.

\begin{proof}[Proof of Theorem~\ref{thm:sd-fock}(i)]
	Let $H\doteq H_{\Upsilon,\Theta}+E\otimes\mathbf 1$ and $\mathrm d\Gamma(\Upsilon)\equiv\mathbf 1\otimes\mathrm d\Gamma(\Upsilon)$. Steps~1 and~2 give self-adjointness and the commutator identity on $D(H)$; Steps~3 and~4 integrate it to (D1) and verify (D2$_0$).

	\emph{Step 1: self-adjointness and cores.}  By \eqref{eq:sd-relbound}, $\phi(g)$ is $\mathrm d\Gamma(\Upsilon)$-bounded with arbitrarily small relative bound for $g\in D(\Upsilon^{-1/2})$. Hence $H-\mathrm d \Gamma(\Upsilon)=E\otimes\mathbf 1+\sum_l\tau_l\otimes\phi(w_l)$ is $\mathrm d \Gamma(\Upsilon)$-bounded with relative bound $0$. By the Kato--Rellich theorem \cite[Thm.~X.12]{RS2}, $H$ is self-adjoint on $D(\mathbf 1 \otimes \mathrm d \Gamma(\Upsilon))=\mathbb C^{n}\otimes D(\mathrm d\Gamma(\Upsilon))$ and bounded from below. Moreover, the graph norms of $H$ and $\mathrm d \Gamma(\Upsilon)$ are equivalent, so that $H$ and $\mathrm d \Gamma(\Upsilon)$ have the same cores. The space $\mathbb C^{n}\otimes\mathcal D_0^{\Upsilon}$ is a core for $\mathrm d \Gamma(\Upsilon)$ (Subsection~\ref{subsec:affine-setting}), and so is $\mathcal E_\Upsilon$ by \eqref{eq:sd-core}. For later use, \eqref{eq:sd-relbound} also yields a constant $c<\infty$ such that, for all $g\in D(\Upsilon^{-1/2})$,
	\begin{equation}
		\|(\mathbf 1\otimes\phi(g))\zeta\|\leq c\,\big(\|g\|+\|\Upsilon^{-1/2}g\|\big)\big(\|H\zeta\|+\|\zeta\|\big),\qquad\zeta\in D(H).
		\label{eq:sd-Hbound}
	\end{equation}

	\emph{Step 2: Heisenberg equation.} Let $\mathcal V'\doteq\{(f,Cf):f\in D(\Upsilon)\cap D(\Upsilon^{-1/2})\}$. It is a dense real subspace of $\Hs_{\Ar}$, invariant under $U_t$, and, for $\xi=(g,Cg)\in\mathcal V'$, $iK\xi=(i\Upsilon g,C(i\Upsilon g))$ with $\Upsilon g\in D(\Upsilon^{-1/2})$. We claim that, for $\xi\in\mathcal V'$ and $\zeta,\zeta'\in D(H)$,
	\begin{equation}
		i\big\langle H\zeta',(\mathbf 1\otimes\Phi(\xi))\zeta\big\rangle-i\big\langle(\mathbf 1\otimes\Phi(\xi))\zeta',H\zeta\big\rangle=\big\langle\zeta',\big(\mathbf 1\otimes\Phi(iK\xi)+\Theta(\xi)\otimes\mathbf 1\big)\zeta\big\rangle .
		\label{eq:sd-commutator}
	\end{equation}
	By \eqref{eq:sd-Hbound}, both sides are continuous for the graph norm of $H$. It thus suffices to prove \eqref{eq:sd-commutator} for $\zeta,\zeta'$ in the core $\mathbb C^{n}\otimes\mathcal D_0^{\Upsilon}$. There, \eqref{eq:sd-commutator} follows from the CCR on $\mathcal D_0^\Upsilon$ once we take care of the following domain issue: $\mathrm d\Gamma(\Upsilon)$ maps $\mathcal D_0^{\Upsilon}$ into $\mathcal D_0$, $\mathcal D_0^{\Upsilon}$ is invariant under $a^{*}(g)$ and $a(g)$, for all $g\in D(\Upsilon)$, while $\mathcal D_0$ is invariant under $\phi(g)$ and the $\phi(w_l)$. Consequently, $(\mathbf 1\otimes\Phi(\xi))\zeta\in D(H)$ and $H\zeta\in D(\mathbf 1\otimes\Phi(\xi))$, so that the left-hand side of \eqref{eq:sd-commutator} equals $\langle\zeta',i[H,\mathbf 1\otimes\phi(g)]\zeta\rangle$.

	\emph{Step 3: integration.} Fix $t\in\mathbb R$, $\psi=(f,Cf)\in\mathcal V'$ and $\eta,\zeta\in D(H)$. For $s\in[0,t]$, set 
	\[\eta_s\doteq e^{-isH}\eta, \qquad \zeta_s\doteq e^{-isH}\zeta\quad \mbox{and}\quad X_s\doteq\mathbf 1\otimes\Phi(U_{t-s}\psi)=\mathbf 1\otimes\phi(u_{t-s}f) \]
	where $u_r\doteq e^{ir\Upsilon}$, $r\in\mathbb R$. Since $f\in D(\Upsilon)\cap D(\Upsilon^{-1/2})$ and $D(\Upsilon)\subseteq D(\Upsilon^{1/2})$, the map $s\mapsto u_{t-s}f$ is differentiable in the topology of the norm $\|\cdot\|+\|\Upsilon^{-1/2}\cdot\|$, with derivatives $-i\Upsilon u_{t-s}f$. As $g\mapsto\mathbf 1\otimes\phi(g)$ is real-linear on $D(H)$, \eqref{eq:sd-Hbound} implies that $s\mapsto X_s$ is differentiable in the norm of bounded operators from $D(H)$, endowed with the graph norm of $H$, to $\mathbb C^{n}\otimes\Fb$. Therefore, in $\mathcal B\big((D(H),\|\cdot\|_H),\mathbb C^{n}\otimes\Fb\big)$,
	\begin{equation}\label{eq:deriveXs}
	\partial_s X_s =-\mathbf 1\otimes\Phi(iKU_{t-s}\psi), \qquad s \in\mathbb R.
	\end{equation} Moreover, $s\mapsto\zeta_s$ is continuous in the graph norm of $H$ and differentiable in $\mathbb C^{n}\otimes\Fb$, with derivative $-iH\zeta_s$, and similarly for $s\mapsto\eta_s$. Writing $q(s)\doteq\langle\eta_s,X_s\zeta_s\rangle$ for all $s\in\mathbb R$, we compute that for all $h\in\mathbb R$,
	\[
	q(s+h)-q(s)=\langle\eta_{s+h}-\eta_s,X_{s+h}\zeta_{s+h}\rangle+\langle\eta_s,(X_{s+h}-X_s)\zeta_{s+h}\rangle+\langle X_s\eta_s,\zeta_{s+h}-\zeta_s\rangle
	\]
	Using the two derivatives above and \eqref{eq:deriveXs}, we see that $q$ is differentiable, with
	\[
	q'(s)=i\langle H\eta_s,X_s\zeta_s\rangle-i\langle X_s\eta_s,H\zeta_s\rangle-\big\langle\eta_s,(\mathbf 1\otimes\Phi(iKU_{t-s}\psi))\zeta_s\big\rangle=\big\langle\eta,\alpha_s\big(\Theta(U_{t-s}\psi)\otimes\mathbf 1\big)\zeta\big\rangle ,
	\]
	by \eqref{eq:sd-commutator} since $\eta_s,\zeta_s\in D(H)$. The right-hand side is continuous in $s$. Let
	\[
	\Delta_t(\psi)\doteq\int_0^{t}\alpha_s\big(\Theta(U_{t-s}\psi)\otimes\mathbf 1\big)\,\mathrm ds .
	\]
	Integrating between $0$ and $t$, and using $e^{-itH}\zeta\in D(H)\subseteq D(\mathbf 1\otimes\Phi(\psi))$, we obtain
	\[
	\big\langle\eta,e^{itH}(\mathbf 1\otimes\Phi(\psi))e^{-itH}\zeta\big\rangle=\big\langle\eta,\big(\mathbf 1\otimes\Phi(U_t\psi)+\Delta_t(\psi)\big)\zeta\big\rangle .
	\]
	As $D(H)$ is dense, the self-adjoint operator $e^{itH}(\mathbf 1\otimes\Phi(\psi))e^{-itH}$ extends the restriction of $\mathbf 1\otimes\Phi(U_t\psi)+\Delta_t(\psi)$ to $D(H)$. Now, $U_t\psi=(u_tf,Cu_tf)$ with $u_tf\in D(\Upsilon)$. Hence $\mathbb C^{n}\otimes\mathcal D_0^{\Upsilon}\subseteq D(H)$ is invariant under $\mathbf 1\otimes\Phi(U_t\psi)$ and consists of analytic vectors for it \cite[Thm.~X.41 and its proof]{RS2}. By Nelson's analytic vector theorem \cite[Thm.~X.39]{RS2}, it is a core for $\mathbf 1\otimes\Phi(U_t\psi)$, hence for $\mathbf 1\otimes\Phi(U_t\psi)+\Delta_t(\psi)$ since it is a bounded perturbation of the former. Taking closures, $e^{itH}(\mathbf 1\otimes\Phi(\psi))e^{-itH}$ extends the self-adjoint operator $\mathbf 1\otimes\Phi(U_t\psi)+\Delta_t(\psi)$, so both are equal. This is \eqref{eq:sd-D1} restricted to $\psi\in\mathcal V'$.

	\emph{Step 4: From $\mathcal V'$ to $\Hs_\Ar$, and (D2$_0$).} The map $\Delta_t$ defined in Step~3 for all $\psi\in\Hs_{\Ar}$ is real-linear and has self-adjoint values. Moreover, as $U_s$ is unitary, \[\|\Delta_t(\psi)\|\leq|t|\,\|\Theta\|_0\,\|\psi\|,\qquad \|\Theta\|_0\doteq\sup\{\|\Theta(\varphi)\|:\varphi\in\Hs_{\Ar},\ \|\varphi\|\leq1\}\leq\sqrt2\,c_\Theta<\infty .\]  Let $\psi\in\Hs_{\Ar}$, and let $\left(\psi_k\right)_{k\in\mathbb N}\subset\mathcal V'$ such that $\psi_k\to\psi$. Since $\xi\mapsto\piF(W(\xi))$ is strongly continuous (Subsection~\ref{subsec:affine-setting}), and $U$ is unitary,
	 \[e^{is\,\mathbf 1\otimes\Phi(U_t\psi_k)}=\piFn(W(sU_t\psi_k))\to e^{is\,\mathbf 1\otimes\Phi(U_t\psi)},\qquad s\in\mathbb R\]
	 in the strong sense. Since moreover $\Delta_t(\psi_k)\to\Delta_t(\psi)$ in norm, Lemma~\ref{lem:sd-bounded}(ii) and Step~3 yield
	\[
	\alpha_t\big(\piFn(W(\psi))\big)=\lim_{k\to\infty}e^{i(\mathbf 1\otimes\Phi(U_t\psi_k)+\Delta_t(\psi_k))}=e^{i(\mathbf 1\otimes\Phi(U_t\psi)+\Delta_t(\psi))},
	\]
	which is the Weyl form of (D1). Finally, for every vector $\zeta$,
	\begin{multline*}
	\big\|\big(t^{-1}\Delta_t(\psi)-\Theta(\psi)\otimes\mathbf 1\big)\zeta\big\| \\
	\leq\sup_{|s|\leq|t|}\Big(\|\Theta\|_0\,\|U_{t-s}\psi-\psi\|\,\|\zeta\|+\big\|\big(\alpha_s(\Theta(\psi)\otimes\mathbf 1)-\Theta(\psi)\otimes\mathbf 1\big)\zeta\big\|\Big).
	\end{multline*}
	 The right-hand side tends to $0$ as $t\to0$, which proves (D2$_0$).
\end{proof}

\begin{proof}[Proof of Theorem~\ref{thm:sd-fock}(ii)]
	Let $H$ $\Theta$-implement $(U_t)_{t\in\mathbb R}$ in $\piF$. By Proposition~\ref{prop:sd-universal}, $H\supseteq\widehat H_E$ for a unique $E\in M_n^{\mathrm{sa}}$, and $\widehat H_E=(H_{\Upsilon,\Theta}+E\otimes\mathbf 1)|_{\mathcal E_\Upsilon}$, since $w_l\in D(\Upsilon^{-1/2})$. By (i), $\mathcal E_\Upsilon$ is a core for the self-adjoint operator $H_{\Upsilon,\Theta}+E\otimes\mathbf 1$, so that $\widehat H_E$ is essentially self-adjoint, with closure $H_{\Upsilon,\Theta}+E\otimes\mathbf 1$. Hence $H=H_{\Upsilon,\Theta}+E\otimes\mathbf 1$ by Proposition~\ref{prop:sd-universal}. Uniqueness of $E$ follows by applying $P_0$ to this last equality tested on vectors $x\otimes\Omega$, since $P_0H_{\Upsilon,\Theta}(x\otimes\Omega)=0$.
\end{proof}

\Needspace{10\baselineskip}
\begin{corollary}[Uniqueness for ultraviolet-regular or commuting couplings]
	\label{cor:sd-uniqueness}
	Let $w_1,\dots,w_m\in\h$, and assume that either
	\begin{enumerate}
		\item[(a)] $w_l\in D(\Upsilon^{1/2})$ for $l=1,\dots,m$, or
		\item[(b)] $[\tau_l,\tau_k]=0$ for all $l,k=1,\dots,m$.
	\end{enumerate}
	Then, for every $E\in M_n^{\mathrm{sa}}$, $\widehat H_E$ is essentially self-adjoint on $\mathcal E_\Upsilon$. Consequently, every self-adjoint operator $H$ which $\Theta$-implements $(U_t)_{t\in\mathbb R}$ in $\piF$ is the closure of $\widehat H_E$, for a unique $E\in M_n^{\mathrm{sa}}$.
\end{corollary}

No infrared condition is imposed: case (a) covers infrared-singular form factors $w_l\in D(\Upsilon^{1/2})\setminus D(\Upsilon^{-1/2})$, for which $H_{\Upsilon,\Theta}$ is not defined by the Kato--Rellich theorem (for $n=1$, the corresponding van Hove Hamiltonians are not even bounded from below, by Remark~\ref{rem:hierarchy}). In case (b), the closure of $\widehat H_E$ $\Theta$-implements $(U_t)$ for every $E$ (Proposition~\ref{prop:sd-commuting}(i)), so that the $\Theta$-implementations are exactly these closures. In case (a), this corollary is a uniqueness statement only: whether the closure of $\widehat H_E$ $\Theta$-implements $(U_t)$ is not addressed here. By Lemma~\ref{lem:sd-commutant}, it suffices to check it for one $E$ (Remark~\ref{rem:sd-commutant}). Uniqueness remains open for non-commuting couplings with form factors in neither $D(\Upsilon^{1/2})$ nor $D(\Upsilon^{-1/2})$, there the spin couples the low and high-frequency modes, and the splitting used in \cite[Prop.~3.11]{Der03} does not apply directly.

\begin{proof}
	By Proposition~\ref{prop:sd-universal}, $H\supseteq\widehat H_E$ for a unique $E\in M_n^{\mathrm{sa}}$, and $H$ is the closure of $\widehat H_E$ as soon as $\widehat H_E$ is essentially self-adjoint. Let $\mathcal E^{(1)}_\Upsilon\doteq\mathrm{span}\{\epsilon(f):f\in D(\Upsilon)\}\subseteq\Fb$, so that $\mathcal E_\Upsilon=\mathbb C^{n}\otimes\mathcal E^{(1)}_\Upsilon$.

	(a) We apply Nelson's commutator theorem \cite[Thm.~X.37]{RS2}, with the comparison operator $N_0\doteq\mathbf 1\otimes(\mathrm d\Gamma(\mathbf 1+\Upsilon)+\mathbf 1)\geq\mathbf 1$, as in the proof of \cite[Prop.~3.11]{Der03}. By \eqref{eq:sd-core} with $\mathsf h=\mathbf 1+\Upsilon$, the space $\mathcal E_\Upsilon$ is contained in $D(N_0)$ and is a core for $N_0$, since $e^{isN_0}=e^{is}\,\mathbf 1\otimes\Gamma(e^{is(\mathbf 1+\Upsilon)})$ (\cite[Prop.~II.1.7]{EN00}). Since $\phi(w)$ is $N^{1/2}$-bounded for $w\in\h$ and $\mathbf 1\otimes\mathrm d\Gamma(\Upsilon)\leq N_0$, there is $c<\infty$ such that $\|\widehat H_E\eta\|\leq c\|N_0\eta\|$ for $\eta\in\mathcal E_\Upsilon$. The operators $E\otimes\mathbf 1$ and $\mathbf 1\otimes\mathrm d\Gamma(\Upsilon)$ commute with $N_0$, so they do not contribute to the commutator form between $\widehat H_E$ and $N_0$. Let $\Upsilon_1\doteq\mathbf 1+\Upsilon$, $w\in D(\Upsilon)$ and $\xi\in\mathcal E^{(1)}_\Upsilon$. A direct computation on exponential vectors, using $[\mathrm d\Gamma(\Upsilon_1),a^{*}(w)]=a^{*}(\Upsilon_1w)$ and $[\mathrm d\Gamma(\Upsilon_1),a(w)]=-a(\Upsilon_1w)$, gives
	\begin{multline*}
	\big|\langle\phi(w)\xi,\mathrm d\Gamma(\Upsilon_1)\xi\rangle-\langle\mathrm d\Gamma(\Upsilon_1)\xi,\phi(w)\xi\rangle\big|=\big|\langle\xi,(a(\Upsilon_1w)-a^{*}(\Upsilon_1w))\xi\rangle\big|\\ \leq2\|a(\Upsilon_1w)\xi\|\,\|\xi\|
	\leq2\|\Upsilon_1^{1/2}w\|\,\big\|(\mathrm d\Gamma(\Upsilon_1)+\mathbf 1)^{1/2}\xi\big\|^{2},
	\end{multline*}
	where we used $\|a(g)\xi\|\leq\|\Upsilon_1^{-1/2}g\|\,\|\mathrm d\Gamma(\Upsilon_1)^{1/2}\xi\|$ for any $g\in\h$ by \cite[Prop.~3.6(1)]{Der03}. Both sides are continuous in $w$ for the norm $\|\Upsilon_1^{1/2}\cdot\|$, so this bound extends to $w\in D(\Upsilon^{1/2})$. Let $\tau\in M_n^{\mathrm{sa}}$, with spectral decomposition $\tau=\sum_a\lambda_aQ_a$, and $\eta\in\mathcal E_\Upsilon$. Write $\eta=\sum_a(Q_a\otimes\mathbf 1)\eta$. The cross terms between different $a$ vanish, since $Q_aQ_b=0$ for $a\neq b$ and $Q_a\otimes\mathbf 1$ commutes with $N_0$. The previous bound, applied to the components of $(Q_a\otimes\mathbf 1)\eta$ in an orthonormal basis of $Q_a\mathbb C^{n}$, gives, via the spectral decomposition of $\tau$,
	\[
	\big|\langle(\tau\otimes\phi(w))\eta,N_0\eta\rangle-\langle N_0\eta,(\tau\otimes\phi(w))\eta\rangle\big|\leq2\|\tau\|\,\|\Upsilon_1^{1/2}w\|\,\|N_0^{1/2}\eta\|^{2}.
	\]
	Applying this bound to every $w_l$, and summing over $l$, the assumptions of \cite[Thm.~X.37]{RS2} hold, and $\widehat H_E$ is essentially self-adjoint on $\mathcal E_\Upsilon$.

	(b) Since the $\tau_l$ commute, there are mutually orthogonal projections $\Pi_1,\dots,\Pi_r$ with $\sum_j\Pi_j=\mathbf 1$, and real numbers $r_{jl}$, such that $\tau_l=\sum_jr_{jl}\Pi_j$. Let $\theta_j\doteq\sum_lr_{jl}w_l\in\h$. Since $w\mapsto\phi(w)$ is real-linear on $D(N^{1/2})$,
	\[
	\widehat H_E=E\otimes\mathbf 1+\sum_{j=1}^{r}\Pi_j\otimes\big(\mathrm d\Gamma(\Upsilon)+\phi(\theta_j)\big)\qquad\text{on }\mathcal E_\Upsilon=\bigoplus_{j=1}^{r}\Pi_j\mathbb C^{n}\otimes\mathcal E^{(1)}_\Upsilon .
	\]
	By Theorem~\ref{thm:affine}(iii), $A_j\doteq\mathrm d\Gamma(\Upsilon)+\phi(\theta_j)$ is essentially self-adjoint on $D(\mathrm d\Gamma(\mathbf 1+\Upsilon))$. There, the graph norm of $A_j$ is dominated by that of $\mathrm d\Gamma(\mathbf 1+\Upsilon)$, since $\phi(\theta_j)$ is $N^{1/2}$-bounded, and $\mathcal E^{(1)}_\Upsilon$ is a core for $\mathrm d\Gamma(\mathbf 1+\Upsilon)$, as in (a) (by \eqref{eq:sd-core}). Hence the closure of $A_j|_{\mathcal E^{(1)}_\Upsilon}$ extends the self-adjoint closure of $A_j|_{D(\mathrm d\Gamma(\mathbf 1+\Upsilon))}$; being symmetric, it is equal to it, so that $A_j$ is essentially self-adjoint on $\mathcal E^{(1)}_\Upsilon$. An orthogonal direct sum of essentially self-adjoint operators is essentially self-adjoint, and so is its sum with the bounded self-adjoint operator $E\otimes\mathbf 1$.
\end{proof}

\begin{example}[Spin--boson models]
	\label{ex:sd-SB}
	Let $n=2$, let $\sigma_x,\sigma_y,\sigma_z$ be the Pauli matrices (not to be confused with the symplectic form $\sigma$), $\sigma_\pm\doteq\frac12(\sigma_x\pm i\sigma_y)$, $\varepsilon\in\mathbb R$, and $v\in D(\Upsilon^{-1/2})$. Theorem~\ref{thm:sd-fock} covers:
	\begin{itemize}
		\item the standard spin--boson model $\frac\varepsilon2\sigma_z\otimes\mathbf 1+\mathbf 1\otimes\mathrm d\Gamma(\Upsilon)+\sigma_x\otimes\phi(v)$, with $m=1$, $\tau_1=\sigma_x$, $w_1=v$, $E=\frac\varepsilon2\sigma_z$, and $\Theta((f,Cf))=2\,\mathrm{Im}\langle f,v\rangle\,\sigma_x$;
		\item the rotating-wave model $\frac\varepsilon2\sigma_z\otimes\mathbf 1+\mathbf 1\otimes\mathrm d\Gamma(\Upsilon)+\sigma_+\otimes a(v)+\sigma_-\otimes a^{*}(v)$, since $\sigma_+\otimes a(v)+\sigma_-\otimes a^{*}(v)=\frac12\sigma_x\otimes\phi(v)-\frac12\sigma_y\otimes\phi(iv)$ on $\mathbb C^{2}\otimes\mathcal D_0$. Here $m=2$, $(\tau_1,w_1)=(\sigma_x,v/2)$ and $(\tau_2,w_2)=(-\sigma_y,iv/2)$.
	\end{itemize}
	In both cases, by Theorem~\ref{thm:sd-fock}, the model is, up to the spin Hamiltonian $E$, the only self-adjoint operator that $\Theta$-implements the free flow $(U_t)$ in $\piF$. For $v\neq0$, the range $\Theta(\Hs_{\Ar})$ is $\mathbb R\sigma_x$ in the first case, and $\mathrm{span}_{\mathbb R}\{\sigma_x,\sigma_y\}$ in the second.
\end{example}

\begin{remark}[Spin-valued displacements]
	\label{rem:sd-nogo}
	A natural attempt to extend Section~\ref{sec:affine} is to replace the scalar displacements in \eqref{eq:sd-scalar-fields} by matrix-valued ones. This means looking for dynamics of the form
	\[
	e^{itH}\,\piFn(W(\psi))\,e^{-itH}=e^{iG_t(\psi)\otimes\mathbf 1}\,\piFn(W(U_t\psi)),\qquad G_t(\psi)\in M_n^{\mathrm{sa}},
	\]
	whose composition is to be computed by a (possibly truncated) Baker--Campbell--Hausdorff formula. In the present language, and assuming $G_t$ real-linear with $t^{-1}G_t(\psi)\to\Theta(\psi)$, these are the spin-dressed implementations with spin-valued displacements, $\Delta_t(\Hs_{\Ar})\subseteq M_n\otimes\mathbf 1$ for all $t$, and then $\Delta_t(\psi)=G_t(\psi)\otimes\mathbf 1$. Since $\chi_{U_t}$ is an automorphism, $(\piFn\circ\chi_{U_t})(\Wa)'=\piFn(\Wa)'=M_n\otimes\mathbf 1$. Lemma~\ref{lem:sd-displaced}(ii), applied to $\varrho=\piFn\circ\chi_{U_t}$, then shows that this forces $[\Delta_t(\psi),\Delta_t(\psi')]=0$ for all $\psi,\psi'\in\Hs_{\Ar}$ and $t\in\mathbb R$. Dividing by $t^{2}$ and letting $t\to0$ gives, by (D2$_0$),
	\[
	[\Theta(\psi),\Theta(\psi')]=0,\qquad\psi,\psi'\in\Hs_{\Ar},
	\]
	i.e., the coupling space $\mathfrak c_\Theta\doteq\Theta(\Hs_{\Ar})\subseteq M_n^{\mathrm{sa}}$ is commutative. This makes precise the heuristic that a truncated Baker--Campbell--Hausdorff product of matrix-valued Weyl operators is associative only if the matrix coefficients commute. Infinitesimally, when the displacements commute with the fields, the compensation identity \eqref{eq:sd-compensation} reduces to $[\Delta(\psi),\Delta(\psi')]=0$. This is why Definition~\ref{def:sd-impl} allows displacements in $\mathcal B(\mathbb C^{n}\otimes\Kc_\pi)$ rather than in $M_n\otimes\mathbf 1$.

	Conversely, for regular form factors $w_l\in D(\Upsilon^{-1/2})$, assume that $\mathfrak c_\Theta$ is commutative and that $[E,\mathfrak c_\Theta]=0$. Let $\Pi_1,\dots,\Pi_r$ be the minimal projections of the commutative $C^{*}$-algebra generated by $\mathbf 1$ and $\mathfrak c_\Theta$. For $j=1,\dots,r$, let
	\[
	\theta_j\doteq\sum_{l=1}^{m}r_{jl}\,w_l\in D(\Upsilon^{-1/2}),\qquad r_{jl}\doteq\frac{\operatorname{tr}(\Pi_j\tau_l)}{\operatorname{tr}\Pi_j}\in\mathbb R .
	\]
	Since $\Theta(\psi)$ lies in the real span of the $\Pi_j$, one has $\Theta(\psi)=\sum_j\sigma(\psi,(\theta_j,C\theta_j))\Pi_j$. Since $H_{\Upsilon,\Theta}$ depends only on $\Theta$ (see after~\eqref{eq:sd-Hhat}), $\sum_l\tau_l\otimes\phi(w_l)=\sum_j\Pi_j\otimes\phi(\theta_j)$ on $\mathbb C^{n}\otimes D(\mathrm d\Gamma(\Upsilon))$. Moreover, as $\sum_j\Pi_j=\mathbf 1$ and $E$ commutes with every $\Pi_j$, $E=\sum_j\Pi_jE\Pi_j$. Therefore,
	\[
	H_{\Upsilon,\Theta}+E\otimes\mathbf 1=\sum_{j=1}^{r}\Big(\Pi_jE\Pi_j\otimes\mathbf 1+\Pi_j\otimes H_{\Upsilon,\theta_j}\Big),
	\]
	with $H_{\Upsilon,\theta_j}=\mathrm d\Gamma(\Upsilon)+\phi(\theta_j)$, self-adjoint on $D(\mathrm d\Gamma(\Upsilon))$ by Theorem~\ref{thm:sd-fock}(i) with $n=1$; by Theorem~\ref{thm:affine}(iii), this is the operator $H_{\Upsilon,\theta_j}$ of Theorem~\ref{thm:affine}, for a suitable additive constant. Since $H_{\Upsilon,\Theta}+E\otimes\mathbf 1$ $\Theta$-implements $(U_t)_{t\in\mathbb R}$ in $\piF$ (Theorem~\ref{thm:sd-fock}(i)) and commutes with every $\Pi_j\otimes\mathbf 1$, \eqref{eq:sd-langevin} gives the spin-valued displacements $\Delta_t(\psi)=\sum_j\sigma(\psi,\nu^{j}_t)\Pi_j\otimes\mathbf 1$, with $\nu^{j}_t$ given by \eqref{eq:affine-data} for $\theta_j$. This is the pure-dephasing case, i.e., a direct sum of van Hove dynamics over the joint spectrum of the coupling matrices.

\end{remark}

\subsection{Singular Form Factors and Commuting Couplings}
\label{subsec:sd-renorm}

In this subsection, $\pi=\piF$, and $\Upsilon\geq0$ with $\Ker\Upsilon=\{0\}$ and $U_t=e^{itK}$ are as in Subsection~\ref{subsec:sd-fock}, but the form factors are only assumed to lie in $\h_{-1}(\Upsilon)$, the most singular case allowed at the scale $L=\Upsilon$, by Theorem~\ref{thm:sd-structure}(iii). The vacuum is then in general not in the domain of the Hamiltonian, as it is already the case for $n=1$ by Proposition~\ref{prop:hierarchy}(i), and Propositions~\ref{prop:sd-vacuum} and~\ref{prop:sd-universal} are not available. We show that the $\Theta$-implementations at the scale $\Upsilon$ nevertheless include the renormalized Hamiltonians with commuting couplings and an arbitrary spin Hamiltonian.

The formal sum $\sum_j\Pi_j\otimes\big(\mathrm d\Gamma(\Upsilon)+\phi(\theta_j)\big)+E\otimes\mathbf 1$ has in general no meaning as an operator when $\theta_j\notin\h$, but its dynamics does. For commuting couplings and without $E$, the Hamiltonian is a block-diagonal sum of van Hove Hamiltonians, which are implementations at the scale $\Upsilon$ by Corollary~\ref{cor:sd-scalar-scale}. Although $E$ may not commute with the coupling, $E\otimes\mathbf 1$ commutes with the Weyl operators, and Lemma~\ref{lem:sd-commutant} below shows that adding it make the displacements deviate at order $2$ in time, so that the coupling map is unchanged.

Under an infrared condition (Proposition~\ref{prop:sd-commuting}(ii)), the model is renormalized in the sense of \cite{HLVM25}: the form factors are cut off at frequency $\Lambda$, a counterterm $e_\Lambda$, a real combination of the $\Pi_j$, is added, and one takes the limit $\Lambda\to\infty$. The ultraviolet part $\Upsilon q_j$ of $\theta_j$ is a coboundary, and a Weyl dressing removes it, which allows for the limit. In general the counterterm diverges (Remark~\ref{rem:sd-dressing}).
\Needspace{10\baselineskip}
\begin{lemma}[Perturbations by the commutant]
	\label{lem:sd-commutant}
	Let $\pi$ be a regular representation of $\Wa$, and let $H$ $\Theta$-implement $(U_t)_{t\in\mathbb R}$ in $\pi$ at scale $L$, with locally bounded displacements, i.e., \[\sup_{|t|\leq T}\|\Delta_t\|<\infty , \qquad T>0.\] Let $B$ be a self-adjoint element of $\pi^{(n)}(\Wa)'$. Then $H+B$ $\Theta$-implements $(U_t)_{t\in\mathbb R}$ in $\pi$ at scale $L$, and its displacements $\Delta^{B}_t$ satisfy
	\begin{equation}
		\|\Delta^{B}_t(\psi)-\Delta_t(\psi)\|\leq2|t|\,\|B\|\sup_{|s|\leq|t|}\|\Delta_s(\psi)\|,\qquad t\in\mathbb R,\ \psi\in\Hs_{\Ar};
		\label{eq:sd-commutant}
	\end{equation}
	in particular, they are locally bounded. For $\pi=\piF$, this applies in particular to $B=E\otimes\mathbf 1$, for every $E\in M_n^{\mathrm{sa}}$.
\end{lemma}

\begin{proof}
	\emph{Step 1:} We compare $e^{is(H+B)}$ with $e^{isH}$ through $G_s$ below, and show that $\|G_te^{irY_\psi}G_t^{*}-e^{irY_\psi}\|\leq|r|\,\ell_t(\psi)$ for $r\in\mathbb R$, where
	\[
	\ell_t(\psi)\doteq2|t|\,\|B\|\sup_{|s|\leq|t|}\|\Delta_s(\psi)\|<\infty .
	\]
	 Let $s\in\mathbb R$ and $\varphi\in\Hs_{\Ar}$, write $\alpha_s\doteq\mathrm{Ad}\,e^{isH}$ and $Y_\varphi\doteq\mathbf 1\otimes\Phi_\pi(\varphi)$. Fix $t\in\mathbb R$ and $\psi\in\Hs_{\Ar}$. Let $G_s\doteq e^{-isH}e^{is(H+B)}$, $s\in\mathbb R$. A direct computation on $D(H+B)=D(H)$, followed by a density argument, shows that $s\mapsto G_s$ and $s\mapsto G_s^{*}$ are strongly differentiable, with derivatives $i\alpha_{-s}(B)G_s$ and $-iG_s^{*}\alpha_{-s}(B)$ respectively.

	Next we conjugate by $e^{irY_\psi}$: for $r\in\mathbb R$, in the same way, $G^{(r)}_s\doteq e^{irY_\psi}G_se^{-irY_\psi}$ and its adjoint are differentiable with derivative $iB^{(r)}_sG^{(r)}_s$ and $-i{G_s^{(r)}}^*B^{(r)}_s$  respectively, where $B^{(r)}_s\doteq e^{irY_\psi}\alpha_{-s}(B)e^{-irY_\psi}$. Differentiating $s\mapsto G^{(r)*}_sG_s$, we obtain
	\[
	\big\|G_te^{irY_\psi}G_t^{*}-e^{irY_\psi}\big\|=\big\|G^{(r)*}_tG_t-\mathbf 1\big\|\leq|t|\sup_{|s|\leq|t|}\big\|B^{(r)}_s-\alpha_{-s}(B)\big\|,\qquad t\in\mathbb R .
	\]
	It remains to estimate $B^{(r)}_s-\alpha_{-s}(B)$. For any $s\in\mathbb R$, we can rewrite
	\[
	B^{(r)}_s=\alpha_{-s}\big(\alpha_s(e^{irY_\psi})\,B\,\alpha_s(e^{irY_\psi})^{*}\big),
	\]
	and, by \eqref{eq:sd-D1-weyl} and the real-linearity of $\Delta_s$, $\alpha_s(e^{irY_\psi})=X\,e^{irY_{U_s\psi}}$, where
	\[
	X\doteq e^{ir(Y_{U_s\psi}+\Delta_s(\psi))}e^{-irY_{U_s\psi}}
	\]
	satisfies $\|X-\mathbf 1\|\leq|r|\,\|\Delta_s(\psi)\|$, by Lemma~\ref{lem:sd-bounded}(i). Since for all $s\in\mathbb R$, $e^{irY_{U_s\psi}}=\pi^{(n)}(W(rU_s\psi))$ commutes with $B$, we get \[\|B^{(r)}_s-\alpha_{-s}(B)\|=\|XBX^{*}-B\|\leq2|r|\,\|B\|\,\|\Delta_s(\psi)\|,\quad s\in\mathbb R.\]
	 Therefore, 
	\[
	\big\|G_te^{irY_\psi}G_t^{*}-e^{irY_\psi}\big\|\leq 2|r||t|\,\|B\|\sup_{|s|\leq|t|}\|\Delta_s(\psi)\|=|r|\,\ell_t(\psi),\qquad r\in\mathbb R .
	\]
	\emph{Step 2: the displacement of $H+B$.} Let $t\in\mathbb R$. Let $A\doteq G_tY_\psi G_t^{*}$, so that $e^{irA}=G_te^{irY_\psi}G_t^{*}$, $r\in\mathbb R$. For $\zeta\in D(Y_\psi)$ and $r\neq0$, this bound gives\[|r|^{-1}\|(e^{irA}-\mathbf 1)\zeta\|\leq\ell_t(\psi)\|\zeta\|+\|Y_\psi\zeta\|,\] so that $\zeta\in D(A)$, by Fatou's lemma in the spectral representation of $A$ (as in the proof of Proposition~\ref{prop:sd-vacuum}(ii)). By symmetry, $D(A)=D(Y_\psi)$, and, on this domain, $A-Y_\psi$ is symmetric with norm at most $\ell_t(\psi)$. Hence $G_tY_\psi G_t^{*}=Y_\psi+R_t(\psi)$ for a bounded self-adjoint operator $R_t(\psi)$ with $\|R_t(\psi)\|\leq\ell_t(\psi)$. Since $e^{it(H+B)}=e^{itH}G_t$, \eqref{eq:sd-D1} for $H$ gives
	\[
	e^{it(H+B)}\,Y_\psi\,e^{-it(H+B)}=Y_{U_t\psi}+\Delta^{B}_t(\psi),\qquad \Delta^{B}_t(\psi)\doteq\Delta_t(\psi)+\alpha_t\big(R_t(\psi)\big),
	\]
	hence \eqref{eq:sd-commutant}.

	\emph{Step 3: (D1) and (D2$_L$) for $H+B$.} Let $t\in\mathbb R$. First, $\Delta^{B}_t$ is real-linear: conjugating by $e^{it(H+B)}$ the relations $Y_\psi+Y_\varphi\subseteq Y_{\psi+\varphi}$, whose left-hand side is densely defined (Remark~\ref{rem:selfdual-ccr}), and $Y_{s\psi}=sY_\psi$, $\varphi\in\Hs_{\Ar}$, $s\in\mathbb R$ \cite[Theorem~5(1)]{DerezinskiCCR}, gives, since the displacements are bounded, $\Delta^{B}_t(\psi+\varphi)=\Delta^{B}_t(\psi)+\Delta^{B}_t(\varphi)$ and $\Delta^{B}_t(s\psi)=s\Delta^{B}_t(\psi)$.

	Second, $\Delta^{B}_t$ is bounded, by \eqref{eq:sd-commutant} and the local boundedness of the $\Delta_s$. Hence (D1) holds for $H+B$, by the discussion following \eqref{eq:sd-D1-weyl}. Finally, for $\psi\in\mathcal V_L$ and $|t|\leq1$, (D2$_L$) for $H$ and \eqref{eq:sd-commutant} give $\|\Delta^{B}_t(\psi)-\Delta_t(\psi)\|\leq2\|B\|\,C_\Delta\,t^{2}\,\|\Lambda_L\psi\|$, and (D2$_L$) for $H+B$ follows, with the same coupling map. For $\pi=\piF$, recall that $\piFn(\Wa)'=M_n\otimes\mathbf 1$.
\end{proof}

\begin{remark}[A converse of the comparison and uniqueness results]
	\label{rem:sd-commutant}
	Lemma~\ref{lem:sd-commutant} is a converse of Lemma~\ref{lem:sd-compare} of Subsection~\ref{subsec:sd-def}: by the latter, two $\Theta$-implementations of the same family $(U_t)_{t\in\mathbb R}$ differ, in the sense of forms on $D(H)\cap D(H')$, by an operator commuting with $\pi^{(n)}(W(\psi))$, $\psi\in\mathcal V_L$. By the former, every self-adjoint element of $\pi^{(n)}(\Wa)'$ can be added to a $\Theta$-implementation with locally bounded displacements without compromising it. In Fock space, $\piFn(\Wa)'=M_n\otimes\mathbf 1$, and Lemma~\ref{lem:sd-commutant} is also the converse of the uniqueness up to a spin Hamiltonian in Theorem~\ref{thm:sd-fock}(ii) and Corollary~\ref{cor:sd-uniqueness}. Indeed, as soon as one $\Theta$-implementation with locally bounded displacements exists, every $E\in M_n^{\mathrm{sa}}$ occurs. At the scale $L=0$, local boundedness follows from (D2$_0$) and \eqref{eq:sd-cocycle} if $\sup_{|t|\leq1}\|U_t\|<\infty$.
\end{remark}

The Hamiltonians with commuting couplings are defined as follows. Let $\Pi_1,\dots,\Pi_m\in M_n$ be mutually orthogonal non-zero projections with $\sum_{j}\Pi_j=\mathbf 1$, let $\theta_1,\dots,\theta_m\in\h_{-1}(\Upsilon)$, and let
\begin{equation}
	\Theta(\psi)\doteq\sum_{j=1}^{m}\sigma\big(\psi,(\theta_j,C\theta_j)\big)\,\Pi_j,\qquad\psi\in\mathcal V_\Upsilon,
	\label{eq:sd-comm-Theta}
\end{equation}
with the notation \eqref{eq:sd-sigma-ext}. Conversely, every coupling map of a $\Theta$-implementation of $(U_t)_{t\in\mathbb R}$ in $\piF$ at the scale $\Upsilon$ with commutative range is of this form. To see this, take for the $\Pi_j$ the minimal projections of the $C^{*}$-algebra generated by $\mathbf 1$ and the range $\Theta(\mathcal V_\Upsilon)$, as in Remark~\ref{rem:sd-nogo}, and apply Theorem~\ref{thm:sd-structure}(iii) to a basis of $M_n^{\mathrm{sa}}$ containing them. The form factors of the remaining basis elements vanish, by the uniqueness in Theorem~\ref{thm:sd-structure}(iii). For $E\in M_n^{\mathrm{sa}}$, let
\begin{equation}
	H_E\doteq\sum_{j=1}^{m}\Pi_j\otimes H_{\Upsilon,\theta_j}+E\otimes\mathbf 1 ,
	\label{eq:sd-HE}
\end{equation}
where $H_{\Upsilon,\theta_j}$ is a van Hove Hamiltonian with data $(\Upsilon,\theta_j)$ (see Theorem~\ref{thm:affine}(ii)). It is self-adjoint, and unique up to the replacement of $E$ by $E+\sum_jc_j\Pi_j$, $c_j\in\mathbb R$. If $\theta_j\in D(\Upsilon^{-1/2})$ for all $j$, then $H_{\Upsilon,\Theta}=\sum_j\Pi_j\otimes(\mathrm d\Gamma(\Upsilon)+\phi(\theta_j))$, and, by Theorem~\ref{thm:affine}(iii), $H_E$ is the operator $H_{\Upsilon,\Theta}+E\otimes\mathbf 1$ of Theorem~\ref{thm:sd-fock}, for a suitable choice of the additive constants.

\Needspace{10\baselineskip}
\begin{proposition}[Commuting singular couplings]
	\label{prop:sd-commuting}
	Let $\Theta$ be given by \eqref{eq:sd-comm-Theta}, and let $E\in M_n^{\mathrm{sa}}$.
	\begin{enumerate}
		\item[(i)] $H_E$ $\Theta$-implements $(U_t)_{t\in\mathbb R}$ in $\piF$ at the scale $\Upsilon$, with locally bounded displacements. If $\theta_j\in\Upsilon^{1/2}\h+\Upsilon\h$ for all $j$, then $H_E$ is bounded from below. If $\theta_j\in\h$ for all $j$, then $H_E$ $\Theta$-implements $(U_t)_{t\in\mathbb R}$ in $\piF$ at the scale $0$, and, for the normalization of Theorem~\ref{thm:affine}(iii), $H_E$ is the closure of the operator~\eqref{eq:sd-Hhat} with $(\tau_j,w_j)=(\Pi_j,\theta_j)$ for all $j\in\{1,\cdots , m\}$.
		\item[(ii)] Let $\kappa>0$ be such that $\mathbf 1_{[0,\kappa]}(\Upsilon)\theta_j\in D(\Upsilon^{-1/2})$ for all $j\in\{1,\cdots , m\}$, and let
		\[
		q_j\doteq\Upsilon^{-1}\mathbf 1_{(\kappa,\infty)}(\Upsilon)\theta_j\in\h,\quad\rho_j\doteq(iq_j,C(iq_j))\in\Hs_{\Ar},\quad\mathbb W\doteq\sum_{j=1}^{m}\Pi_j\otimes\piF\big(W(\rho_j)\big),
		\]
		and $H^{<}\doteq\mathbf 1\otimes\mathrm d\Gamma(\Upsilon)+\sum_j\Pi_j\otimes\phi(\mathbf 1_{[0,\kappa]}(\Upsilon)\theta_j)$ on $\mathbb C^{n}\otimes D(\mathrm d\Gamma(\Upsilon))$. Then, for a suitable choice of the additive constants in \eqref{eq:sd-HE},
		\begin{equation}
			H_E=\mathbb W\,H^{<}\,\mathbb W^{*}+E\otimes\mathbf 1 ,
			\label{eq:sd-dressed}
		\end{equation}
		and, up to the unitary equivalence $\h\cong L^{2}(\mathcal M,\mu)$ given by the spectral theorem, $H_E$ is the renormalized Hamiltonian of \cite[Theorem~2.5]{HLVM25} with spin Hamiltonian $E\otimes\mathbf 1$ and form factor $\sum_j\theta_j\Pi_j$. In particular, for $\Lambda>\kappa$, let $\Theta_\Lambda$ be given by \eqref{eq:sd-comm-Theta} with $\theta_j$ replaced by $\theta^{\Lambda}_j\doteq\mathbf 1_{[0,\Lambda]}(\Upsilon)\theta_j\in D(\Upsilon^{-1/2})$, and let $e_\Lambda\doteq\sum_j\|\Upsilon^{-1/2}\mathbf 1_{(\kappa,\Lambda]}(\Upsilon)\theta_j\|^{2}\,\Pi_j$. Then \[H_{\Upsilon,\Theta_\Lambda}+(E+e_\Lambda)\otimes\mathbf 1\to H_E\]
		 in the norm resolvent sense, as $\Lambda\to\infty$.
	\end{enumerate}
\end{proposition}

Part~(i) needs no infrared condition. In Part~(ii), the renormalized Hamiltonian and its approximation by cutoff Hamiltonians are due to \cite{HLVM25}, the point of Proposition~\ref{prop:sd-commuting} is that this operator is a $\Theta$-implementation at the scale $\Upsilon$. In Part~(ii), the additive constants in \eqref{eq:sd-HE} are those for which \eqref{eq:sd-dressed} holds, namely $H_{\Upsilon,\theta_j}=\piF(W(\rho_j))H^{<}_j\piF(W(\rho_j))^{*}$ (Step~1 of the proof).

\begin{proof}
	Part~(i) treats the blocks by Corollary~\ref{cor:sd-scalar-scale} and adds $E$ by Lemma~\ref{lem:sd-commutant}. Part~(ii) proves \eqref{eq:sd-dressed} by a conjugation, then checks the hypotheses of \cite[Theorem~2.5]{HLVM25}.

	(i) \emph{Step 1: the case without spin Hamiltonian.} Let $j\in\{1,\cdots , m\}$. Let $H^{(0)}\doteq\sum_j\Pi_j\otimes H_{\Upsilon,\theta_j}$, and let $\nu^{j}_t\doteq(g^{j}_t,Cg^{j}_t)$, $g^{j}_t\doteq f_t(\Upsilon)\theta_j\in\h$ (see \eqref{eq:ft} and \eqref{eq:affine-data}). By Corollary~\ref{cor:sd-scalar-scale} with $L=\Upsilon$, $H_{\Upsilon,\theta_j}$ $\sigma(\cdot,(\theta_j,C\theta_j))$-implements $(U_t)_{t\in\mathbb R}$ in $\piF$ at the scale $\Upsilon$, with displacements $\sigma(\cdot,\nu^{j}_t)\mathbf 1$. Since $e^{itH^{(0)}}=\sum_j\Pi_j\otimes e^{itH_{\Upsilon,\theta_j}}$, \eqref{eq:sd-D1-weyl} for each $H_{\Upsilon,\theta_j}$ gives
	\[
	e^{itH^{(0)}}\,\piFn(W(\psi))\,e^{-itH^{(0)}}=\sum_{j=1}^{m}\Pi_j\otimes e^{i(\Phi(U_t\psi)+\sigma(\psi,\nu^{j}_t))}=e^{i(\mathbf 1\otimes\Phi(U_t\psi)+\Delta^{(0)}_t(\psi))},
	\]
	for $t\in\mathbb R$ and $\psi\in\Hs_{\Ar}$, where 
	\[\Delta^{(0)}_t(\psi)\doteq\sum_{j}\sigma(\psi,\nu^{j}_t)\,\Pi_j\otimes\mathbf 1,\]
	 i.e., (D1) for $H^{(0)}$. As the $\Pi_j$ are mutually orthogonal, $\|\Delta^{(0)}_t(\psi)\|=\max_j|\sigma(\psi,\nu^{j}_t)|$, so that (D2$_\Upsilon$) for $H^{(0)}$ follows from (D2$_\Upsilon$) for each $H_{\Upsilon,\theta_j}$, and $\|\Delta^{(0)}_t\|\leq\sqrt2\,(|t|+2)\max_j\|\theta_j\|_{-1}$, since $|f_t(x)|(|x|+1)\leq|t|+2$. 

	\emph{Step 2: adding $E$.} Lemma~\ref{lem:sd-commutant} with $B=E\otimes\mathbf 1$ applies to $H_E=H^{(0)}+E\otimes\mathbf 1$ and yields (D1) and (D2$_\Upsilon$) for $H_E$, with locally bounded displacements.

	\emph{Step 3: lower bound and scale $0$.} If $\theta_j\in\Upsilon^{1/2}\h+\Upsilon\h$, then $H_{\Upsilon,\theta_j}$ is bounded from below \cite[Thm.~3.2(3)]{Der03} (Remark~\ref{rem:hierarchy}), and so is $H_E$. If $\theta_j\in\h$ for all $j\in\{1,\cdots , m\}$, then the same argument applies at the scale $0$, by Corollary~\ref{cor:sd-scalar}, and the identification of $H_E$ with the closure of \eqref{eq:sd-Hhat} is shown in the proof of Corollary~\ref{cor:sd-uniqueness}(b).

	(ii) \emph{Step 1: the dressing gives \eqref{eq:sd-dressed}.} Let $u_t\doteq e^{it\Upsilon}$ and $\theta^{<}_j\doteq\mathbf 1_{[0,\kappa]}(\Upsilon)\theta_j$. By the functional calculus on $\h_{-1}(\Upsilon)$ (Subsection~\ref{subsec:affine-ham}), $q_j\in\h$, since $x^{-1}\mathbf 1_{(\kappa,\infty)}(x)(x+1)\leq1+\kappa^{-1}$, and $\theta_j=\theta^{<}_j+\Upsilon q_j$ in $\h_{-1}(\Upsilon)$. By Theorem~\ref{thm:sd-fock}(i) with $n=1$ and Corollary~\ref{cor:sd-scalar}, $H^{<}_j\doteq\mathrm d\Gamma(\Upsilon)+\phi(\theta^{<}_j)$ is self-adjoint on $D(\mathrm d\Gamma(\Upsilon))$ and is an $N$-diagonal affine quadratic Hamiltonian with data $(\Upsilon,\theta^{<}_j)$. By Lemma~\ref{lem:type-S} and Theorem~\ref{thm:affine}(i), $e^{itH^{<}_j}\in\mathbb T\,\Gamma(u_t)\piF(W(\nu^{<,j}_t))$, where $\nu^{<,j}_t$ are the vectors~\eqref{eq:affine-data} associated with $\theta^{<}_j$. By \eqref{eq:bogotransfo} applied to $u_t^{*}$, $\piF(W(\rho_j))\Gamma(u_t)=\Gamma(u_t)\piF(W(U_{-t}\rho_j))$, so that, by \eqref{eq:weyl},
	\[
	\piF(W(\rho_j))\,e^{itH^{<}_j}\,\piF(W(\rho_j))^{*}\in\mathbb T\,\Gamma(u_t)\,\piF\big(W(\nu^{<,j}_t+U_{-t}\rho_j-\rho_j)\big),\qquad t\in\mathbb R .
	\]
	Now $U_{-t}\rho_j-\rho_j=(g,Cg)$ with $g=i(u_t^{*}-\mathbf 1)q_j=f_t(\Upsilon)\Upsilon q_j$ (see the proof of Lemma~\ref{lem:cocycle}(iii)), so that $\nu^{<,j}_t+U_{-t}\rho_j-\rho_j=\nu^{j}_t$. By Corollary~\ref{cor:implementation} and Theorem~\ref{thm:affine}(i), $\piF(W(\rho_j))H^{<}_j\piF(W(\rho_j))^{*}$ is thus an $N$-diagonal affine quadratic Hamiltonian with data $(\Upsilon,\theta_j)$, which we choose as $H_{\Upsilon,\theta_j}$ in \eqref{eq:sd-HE}. Since $H^{<}=\sum_j\Pi_j\otimes H^{<}_j$, this gives \eqref{eq:sd-dressed}. Note that $\piF(W(\rho_j))=\piF(W(\rho))^{*}$, with $\rho$ as in Proposition~\ref{prop:hierarchy}(ii) for the form factor $\Upsilon q_j$.

	\emph{Step 2: \cite[Theorem~2.5]{HLVM25} applies to $H_E$.} We first recall the result of \cite{HLVM25} that is used. By the spectral theorem, we may assume that $\h=L^{2}(\mathcal M,\mu)$, with $(\mathcal M,\mu)$ $\sigma$-finite, and that $\Upsilon$ is the multiplication by a measurable function $\omega>0$ (a.e., since $\Ker\Upsilon=\{0\}$). In this situation, the $\theta_j$ are then functions with $(1+\omega)^{-1}\theta_j\in L^{2}(\mathcal M,\mu)$. We use the notation of \cite[Sect.~2]{HLVM25} with spin space $\mathbb C^{n}$: $\mathfrak b_s\doteq L^{2}(\mathcal M,\omega^{-s}\mathrm d\mu;M_n)$, with the $M_n$-valued pairing $\langle F,G\rangle_{\mathfrak b_s}\doteq\int_{\mathcal M}F^{*}G\,\omega^{-s}\,\mathrm d\mu$, $F_{>}\doteq\mathbf 1_{\{\omega>\kappa\}}F$, where the infrared cutoff of \cite{HLVM25} is taken equal to the $\kappa$ of Proposition~\ref{prop:sd-commuting}(ii), and, for $F=\sum_jf_j\Pi_j\in\mathfrak b_0$, the field and the Weyl operator of \cite[(2.8)]{HLVM25} are \[\phi(F)=\sum_j\Pi_j\otimes\phi(f_j)\quad\mbox{and}\quad\mathrm W(F)=e^{i\phi(iF)}=\sum_j\Pi_j\otimes\piF(W((if_j,C(if_j)))),\] 
	respectively. In this notation, and in the case without $2$-nilpotent part ($V_{\mathsf N}=0$ in \cite{HLVM25}), \cite[Theorem~2.5]{HLVM25} reads as follows: For a bounded operator $\mathsf S$ on $\mathbb C^{n}\otimes\Fb$ and a measurable $M_n$-valued function $V$ on $\mathcal M$ with a decomposition $V=V_{\leq}+V_{\mathsf D}$, consider the assumption
\begin{itemize}
	\item[(HLV)] $\mathsf S$ is symmetric, $V_{\leq}\in\mathfrak b_1$, $V_{\mathsf D}\in\mathfrak b_2$, $V_{\mathsf D}(x)$ is normal for $\mu$-a.e.\ $x\in\mathcal M$, and $[V_{\mathsf D}(x),V_{\mathsf D}(y)]=0$ for $\mu\otimes\mu$-a.e.\ $(x,y)\in\mathcal M^{2}$.
\end{itemize}
Under (HLV), the operator
\[
H(\mathsf S,V)\doteq \mathsf S+\mathrm W(\omega^{-1}V_{\mathsf D})\big(\mathrm d\Gamma(\omega)+\phi(V_{\leq})\big)\mathrm W(\omega^{-1}V_{\mathsf D})^{*},
\]
with domain $\mathrm W(\omega^{-1}V_{\mathsf D})\,D(\mathrm d\Gamma(\omega))$, is self-adjoint and bounded from below. Compare with $H_{\mathsf{IBC},\lambda}(0)$ of \cite[(2.7)]{HLVM25}. Moreover, let $V_k=V_{k,\leq}+V_{k,\mathsf D}\in\mathfrak b_0\cap\mathfrak b_1$, $k\in\mathbb N$, be such that every pair $(\mathsf S,V_k)$ satisfies (HLV), $V_{k,\leq}\to V_{\leq}$ in $\mathfrak b_1$, and $V_{k,\mathsf D}\to V_{\mathsf D}$ in $\mathfrak b_2$. Then
\[
\mathsf S+\mathrm d\Gamma(\omega)+\phi(V_k)+\langle V_{k,>},V_{k,>}\rangle_{\mathfrak b_1}\longrightarrow H(\mathsf S,V)\qquad\text{as }k\to\infty,
\]
in the norm resolvent sense, where the regularized operators on the left-hand side are self-adjoint on $\mathbb C^{n}\otimes D(\mathrm d\Gamma(\omega))$. (In general, \cite[Theorem~2.5]{HLVM25} gives strong resolvent convergence, but here we have norm resolvent convergence since $V_{\mathsf N}=V_{k,\mathsf N}=0$ satisfies the condition of its last assertion with $s_{\mathsf N}=1$.)

	To apply this result, we verify (HLV) and the convergence hypotheses for the data below and for an approximating sequence. We take $\mathsf S\doteq E\otimes\mathbf 1$ and, with $\theta^{<}_j$ and $q_j$ as in Step~1,
	\[
	V\doteq\sum_{j=1}^{m}\theta_j\Pi_j,\qquad V_{\leq}\doteq\sum_{j=1}^{m}\theta^{<}_j\Pi_j\in\mathfrak b_0\cap\mathfrak b_1,\qquad V_{\mathsf D}\doteq V_{>}=\sum_{j=1}^{m}\omega q_j\,\Pi_j\in\mathfrak b_2 .
	\]
	Here $V_{\leq}\in\mathfrak b_0\cap\mathfrak b_1$, since $\theta^{<}_j\in\h$ and $\mathbf 1_{[0,\kappa]}(\Upsilon)\theta_j\in D(\Upsilon^{-1/2})$, and $V_{\mathsf D}\in\mathfrak b_2$, since $\omega^{-1}V_{\mathsf D}=\sum_jq_j\Pi_j$ with $q_j\in\h$. The values of $V_{\mathsf D}$ lie in the commutative algebra spanned by the self-adjoint projections $\Pi_j$, so that $(\mathsf S,V)$ satisfies (HLV). As $\omega^{-1}V_{\mathsf D}=\sum_jq_j\Pi_j$, one has $\mathrm W(\omega^{-1}V_{\mathsf D})=\mathbb W$ and $\mathrm d\Gamma(\omega)+\phi(V_{\leq})=H^{<}$, so that $H(\mathsf S,V)=H_E$, by \eqref{eq:sd-dressed}. Next, let $(\Lambda_k)_{k\in\mathbb N}$ be a sequence with $\Lambda_k>\kappa$ and $\Lambda_k\to\infty$, and recall the cutoff form factors $\theta^{\Lambda}_j=\mathbf 1_{[0,\Lambda]}(\Upsilon)\theta_j$, $\Lambda>\kappa$, of Proposition~\ref{prop:sd-commuting}(ii), i.e., $\theta^{\Lambda}_j=\mathbf 1_{\{\omega\leq\Lambda\}}\theta_j$ in the present representation. Let
	\[
	V_k\doteq\sum_{j=1}^{m}\theta^{\Lambda_k}_j\Pi_j,\qquad V_{k,\leq}\doteq V_{\leq},\qquad V_{k,\mathsf D}\doteq V_k-V_{\leq}=\mathbf 1_{\{\kappa<\omega\leq\Lambda_k\}}V,\qquad k\in\mathbb N .
	\]
	Then $V_k\in\mathfrak b_0\cap\mathfrak b_1$, every pair $(\mathsf S,V_k)$ satisfies (HLV) for the same reason as $(\mathsf S,V)$, and $V_{k,\mathsf D}\to V_{\mathsf D}$ in $\mathfrak b_2$ by dominated convergence. Moreover, $\mathrm d\Gamma(\omega)+\phi(V_k)=H_{\Upsilon,\Theta_{\Lambda_k}}$, by \eqref{eq:sd-HUT} with $(\tau_j,w_j)=(\Pi_j,\theta^{\Lambda_k}_j)$, and $\langle V_{k,D},V_{k,D}\rangle_{\mathfrak b_1}=e_{\Lambda_k}$, since $V_{k,D}^{*}V_{k,D}=\sum_j|\mathbf 1_{\{\kappa<\omega\leq\Lambda_k\}}\theta_j|^{2}\,\Pi_j$, the $\Pi_j$ being mutually orthogonal projections. Hence $H_{\Upsilon,\Theta_{\Lambda_k}}+(E+e_{\Lambda_k})\otimes\mathbf 1\to H_E$ in the norm resolvent sense. Since the sequence $(\Lambda_k)_{k\in\mathbb N}$ is arbitrary, this proves the last assertion of Proposition~\ref{prop:sd-commuting}(ii).
\end{proof}

\begin{example}[Renormalized spin--boson model]
	\label{ex:sd-SB-ren}
	Let $n=2$, $\varepsilon\in\mathbb R$, $v\in\h_{-1}(\Upsilon)$, $\Pi_\pm\doteq\frac12(\mathbf 1\pm\sigma_x)$ and $\theta_\pm\doteq\pm v$. Then \eqref{eq:sd-comm-Theta} is the coupling map $\Theta((f,Cf))=2\,\mathrm{Im}\langle f,v\rangle\,\sigma_x$ of the standard spin--boson model (Example~\ref{ex:sd-SB}). By Proposition~\ref{prop:sd-commuting}(i), the operator \[H^{\mathrm{ren}}_{\mathrm{SB}}\doteq\Pi_+\otimes H_{\Upsilon,v}+\Pi_-\otimes H_{\Upsilon,-v}+\frac\varepsilon2\sigma_z\otimes\mathbf 1\]
	 $\Theta$-implements $(U_t)_{t\in\mathbb R}$ in $\piF$ at the scale $\Upsilon$, for every tunnelling parameter $\varepsilon$, although $\sigma_z$ does not commute with the coupling. If $\mathbf 1_{[0,\kappa]}(\Upsilon)v\in D(\Upsilon^{-1/2})$, it is, by Proposition~\ref{prop:sd-commuting}(ii) and for a suitable normalization, the norm resolvent limit of $\frac\varepsilon2\sigma_z\otimes\mathbf 1+\mathbf 1\otimes\mathrm d\Gamma(\Upsilon)+\sigma_x\otimes\phi(\mathbf 1_{[0,\Lambda]}(\Upsilon)v)$ plus the scalar counterterm $\|\Upsilon^{-1/2}\mathbf 1_{(\kappa,\Lambda]}(\Upsilon)v\|^{2}$. For $\Upsilon$ the multiplication by $\omega$ on $L^{2}(\mathbb R^{d})$, the ultraviolet part $\omega^{-1}\mathbf 1_{\{\omega>\kappa\}}v\in L^{2}(\mathbb R^{d})$ of the condition $v\in\h_{-1}(\Upsilon)$ is sharp in the sense of \cite[Theorem~2.8]{HLVM25}. Moreover, Theorem~\ref{thm:sd-structure}(iv) yields the quantum Langevin equation \eqref{eq:sd-langevin} of the renormalized model, with the singular coupling map $\Theta$, and Proposition~\ref{prop:sd-asymptotic} yields its asymptotic Weyl operators on $\mathcal D_\pm$, which is dense under the assumptions of Lemma~\ref{lem:sd-density}.

	Let $P\doteq(-1)^{\mathrm d\Gamma(\mathbf 1)}=\Gamma(-\mathbf 1)$ be the parity operator. Since $P\phi(f)P=\phi(-f)$ for $f\in\h$, $PH_{\Upsilon,v}P$ is a van Hove Hamiltonian with data $(\Upsilon,-v)$, so that one may take $H_{\Upsilon,-v}=PH_{\Upsilon,v}P$, as in Theorem~\ref{thm:intro-ren}. This normalization is compatible with the additive constants chosen in Step~1 of the proof of Proposition~\ref{prop:sd-commuting}(ii): there, with $\theta_\pm=\pm v$, the dressing vectors are $\rho_\pm=(iq_\pm,C(iq_\pm))$ with $q_\pm=\pm\Upsilon^{-1}\mathbf 1_{(\kappa,\infty)}(\Upsilon)v$, and the infrared parts of $H_{\Upsilon,\pm v}$ are $H^{<}_\pm=\mathrm d\Gamma(\Upsilon)+\phi(\pm\mathbf 1_{[0,\kappa]}(\Upsilon)v)$. Thus $\rho_-=-\rho_+$, so that $\piF(W(\rho_-))=P\piF(W(\rho_+))P$ by \eqref{eq:bogotransfo} with $u=-\mathbf 1$, and $H^{<}_-=PH^{<}_+P$, since $P$ commutes with $\mathrm d\Gamma(\Upsilon)$. Hence $\piF(W(\rho_-))H^{<}_-\piF(W(\rho_-))^{*}=P\,\piF(W(\rho_+))H^{<}_+\piF(W(\rho_+))^{*}P$.
\end{example}

\begin{remark}[Dressing, coboundaries and non-commuting couplings]
	\label{rem:sd-dressing}
	The proof of Proposition~\ref{prop:sd-commuting}(ii) splits each form factor as $\theta_j=\theta^{<}_j+\Upsilon q_j$, according to $\h_{-1}(\Upsilon)=\h+\Upsilon\h$. In the language of Remark~\ref{rem:vanhove}, the ultraviolet part $\Upsilon q_j$ is a coboundary (Lemma~\ref{lem:cocycle}(iii)), and it is removed by the Weyl dressing $\mathbb W$, as the whole form factor is removed in Proposition~\ref{prop:hierarchy}(ii). Formally, \[\mathbb W(\mathbf 1\otimes\mathrm d\Gamma(\Upsilon))\mathbb W^{*}=\mathbf 1\otimes\mathrm d\Gamma(\Upsilon)+\sum_j^m\Pi_j\otimes\phi(\Upsilon q_j)+\sum_j^m\|\Upsilon^{1/2}q_j\|^{2}\,\Pi_j\otimes\mathbf 1,\] where $\|\Upsilon^{1/2}q_j\|^{2}$ is the limit of the coefficient of $\Pi_j$ in $e_\Lambda$: it is finite if and only if $q_j\in D(\Upsilon^{1/2})$, whereas the dressing only requires $q_j\in\h$. By \eqref{eq:segal-weyl}, $\mathrm{Ad}\,\mathbb W^{*}\circ\piFn$ is the displacement of $\piFn$ (Definition~\ref{def:sd-displaced}) by the static, spin-valued map $\psi\mapsto-\sum_j\sigma(\psi,\rho_j)\Pi_j\otimes\mathbf 1$. By Lemma~\ref{lem:sd-displaced}(ii), a static displacement of $\piFn$ with values in $\piFn(\Wa)'=M_n\otimes\mathbf 1$ must have commuting values, so that, formally, a spin-valued dressing can only remove a singular part of the coupling map with commutative range, as in Remark~\ref{rem:sd-nogo}. This is consistent with \cite{HLVM25}, where the dressing $\mathrm W(\omega^{-1}V_{\mathsf D})$ is used for normal interactions with pointwise commuting values, and an interior boundary condition for $2$-nilpotent ones, such as the rotating-wave model of Example~\ref{ex:sd-SB}.
\end{remark}

\begin{remark}[Open problems]
	\label{rem:sd-open}
	Let $\Theta$ be a coupling map at the scale $\Upsilon$ which does not extend to a bounded map on $\Hs_{\Ar}$, i.e., by Theorem~\ref{thm:sd-structure}(iii), $w_l\notin\h$ for some $l$, for a basis $(\tau_l)$ of $M_n^{\mathrm{sa}}$. \emph{Uniqueness:} are the $\Theta$-implementations of $(U_t)_{t\in\mathbb R}$ in $\piF$ at the scale $\Upsilon$ unique up to a spin Hamiltonian, at least for commuting couplings? The proof of Theorem~\ref{thm:sd-fock}(ii) does not apply, since the bound of Proposition~\ref{prop:sd-vacuum}(i) is not available and the vacuum need not be in the domain. \emph{Other singular couplings:} are the renormalized Hamiltonians of \cite[Theorem~2.5]{HLVM25} with $V_{\mathsf N}\neq0$, or with $V_{\leq}$ not commuting with $V_{\mathsf D}$, and those of \cite{ALLVM25}, with commuting but possibly non-normal coupling matrices, $\Theta$-implementations at the scale $\Upsilon$? The main difficulty is the boundedness of each $\Delta_t$ on the whole of $\Hs_{\Ar}$, required by (D1). Form factors beyond the threshold \cite{DM20a,FHVM26} lie outside the scale $\Upsilon$, by Theorem~\ref{thm:sd-structure}(iii).
\end{remark}

\appendix
\section{Cohomological Vocabulary and Dictionary}
\label{app:dictionary}

This appendix collects the vocabulary of group cohomology used in Section~\ref{sec:affine}, see also the summary in Subsection~\ref{subsec:affine-ham}, and relates the notation of \cite{Gui72} and \cite{Der03} to ours.

\subsection{Cocycles, Coboundaries and Multipliers}
\label{app:cohomology}

Let $G$ be a topological group acting continuously by automorphisms on an abelian topological group $M$. Here $M$ is either $\mathfrak{h}$, with $g\cdot w\doteq A(g)w$ for a strongly continuous unitary representation $A$, or $\mathbb T$ with the trivial action, written multiplicatively. Let $\mathsf C^{k}$ be the space of continuous maps $G^{k}\to M$, with $\mathsf C^{0}\doteq M$, and define \cite[Appendix B.1]{Gui72}
\begin{align*}
	\delta_0&:\mathsf C^{0}\to \mathsf C^{1}, & (\delta_0w)(g)&\doteq g\cdot w-w,\\
	\delta_1&:\mathsf C^{1}\to \mathsf C^{2}, & (\delta_1b)(g,g')&\doteq b(gg')-b(g)-g\cdot b(g'),\\
	\delta_2&:\mathsf C^{2}\to \mathsf C^{3}, & (\delta_2\mathsf m)(g,g',g'')&\doteq g\cdot\mathsf m(g',g'')-\mathsf m(gg',g'')+\mathsf m(g,g'g'')-\mathsf m(g,g').
\end{align*}
Then $\delta_1\delta_0=0$ and $\delta_2\delta_1=0$.
\begin{itemize}
	\item For $M=\h$: the \emph{$1$-cocycles} are $Z^{1}(G,A)\doteq\Ker\delta_1$, the \emph{$1$-coboundaries} are $B^{1}(G,A)\doteq\operatorname{Ran}\delta_0$, and $H^{1}(G,A)\doteq\Ker\delta_1/\operatorname{Ran}\delta_0$. For $G=\mathbb R$, these are the notions of Subsection~\ref{subsec:affine-ham}, see \eqref{eq:cocycle-std}.
	\item For $M=\mathbb T$: the \emph{multipliers} are the elements of $\Ker\delta_2$. A multiplier $\mathsf m$ is a \emph{coboundary} if $\mathsf m\in\operatorname{Ran}\delta_1$, i.e., $c(gg')=c(g)\,c(g')\,\mathsf m(g,g')$ for some $c\in \mathsf C^{1}$; we then write $\delta c=\mathsf m$. Such a $c$ is unique up to multiplication by an element of $\Ker\delta_1$, the group of continuous characters of $G$.
\end{itemize}

\subsection{The Notation of Guichardet}
\label{app:dico-Gui}

The symmetric Hilbert space $S\h$ of \cite{Gui72} is $\Fb$, and $\mathrm{EXP}\,x$ is the exponential vector $\epsilon(x)$ of \eqref{eq:defexpvector}. Note that the scalar product $(\cdot|\cdot)$ of \cite{Gui72} is linear in the first variable. The operator $U_{A,0,1}$ of \cite[Sect.~2.2]{Gui72} is $\Gamma(A)$, while the map $b\mapsto U_{1,b,1}$ is the Fock representation of the canonical commutation relations, \cite[Sect.~2.2 and App.~F]{Gui72}. Comparing \cite[(2.2)]{Gui72},
\[
U_{1,b,1}\epsilon(x)=e^{-\|b\|^{2}/2-(x|b)}\epsilon(x+b),\qquad b,x\in\h ,
\]
with the displacement operators
\[
\mathrm D_b\,\epsilon(x)\doteq e^{\overline{a^{*}(b)-a(b)}}\epsilon(x)=e^{-\|b\|^{2}/2-\langle b,x\rangle}\epsilon(x+b),\qquad b,x\in\h ,
\]
shows that $U_{1,b,1}=\mathrm D_b$. Hence $U_{A,b,c}=c\,\mathrm D_b\Gamma(A)$, by the definition of $U_{A,b,c}$ in \cite[Sect.~2.2]{Gui72}. Since $\piF(W(\nu))=\mathrm D_{ig}$ for $\nu=(g,Cg)$, by \eqref{eq:Weyl-exp}, Corollary~\ref{cor:implementation} gives
\begin{equation}
	c\,\Gamma(u)\,\piF\big(W(\nu)\big)=U_{u,\,iug,\,c},\qquad \nu=(g,Cg)\in\Hs_{\Ar},\ u\in\mathcal U(\h),\ c\in\mathbb T.
	\label{eq:dictionary}
\end{equation}
Thus the pair $(A,b)=(u,iug)$ is the rigid motion $x\mapsto ux+iug$ of $\h$, i.e., the phase-space displacement $U\nu$ of Lemma~\ref{lem:block} up to the factor $i$, and $c$ is the phase of the implementer. For $G=\mathbb R$ this is the notion of Subsection~\ref{subsec:affine-ham}: a continuous unitary representation of $G$ on $\Fb$ by operators of the form $U_{A,b,c}$ is a \emph{representation of type~(S)} \cite[Def.~3.3]{Gui72}.

\subsection{The Notation of Dereziński}
\label{app:dico-Der}

In \cite{Der03}, the one-particle space is our $\h$, the one-particle Hamiltonian $h$ is our $\Upsilon$, with $\Upsilon\geq0$ and $\mathrm{Ker}\,\Upsilon=\{0\}$ as assumed in \cite[Sect.~3.1]{Der03}, and the form factor $z$ is our $\theta$. The space of admissible $z$ in \cite[(3.13)]{Der03} is the extrapolation space $\h_{-1}(\Upsilon)$ of Subsection~\ref{subsec:affine-ham}. The classical dynamics
\[
w\mapsto e^{ith}w+(e^{ith}-1)h^{-1}z,\qquad w\in\mathfrak{h},\ t\in\mathbb R,
\]
of \cite[Sect.~3.1]{Der03} (denoted $\alpha_t$ in \cite{Der03}) is the rigid motion $(A_t,b_t)$ of Lemma~\ref{lem:type-S}. Note that the fields are normalized differently, so that $\piF(W((g,Cg)))$ is the Weyl operator $W(\sqrt2\,g)$ of \cite[(2.12)]{Der03}. The unitary $V(t)$ of \cite[Sect.~3.3]{Der03} is exactly $\Gamma(e^{it\Upsilon})\piF(W(\nu_t))$, with $\nu_t$ as in Theorem~\ref{thm:affine}, since $ig_t=(1-e^{-it\Upsilon})\Upsilon^{-1}\theta$. The dressing operator of \cite[Thm.~3.2(4)]{Der03} is $\piF(W(\rho))^{*}$, with $\rho$ as in Proposition~\ref{prop:hierarchy}(ii).

\section{Bounded Perturbations of Operators}
\label{app:bounded}

The following lemma gives Duhamel's formula and the continuity of $e^{i(Y+B)}$ in $Y$ and in the bounded perturbation $B$; it is used in Section~\ref{sec:spin}, for instance in Lemma~\ref{lem:sd-commutant}.

\begin{lemma}[Bounded perturbations]
	\label{lem:sd-bounded}
	Let $Y$ be self-adjoint and $B,B'$ be bounded self-adjoint operators on a Hilbert space.
	\begin{enumerate}
		\item[(i)] (Duhamel's formula.) In the strong sense,
		\begin{equation}
			e^{i(Y+B)}-e^{i(Y+B')}=i\int_0^{1}e^{is(Y+B)}\,(B-B')\,e^{i(1-s)(Y+B')}\,\mathrm ds .
			\label{eq:sd-duhamel}
		\end{equation}
		In particular, $\|e^{i(Y+B)}-e^{i(Y+B')}\|\leq\|B-B'\|$.
		\item[(ii)] Let $(Y_k)_{k\in\mathbb N}$ be self-adjoint operators such that $e^{isY_k}\to e^{isY}$ strongly for every $s\in\mathbb R$, and let $(B_k)_{k\in\mathbb N}$ be bounded self-adjoint operators such that $\sup_k\|B_k\|<\infty$ and $B_k\to B$ strongly. Then $e^{i(Y_k+B_k)}\to e^{i(Y+B)}$ strongly.
	\end{enumerate}
\end{lemma}

\begin{proof}
	(i) For $\zeta\in D(Y)$, the map $s\mapsto e^{is(Y+B)}e^{i(1-s)(Y+B')}\zeta$ is differentiable on $[0,1]$, with derivative $ie^{is(Y+B)}(B-B')e^{i(1-s)(Y+B')}\zeta$, since $D(Y+B)=D(Y+B')=D(Y)$. Integrate over $[0,1]$ and use the density of $D(Y)$.
	
	(ii) Iterating \eqref{eq:sd-duhamel} with $B'=0$ gives the Dyson expansion
	\[
	e^{i(Y+B)}=\sum_{N\geq0}i^{N}\int_{0\leq s_1\leq\cdots\leq s_N\leq1}e^{i(1-s_N)Y}B\,e^{i(s_N-s_{N-1})Y}B\cdots B\,e^{is_1Y}\,\mathrm ds_1\cdots\mathrm ds_N ,
	\]
	whose $N$-th term has norm at most $\|B\|^{N}/N!$. The remainder after $N$ iterations has norm at most $\|B\|^{N+1}/(N+1)!$. The same holds for $Y_k$ and $B_k$, uniformly in $k$, so that the tails of the series are small uniformly in $k$. Each term converges strongly as $k\to\infty$, by dominated convergence, since products of uniformly bounded, strongly convergent sequences of operators converge strongly.
\end{proof}

\noindent \textit{Acknowledgments:} This work was supported by the Grant PID2020-112948GB-I00 funded by MCIN/ AEI/10.13039/501100011033 and by "ERDF A way of making Europe" and benefited from the FAPESP grant 26/04204-0\\

\noindent \textit{Data availability statement: } Data sharing is not applicable to this article as no datasets were generated or analyzed during the current study.\\

\noindent \textit{Conflict of Interest Statement: }
The author declares that he has no conflicts of interest or competing financial interests that are relevant to the content of this article.

\end{document}